\documentclass[a4paper,11pt]{article}

\usepackage[a4paper,left=2.73cm,right=2.7cm,top=3cm,bottom=3.5cm]{geometry}

\usepackage{amsfonts,amsmath,amssymb,amsthm, mathtools, tensor, tabstackengine, mathrsfs}
\usepackage[usenames, dvipsnames]{color}
\usepackage{array}
\usepackage{multirow}
\usepackage{makecell}
\usepackage{yfonts}
\allowdisplaybreaks
\usepackage{physics}
\usepackage{bm, bbm}
\usepackage{color, colortbl,xcolor}
\definecolor{mygray}{gray}{0.90}

\usepackage[colorlinks=true,linktocpage=true,linkcolor=red,citecolor=blue]{hyperref}
\usepackage{graphicx}

\usepackage{comment}

\numberwithin{equation}{section}

\newcommand{\be}{\begin{eqnarray}}
\newcommand{\ee}{\end{eqnarray}}

\begin{document}

\begin{titlepage}

\thispagestyle{empty}

\begin{center}


{\Large \textbf{Open strings in type IIB AdS$_3$ flux vacua}}

\vspace{40pt}

{\large \bf \'Alvaro Arboleya}$^{(1,2)}$ \,  \large{,} \,  {\large \bf Adolfo Guarino}$^{(1,2)}$

\vspace{8pt}

\,\,\quad\quad {\large \bf Matteo Morittu}$^{(1,2)}$  \,  \large{,} \, {\large \bf Giuseppe Sudano}$^{(1,2,3,4)}$

\vspace{25pt}

{\normalsize  
${}^{1}$ Departamento de F\'isica, Universidad de Oviedo,\\
Avda. Federico Garc\'ia Lorca 18, 33007 Oviedo, Spain.}
\\[3mm]

{\normalsize  
${}^{2}$ Instituto Universitario de Ciencias y Tecnolog\'ias Espaciales de Asturias (ICTEA) \\
Calle de la Independencia 13, 33004 Oviedo, Spain.}
\\[3mm]

{\normalsize
${}^{3}$ Dipartimento di Fisica, Universit\`a di Roma ``Tor Vergata''\\
Via della Ricerca Scientifica 1, 00133 Rome, Italy.}
\\[3mm]

{\normalsize
${}^{4}$ INFN Sezione di Roma ``Tor Vergata''\\
Via della Ricerca Scientifica 1, 00133 Roma, Italy.}
\\[12mm]

\texttt{alvaroarbo@gmail.com}  ,  \,\,
\texttt{adolfo.guarino@uniovi.es} \,\,
\texttt{morittumatteo@uniovi.es} \,\,, \, \texttt{giuseppe.sudano@roma2.infn.it}

\vspace{20pt}

\vspace{20pt}
				
\abstract{The role of open strings is investigated in the context of AdS$_{3}$ flux vacua arising from type IIB orientifold reductions. On the one hand, contrary to expectations, the perturbative stability of certain recently found non-supersymmetric AdS$_{3}$ flux vacua is shown to be robust under fluctuations of the open-string moduli. On the other hand, a wealth of new perturbatively stable AdS$_{3}$ flux vacua appears as a consequence of the non-trivial dynamics of open-string moduli. We also comment on the issue of scale separation in these AdS$_{3}$ flux vacua.}

\end{center}

\end{titlepage}

\tableofcontents

\hrulefill
\vspace{10pt}

\baselineskip 4.55mm

\newpage

\section{Motivation}

String theory consists not only of strings but also D$p$-branes as fundamental constituents \cite{Polchinski:1995mt}. These are extended objects in spacetime on which open strings can end, and they play a central role in both the perturbative and non-perturbative regimes of the theory. In the perturbative regime where the string coupling is small, \textit{i.e.} $g_{s} \ll 1\,$, D$p$-branes become very massive objects (their tension $\sim 1/g_{s}$) and their dynamics is often ignored in many phenomenological applications. This is the case, for example, in most of the string compactification scenarios including fluxes and D-branes put forward in the literature. However, it was argued in \cite{Danielsson:2016mtx} that considering D-branes dynamical is crucial in order to make non-supersymmetric vacua decay, as expected from the AdS swampland conjecture \cite{Ooguri:2016pdq}. More concretely, ref.~\cite{Danielsson:2016mtx} argued that the anticipated instabilities are perturbative in nature and originate from the interplay between the closed- and open-string sectors of the theory. Further support for this proposal came from \cite{Danielsson:2017max}, where it was shown that open strings introduce modes\footnote{These open-string modes are identified with scalars that parameterise the D-brane position, and also with Wilson lines for the gauge fields on the D-brane wrapping the internal space.} with masses below the Breitenlohner--Freedman (BF) bound \cite{Breitenlohner:1982bm} for the non-supersymmetric and perturbatively stable $\,\textrm{AdS}_{7}\,$ \cite{Passias:2015gya,Danielsson:2017max} and $\,\textrm{AdS}_{4}\,$ \cite{Dibitetto:2011gm,Dibitetto:2012ia} vacua of massive IIA supergravity, thus destabilising them. Ref.~\cite{Balaguer:2024cyb} then showed that, apart from destabilising the AdS$_{4}$ vacua of \cite{Dibitetto:2011gm,Dibitetto:2012ia}, open-string modes can also generate new AdS$_{4}$ vacua which, on the contrary, are perturbatively stable. A decay channel for these new vacua, in light of the AdS swampland conjecture, has not been envisaged yet.

Motivated by the above results, we will investigate how open strings affect the stability of the non-supersymmetric and perturbatively stable AdS$_{3}$ vacua of type IIB supergravity recently found in \cite{Arboleya:2024vnp} (see \cite{Arboleya:2025ocb} for a reader-friendly version of \cite{Arboleya:2024vnp}). These AdS$_{3}$ vacua arise from type IIB orientifold reductions down to three dimensions (3D) that include a seven-dimensional twisted torus $\,\mathcal{M}_{7}\,$ as internal space, $F_{(3)}$ and $F_{(7)}$ background fluxes in the RR sector, and a \textit{single type} of spacetime-filling D$5$-branes (and O$5$-planes) in the smeared limit.\footnote{This setup lies within the class of G$_{2}$ orientifold reductions of type IIB supergravity investigated in \cite{Emelin:2021gzx}.} In order to capture the dynamics of open strings, we will adapt the procedure in \cite{Balaguer:2023jei} to 3D, and dimensionally reduce the (non-abelian) DBI action
\begin{equation} 
\label{D5_DBI_intro}
\begin{aligned}
S^{\textrm{DBI}}_{\textrm{D}5} & = -T_{\textrm{D}5} \displaystyle\int_{\textrm{WV}(\textrm{D}5)} \textrm{d}^6\sigma \; \textrm{Tr}\left( e^{-\Phi} \sqrt{-\det[\mathbb{M}] \det[\mathbb{Q}]} \; \right) \ , 
\end{aligned}
\end{equation}
for D$5$-branes embedded in a curved background with fluxes \cite{Myers:1999ps,Martucci:2005rb}. The action (\ref{D5_DBI_intro}) depends on the quantities $\,\mathbb{M}\,$ and $\,\mathbb{Q}\,$ which account for (matrix-valued) gauge fields and scalars living in the six-dimensional worldvolume of the D$5$-branes. Upon dimensional reduction to 3D, the DBI action (\ref{D5_DBI_intro}) gives rise to a set of scalar fields -- we will refer to them as open-string moduli -- which interact with the closed-string moduli arising from the reduction of the type IIB bulk action.\footnote{Together with the DBI action in (\ref{D5_DBI_intro}), there is also the topological WZ term (see eq.~(\ref{D5_DBI_WZ})). Its dimensional reduction introduces various subtleties and technicalities to be addressed in Section~\ref{sec:reduction_open-sector}.} The AdS$_{3}$ vacua of \cite{Arboleya:2024vnp,Arboleya:2025ocb} are then consistently recovered by setting the open-string moduli to zero. However, even when these moduli are trivial, they still contribute to the spectrum of scalar fluctuations. And these are precisely the perturbative instabilities predicted in \cite{Danielsson:2016mtx}. As we will show, and unlike for the non-supersymmetric AdS$_{7}$ and AdS$_{4}$ vacua analysed in \cite{Danielsson:2017max}, we find no perturbative instabilities arising from open strings in the non-supersymmetric AdS$_{3}$ vacua of \cite{Arboleya:2024vnp,Arboleya:2025ocb}. The identification of a decay channel for these vacua, as suggested by the AdS swampland conjecture, then remains an open question.

As in the AdS$_4$ analysis of \cite{Balaguer:2024cyb}, we will show that including open strings generates a wealth of new non-supersymmetric and perturbatively stable AdS$_{3}$ flux vacua which simply do not exist when the dynamics of open strings is ignored. In order to fully chart the landscape of AdS$_{3}$ flux vacua, we will exploit the fact that the 3D supergravity we obtained upon dimensional reduction has half-maximal $\,\mathcal{N}=8\,$ supersymmetry ($16$ supercharges). This was somehow to be expected, as our compactification scheme only contains a single type of coincident O$5$/D$5$ sources and these break half of the supersymmetries. We will demonstrate this by reformulating the three-dimensional supergravity obtained via dimensional reduction within the general framework for describing half-maximal gauged supergravities developed in \cite{Nicolai:2001ac,deWit:2003ja}. As anticipated in \cite{Danielsson:2016mtx}, the closed-string moduli arising from the reduction of the type IIB bulk action go into the 3D supergravity multiplet, whereas the open-string moduli arising from the reduction of the DBI action in (\ref{D5_DBI_intro}) go into matter multiplets.\footnote{In 3D, each matter multiplet contains $\,8\,$ real scalars to be identified as open-string moduli. These split as $\,4\,$ moduli parameterising the position of the D$5$-brane in its transverse space, $\,3\,$ moduli associated with the internal components of the gauge field living in the six-dimensional worldvolume of the D$5$-brane, and $\,1\,$ modulus that appears upon dualisation of the external components of such a gauge field.} Importantly, this is not merely a reformulation of the three-dimensional supergravity obtained via dimensional reduction. Recasting it in the framework of \cite{Nicolai:2001ac,deWit:2003ja} enables the application of specific supergravity techniques which, when combined with tools from algebraic geometry \cite{DGPS}, allow for a systematic classification of AdS$_3$ flux vacua. It will also allow us to determine the amount of supersymmetry preserved by a given AdS$_{3}$ vacuum, and to obtain the complete spectrum of scalar and gravitino fluctuations about such a vacuum within half-maximal supergravity.

The paper is organised as follows. In Section~\ref{sec:IIB_reduction} we perform the dimensional reduction of the type IIB bulk action as well as the DBI action and the topological WZ term for a stack of spacetime-filling D$5$-branes. For the sake of simplicity, we will impose an $\textrm{SO}(3)$ symmetry in the dimensional reduction and restrict to the set of $\textrm{SO}(3)$-invariant fields and fluxes. In this manner, we will obtain a simple 3D supergravity theory containing both closed-string and open-string moduli interacting with each other. In Section~\ref{sec:3D_supergravity} we will first recast the 3D supergravity obtained upon dimensional reduction in the form of a (half-maximal) $\,\mathcal{N}=8$, $\,D=3$ gauged supergravity \cite{Nicolai:2001ac,deWit:2003ja}. Then, making a combined use of supergravity and algebraic geometry techniques \cite{DGPS}, we will exhaustively chart the landscape of three-dimensional flux vacua and analyse their supersymmetry and perturbative stability. In Section~\ref{sec:flux_vacua_phenomenology} we investigate the phenomenon of scale separation between the volume of the seven-dimensional internal space and the characteristic size of the external spacetime in the set of AdS$_{3}$ flux vacua previously found, and identify particular examples that feature this property. In Section~\ref{sec:conclusions} we summarise the main results and discuss some issues and open questions for future work. Two appendices accompany the main text. The first one can be viewed as a crash course on 3D half-maximal gauged supergravity based on \cite{Nicolai:2001ac,deWit:2003ja,Deger:2019tem}. The second appendix contains various tables that collect the gravitino mass spectra for all the three-dimensional flux vacua presented in this work.

\section{Type IIB reduction with O$5$/D$5$ sources}
\label{sec:IIB_reduction}

As stated in the introduction, we want to analyse how the inclusion of open strings affects the non-supersymmetric and perturbatively stable type IIB AdS$_{3}$ flux vacua put forward in \cite{Arboleya:2024vnp}, and also their supersymmetric cousin of \cite{VanHemelryck:2025qok}. These flux vacua arise from the class of type IIB orientifold reductions investigated in \cite{Emelin:2021gzx}, which in turn fits into the class of Scherk--Schwarz (SS) reductions \cite{Scherk:1979zr}: the internal space is (locally) a seven-dimensional group manifold associated with a Lie group $\textrm{G}$. This manifold is often referred to in the literature as a \textit{twisted torus} (see \textit{e.g.} \cite{Hull:2005hk}), and we will denote it $\,\mathbb{T}_{\omega}^{7}\,$ in the rest of the work.

\subsection{O$5$-plane orientifold action}

We focus on the class of orientifold reductions of \cite{Arboleya:2024vnp}, which include just a single type of spacetime-filling O$5$-planes and D$5$-branes. For the sake of concreteness, we choose to place the sources as
\begin{equation}
\label{O5/D5_location}
\begin{array}{lll|lc|lc|lc|c}
x^{0} & x^{1} & x^{2} & y^{1}  & y^{2} & y^{3}  & y^{4}& y^{5} & y^{6} & y^{7}   \\
\hline
\times & \times & \times &  & \times &  & \times &  & \times & 
\end{array}   
\end{equation}
thus inducing a splitting of the internal space coordinates $\,y^{m}\,$ ($m=1,\dots,7$) of the form 
\begin{equation}
y^{m}=(y^{i},y^{\hat{a}})
\hspace{10mm} \textrm{ with } \hspace{10mm}
i=2,4,6
\hspace{8mm} , \hspace{8mm}
\hat{a}=1,3,5,7 \ .
\end{equation}
Following standard convention, we denote $\,x^{\mu}\,$ ($\mu=0,1,2$) the coordinates on the external 3D spacetime. The O$5$-plane orientifold action $\mathcal{O}_{\mathbb{Z}_{2}} = \Omega_{P} \, \sigma_{\textrm{O}5}$ combines a worldsheet orientation-reversal parity transformation $\,\Omega_{p}\,$ and an internal target space involution $\sigma_{\textrm{O}5}$ \cite{Koerber:2007hd}. The latter reflects the four coordinates in (\ref{O5/D5_location}) transverse to the O$5$-planes, namely,
\begin{equation}
\label{sigma_O5}
\sigma_{\textrm{O}5}: \hspace{5mm} 
x^{\mu} \rightarrow x^{\mu} 
\hspace{8mm} , \hspace{8mm} 
y^{i} \rightarrow y^{i} 
\hspace{8mm} , \hspace{8mm}
y^{\hat{a}} \rightarrow - y^{\hat{a}}  \ ,
\end{equation}
and so does it with the associated derivatives. As a result, the O$5$-planes (and D$5$-branes) extend along a total of six coordinates $\,X^{\mathsf{\bar M}}=(x^{\mu},y^{i})$, breaking the group of internal diffeomorphisms down to a subgroup $\,\textrm{GL}(3) \times \textrm{GL}(4) \subset \textrm{GL}(7)$. The various internal components of the fields (gauge potentials) of type IIB supergravity acquire an even ($+$) \textit{vs.} odd ($-$) character under the O$5$-plane orientifold action $\,\mathcal{O}_{\mathbb{Z}_{2}} = \Omega_{P} \, \sigma_{\textrm{O}5}\,$ summarised in Table~\ref{Table:O5_Fields}. The corresponding fluxes (field strengths) are obtained by applying exterior derivatives on the fields (gauge potentials). The set of $\mathcal{O}_{{\mathbb{Z}_{2}}}$-even fluxes is summarised in Table~\ref{Table:O5_Fluxes_even}. For further details on this, we refer the reader to Section~$3.5$ of \cite{Arboleya:2024vnp}.

\begin{table}[t]
\begin{center}
\renewcommand{\arraystretch}{1.7}
\begin{tabular}{|c|c|c|c|}
\hline
Fields & $\Omega_{P}$   &  $\sigma_{\textrm{O}5}$ & $\mathcal{O}_{\mathbb{Z}_{2}}$  \\
\hline
\hline
$e_{\hat{a}}{}^{\hat{b}} \in \frac{\textrm{GL}(4)}{\textrm{SO}(4)} \, , \, e_{i}{}^{j} \in \frac{\textrm{GL}(3)}{\textrm{SO}(3)}$ & \multirow{2}{*}{$+$}  & $+$  & $+$  \\
\cline{1-1}\cline{3-4}
$e_{\hat{a}}{}^{i}  \, , \, e_{i}{}^{\hat{a}}$ &  & $-$  & $-$ \\
\hline
\hline
$\Phi$ & $+$ & $+$  & $+$ \\
\hline
\hline
$B_{ij}\, , \, B_{\hat{a}\hat{b}}$ & \multirow{2}{*}{$-$}  & $+$  & $-$  \\
\cline{1-1}\cline{3-4}
$B_{i\hat{a}}$ &  & $-$  & $+$ \\
\hline
\hline
$B_{ij\hat{a}\hat{b}\hat{c}\hat{d}}$ & \multirow{2}{*}{$-$}  & $+$  & $-$  \\
\cline{1-1}\cline{3-4}
$B_{ijk\hat{a}\hat{b}\hat{c}}$ &  & $-$  & $+$ \\
\hline
\end{tabular}
\hspace{10mm}
\begin{tabular}{|c|c|c|c|}
\hline
Fields & $\Omega_{P}$   &  $\sigma_{\textrm{O}5}$ & $\mathcal{O}_{\mathbb{Z}_{2}}$  \\
\hline
\hline
$C_{\hat{a}\hat{b}\hat{c}\hat{d}} \, , \, C_{ij\hat{a}\hat{b}}$ & \multirow{2}{*}{$-$}  & $+$  & $-$  \\
\cline{1-1}\cline{3-4}
$C_{i\hat{a}\hat{b}\hat{c}} \, , \, C_{ijk\hat{a} }$ &  & $-$  & $+$ \\
\hline
\hline
$C_{(0)}$ & $-$ & $+$  & $-$ \\
\hline
\hline
$C_{ij}\, , \, C_{\hat{a}\hat{b}}$ & \multirow{2}{*}{$+$}  & $+$  & $+$  \\
\cline{1-1}\cline{3-4}
$C_{i\hat{a}}$ &  & $-$  & $-$ \\
\hline
\hline
$C_{ij\hat{a}\hat{b}\hat{c}\hat{d}}$ & \multirow{2}{*}{$+$}  & $+$  & $+$  \\
\cline{1-1}\cline{3-4}
$C_{ijk\hat{a}\hat{b}\hat{c}}$ &  & $-$  & $-$ \\
\hline
\end{tabular}
\caption{Even ($+$) \textit{vs.} odd ($-$) character of type~IIB fields under the O$5$-plane orientifold action $\mathcal{O}_{\mathbb{Z}_{2}} = \Omega_{P} \, \sigma_{\textrm{O}5}$. The $\mathcal{O}_{\mathbb{Z}_{2}}$-even fields add up to $\,64-3=61\,$ scalars. The missing three scalars in half-maximal $\,D=3\,$ supergravity are dual to the vectors $\,e_{\mu}{}^{i}$.} 
\label{Table:O5_Fields}
\end{center}
\end{table}

\begin{table}[t]
\begin{center}
\renewcommand{\arraystretch}{1.8}
\begin{tabular}{|c|c|}
\hline
Fluxes &  $\Theta$-irrep $\leftrightarrow$ Flux component \\
\hline
\hline
\multirow{3}{*}{$\omega$} & $\theta^{ij8}{}_{k}=\omega_{ij}{}^{k}$  \\
\cline{2-2}
 & $\theta^{\hat{a}\hat{b}ij} = \frac{1}{2!} \, \epsilon^{\hat{a}\hat{b}\hat{c}\hat{d}} \, \epsilon_{ijk}\, \omega_{\hat{c}\hat{d}}{}^{k}  $  \\
\cline{2-2}
 & $\theta^{i\hat{b}8}{}_{\hat{a}} = \omega_{\hat{a}i}{}^{\hat{b}}$ \\
\hline
\hline
\multirow{2}{*}{$H_{(3)}$} & $ \theta^{ijk\hat{d}} = -\frac{1}{3!} \, \epsilon_{ijk} \, \epsilon^{\hat{a}\hat{b}\hat{c}\hat{d}} \, H_{\hat{a}\hat{b}\hat{c}}$ \\
\cline{2-2}
 & $\theta^{ij8}{}_{\hat{c}} = -H_{ij\hat{c}}$ \\
\hline
\end{tabular}
\hspace{10mm}
\begin{tabular}{|c|c|}
\hline
Fluxes &   $\Theta$-irrep $\leftrightarrow$ Flux component   \\
\hline
\hline
$F_{(1)}$ & $\theta^{\hat{a}\hat{b}\hat{c}8} =  - \epsilon^{\hat{a}\hat{b}\hat{c}\hat{d}} \, F_{\hat{d}}$ \\
\hline
\hline
\multirow{2}{*}{$F_{(3)}$} & $\theta^{88}  = - \frac{1}{3!} \, \epsilon^{ijk} \, F_{ijk}  $  \\
\cline{2-2}
 & $\theta^{\hat{a}\hat{b}k8} = \frac{1}{2!} \, \epsilon^{\hat{a}\hat{b}\hat{c}\hat{d}} \, F_{\hat{c}\hat{d}k} $  \\
\hline
\hline
$F_{(5)}$ & $\theta^{\hat{d}ij8} = \frac{1}{3!} \, \epsilon^{\hat{a}\hat{b}\hat{c}\hat{d}} \, F_{\hat{a}\hat{b}\hat{c}ij}$ \\
\hline
\hline
$F_{(7)}$ & $ \theta^{ijk8} =  \frac{1}{4!} \, \epsilon^{\hat{a}\hat{b}\hat{c}\hat{d}} \, F_{\hat{a}\hat{b}\hat{c}\hat{d}ijk}$ \\
\hline
\end{tabular}
\caption{Type~IIB ($\mathcal{O}_{\mathbb{Z}_{2}}$-even) fluxes allowed by the orientifold action $\,\mathcal{O}_{\mathbb{Z}_{2}} = \Omega_{P} \, \sigma_{\textrm{O}5}\,$ and their identification with embedding tensor irreps of half-maximal $D=3$ supergravity. An internal $\,H_{(7)}\,$ flux is projected out by the orientifold action.} 
\label{Table:O5_Fluxes_even}
\end{center}
\end{table}

\subsection{$\textrm{SO}(3)$ symmetry}

In ref.~\cite{Arboleya:2024vnp}, an additional $\textrm{SO}(3)$ symmetry was imposed in the compactification scheme in order to obtain simple $\,\mathcal{N}=2$, $\,D=3\,$ supergravity models upon dimensional reduction. These were dubbed RSTU-models therein.

To understand the action of the $\,\textrm{SO}(3)\,$ symmetry, we must first split $\,y^{\hat{a}}\,$ as 
\begin{equation}
y^{\hat{a}}=(y^{a},y^{7})
\hspace{10mm} \textrm{ with } \hspace{10mm}
a=1,3,5 \ .
\end{equation}
Then, the $\,\textrm{SO}(3)\,$ symmetry simply rotates the two triplets $\,y^{i}\,$ and $\,y^{a}\,$ \textit{simultaneously} and leaves $\,y^{7}\,$ invariant (see Section~$4$ of \cite{Arboleya:2024vnp} for additional group-theoretic details). The set of $\,\textrm{SO}(3)$-invariant tensors is therefore given (up to permutation of indices of different type) by
\begin{equation}
\label{inv-tensors_closed}
\delta_{ij} \,\,,\,\, \delta_{ab} \,\,,\,\, \delta_{ia} 
\hspace{10mm} \textrm{ and } \hspace{10mm}
\epsilon_{ijk} \,\,,\,\, \epsilon_{ijc} \,\,,\,\, \epsilon_{ibc} \,\,,\,\, \epsilon_{abc} \ .    
\end{equation}

In this work we will extend the compactification scheme of \cite{Arboleya:2024vnp} and consider also fields and fluxes that arise from the open-string sector living on the worldvolume of the D$5$-branes. This implies having to extend the action of the $\,\textrm{SO}(3)\,$ symmetry of \cite{Arboleya:2024vnp} also to the open-string sector. The dynamics of a stack of D$5$-branes is described by a non-abelian (super) Yang--Mills (YM) theory, with gauge group $\,\textrm{G}_{\textrm{YM}}$, that lives in the D$5$-branes' worldvolume. Both gauge fields and matter fields in the (super) YM theory transform in the adjoint representation of $\,\textrm{G}_{\textrm{YM}}$. Therefore, a generic open-string field $\,\Psi^{I}\,$ will carry an adjoint index $\,I=1,\ldots,\textrm{dim}\left(\textrm{G}_{\textrm{YM}}\right)$. In this work we will restrict to the case of
\begin{equation}
\label{G_YM_SO(3)}
\textrm{G}_{\textrm{YM}} = \textrm{SO}(3) \ , 
\end{equation}
so that $\,I=1,2,3$. The YM gauge group in (\ref{G_YM_SO(3)}) requires the O$5$-planes variant to be O$5^{-}$ \cite{Gimon:1996rq,Giveon:1998sr,Bergman:2001rp}. A natural choice is then to extend the action of the $\,\textrm{SO}(3)\,$ symmetry to \textit{simultaneously} rotate the indices $\,i\,$, $\,a\,$ and $\,I\,$. The new set of $\,\textrm{SO}(3)\,$ invariant tensors includes those in (\ref{inv-tensors_closed}) but also (up to permutation of indices of different type)
\begin{equation}
\delta_{Ii} \,\, , \,\, \delta_{Ia} \,\, , \,\, \delta_{IJ} 
\hspace{10mm} \textrm{ and } \hspace{10mm}
\epsilon_{Ijk} \,\,,\,\, \epsilon_{Ijc} \,\,,\,\, \epsilon_{Ibc} \,\,,\,\, \epsilon_{IJk} \,\,,\,\,  \epsilon_{IJc} \,\,,\,\, \epsilon_{IJK}  \ .  
\end{equation}
The gauge group in (\ref{G_YM_SO(3)}) can be straightforwardly generalised to a (non necessarily direct) product of $\,k\,$ factors $\,\textrm{SO}(3)_{(k)}$, each one with its own adjoint index $\,I_{(k)}=1,2,3$.

It is worth emphasising that the $\,\textrm{SO}(3)\,$ symmetry we are modding out by acts diagonally (or simultaneously) on both the spacetime (it acts on the coordinates $\,y^{i}$) and the internal YM space (it acts on the generators $t_{I}$). Namely, the particular $\,\textrm{SO}(3)\,$ symmetry we are imposing on our setup is not a pure spacetime symmetry or a pure YM internal symmetry, but a combination thereof.

\subsection{Closed-string sector: fields and fluxes}

Let us first present the set of $\mathcal{O}_{\mathbb{Z}_{2}}$-even and $\,\textrm{SO}(3)$-invariant fields and fluxes in the closed-string sector of type IIB supergravity.

\subsubsection*{Fields}

The set of $\mathcal{O}_{\mathbb{Z}_{2}}$-even and $\,\textrm{SO}(3)$-invariant closed-string IIB fields in Table~\ref{Table:O5_Fields} is the following.  From the internal seven-dimensional geometry (in string frame, thus the superscript $^{s}$) and the ten-dimensional dilaton $\,\Phi$, one finds four fields
\begin{equation}
\label{dilatons_def}
e_{i}{}^{j} = L^{s}_{i} \,\, \delta_{i}^{j} 
\hspace{8mm} , \hspace{8mm} 
e_{a}{}^{b} = L_{a}^{s} \,\,\delta_{a}^{b} 
\hspace{8mm} , \hspace{8mm} 
L^{s}_{7} 
\hspace{8mm} , \hspace{8mm} 
\Phi \ ,
\end{equation}
which will be generically referred to as \textit{dilatons}. From the NS-NS and R-R gauge potentials, we can identify another four fields
\begin{equation}
\label{axions_def}
B_{ia} =  b_{2} \, \delta_{ia}
\hspace{5mm} , \hspace{5mm} 
B_{ijkabc} = b_{6} \, \epsilon_{ijk} \, \epsilon_{abc}
\hspace{5mm} , \hspace{5mm} 
C_{ibc7} =  c_{4} \, \epsilon_{ibc7}
\hspace{5mm} , \hspace{5mm} 
C_{ijk7} =  \tilde{c}_{4} \, \epsilon_{ijk7} \ ,
\end{equation}
which will be generically referred to as \textit{axions}. One is therefore left with $\,8\,$ fields: $\,4\,$ dilatons $\,(L^{s}_{i}\,,\,L^{s}_{a}\,,\,L^{s}_{7} \,,\, \Phi)\,$ and $\,4\,$ axions $\,(b_{2}\,,\,b_{6}\,,\,c_{4}\,,\,\tilde{c}_{4})$. However, in order to be compatible with the type IIB flux vacua of \cite{Arboleya:2024vnp}, and also to connect with the type IIB co-calibrated $\textrm{G}_{2}$ orientifolds of \cite{Emelin:2021gzx}, we will set the axions in (\ref{axions_def}) to zero, namely,
\begin{equation}
\label{axions_def_null}
b_{2} = 0
\hspace{5mm} , \hspace{5mm} 
b_{6} = 0
\hspace{5mm} , \hspace{5mm} 
c_{4} = 0
\hspace{5mm} , \hspace{5mm} 
\tilde{c}_{4} = 0 \ ,
\end{equation}
in our compactification scheme.

\subsubsection*{Fluxes}

The set of $\mathcal{O}_{\mathbb{Z}_{2}}$-even and SO(3)-invariant closed-string IIB fluxes in Table~\ref{Table:O5_Fluxes_even} is the following. From the geometry of the internal seven-dimensional twisted torus $\,\mathbb{T}^{7}_{\omega}$, one finds $\,6\,$ metric fluxes
\begin{equation}
\label{metric_fluxes}
\begin{array}{lclclc}
\omega_{ab}{}^{k}=\omega_{1} \, \epsilon_{ab}{}^{k}
& \hspace{5mm} , & \hspace{5mm} 
\omega_{7a}{}^{i}=\omega_{2} \, \delta_{a}^{i}
& \hspace{5mm} , & \hspace{5mm} 
\omega_{i7}{}^{a}=\omega_{3} \, \delta_{i}^{a} & , \\[2mm]
\omega_{ai}{}^{7}=\omega_{4} \, \delta_{ai}
& \hspace{5mm} , & \hspace{5mm} 
\omega_{aj}{}^{c}=-\omega_{5} \, \epsilon_{aj}{}^{c}
& \hspace{5mm} , & \hspace{5mm} 
\omega_{ij}{}^{k}=\omega_{6} \, \epsilon_{ij}{}^{k} & .
\end{array}
\end{equation}
From the NS-NS and R-R sectors, one identifies $\,2+5\,$ gauge fluxes of the form
\begin{equation}
\label{gauge_fluxes_NSNS}
\begin{array}{c}
H_{abc}= h_{31}  \, \epsilon_{abc}
\hspace{10mm} , \hspace{10mm} 
H_{aij}= h_{32}  \, \epsilon_{aij}  \ ,
\end{array}
\end{equation}
and
\begin{equation}
\label{gauge_fluxes_RR}
\begin{array}{c}
F_{ijk}= -f_{31}  \, \epsilon_{ijk}
\hspace{8mm} , \hspace{8mm} 
F_{ia7}= f_{32}  \, \delta_{ia}
\hspace{8mm} , \hspace{8mm} 
F_{ibc}= f_{33}  \, \epsilon_{ibc} \ , \\[4mm]
F_{7}= f_{1}
\hspace{8mm} , \hspace{8mm} 
F_{abij7}= -f_{5}  \, \delta_{ai} \, \delta_{bj}
\hspace{8mm} , \hspace{8mm} 
F_{abcijk7}= f_{7}  \, \epsilon_{abc} \, \epsilon_{ijk} \ ,
\end{array}
\end{equation}
where a proper antisymmetrisation of indices is understood in \eqref{metric_fluxes}, \eqref{gauge_fluxes_NSNS} and \eqref{gauge_fluxes_RR}. All together, they add up to $\,6+2+6=14\,$ flux parameters coming from the closed-string sector of type IIB supergravity.

\subsection{Open-string sector: fields, fluxes and non-abelian structure constants}

Let us now present the set of $\mathcal{O}_{\mathbb{Z}_{2}}$-even and $\,\textrm{SO}(3)$-invariant fields and fluxes in the open-string sector of type IIB supergravity.

\subsubsection*{Fields}

Adding D5-branes to the compactification scheme results in the appearance of additional fields. In the static gauge, and performing a weak-field expansion, a stack of D$5$-branes accommodates a non-abelian (maximally supersymmetric) Yang--Mills (YM) theory in its worldvolume that includes (at the bosonic level) scalars and non-abelian vector fields. Placing the stack of D$5$-branes as in (\ref{O5/D5_location}), there will be a set of scalars $\,Y^{\hat{a}}(x^{\mu}, y^i)\,$ parameterising the stack's position within its transverse $y^{\hat{a}}$-space. The scalars $\,Y^{\hat{a}}\,$ transform in the adjoint representation of the YM gauge group $\,\textrm{G}_{\textrm{YM}}\,$ and so they are algebra-valued
\begin{equation}
\label{Y-scalars_def}
Y^{\hat{a}} = Y^{\hat{a} I} \, t_{I} \ ,
\end{equation}
where $\,t_{I}$, with $\,I=1,\ldots, \textrm{dim}(\textrm{G}_{\textrm{YM}})$, are the generators of the YM gauge algebra. These generators satisfy commutation relations and a normalisation condition of the form
\begin{equation}
\label{eq:GeneratorRelations}
[t_I, t_J] = -i \ {\mathcal{G}_{IJ}}^K t_K \hspace{10mm}  , \hspace{10mm} \textrm{Tr}\left( t_I  \, t_J \right) = N_{\textrm{D}5} \, \delta_{IJ} \ ,
\end{equation}
where $\,\mathcal{G}_{IJ}{}^{K}\,$ are the structure constant of the YM gauge algebra. In addition to the scalars in (\ref{Y-scalars_def}), the YM theory living on the worldvolume of the D$5$-branes -- which extend along coordinates $\,X^{\mathsf{\bar M}}=(x^{\mu},y^{i})\,$ -- also contains non-abelian vector fields $\,\mathcal{A}_{\bar{\mathsf{M}}}(x^{\mu}, y^i)\,$ in the adjoint representation of the YM gauge group, namely, these are also algebra-valued
\begin{equation}
\label{A-vectors_def}
\mathcal{A}_{\bar{\mathsf{M}}} = \mathcal{A}_{\bar{\mathsf{M}}}{}^{I} \, t_{I} \ .
\end{equation}
We can therefore split the non-abelian vectors in (\ref{A-vectors_def}) as
\begin{equation}
\label{A-vectors_def_blocks}
\mathcal{A}_{\mu} = \mathcal{A}_{\mu}{}^{I}  \, t_{I}
\hspace{8mm} \textrm{ and } \hspace{8mm}
\mathcal{A}_{i} = \mathcal{A}_{i}{}^{I} \, t_{I} \ .
\end{equation}
From the point of view of the reduced three-dimensional theory, $\,\mathcal{A}_{\mu}\,$ describes and algebra-valued vector field whereas $\,\mathcal{A}_{i}\,$ describes three algebra-valued scalars. Considering only the internal component of the various fields in the open-string sector -- as we did for the closed-string sector -- we are left with 
\begin{equation}
\label{scalars_open-sector}
Y^{\hat{a} I} 
\hspace{10mm} \textrm{ and } \hspace{10mm}
\mathcal{A}_{i}{}^{I} \ .
\end{equation}
As shown in  Table~\ref{Table:O5_Fields_open}, the open-string fields in (\ref{scalars_open-sector}) are both $\mathcal{O}_{\mathbb{Z}_{2}}$-even ($+$) under the O$5$-plane orientifold action.

Demanding also invariance under the $\,\textrm{SO}(3)\,$ symmetry discussed above, one is left with two fields
\begin{equation}
\label{scalars_open-sector_SO(3)}
Y^{a I} = \psi \, \delta^{aI}
\hspace{10mm} \textrm{ and } \hspace{10mm}
\mathcal{A}_{i}{}^{I} = \alpha  \, \delta_{i}^{I} \ .
\end{equation}
It is worth emphasising here again that the internal component $\,\mathcal{A}_{i}{}^{I} = \alpha(x) \,  \delta_{i}^{I} \,$ is left invariant by an $\,\textrm{SO}(3)\,$ group that acts simultaneously on both the spacetime and the internal YM space.

\begin{table}[t]
\begin{center}
\renewcommand{\arraystretch}{1.7}
\begin{tabular}{|c|c|c|c|}
\hline
Fields & $\Omega_{P}$   &  $\sigma_{\textrm{O}5}$ & $\mathcal{O}_{\mathbb{Z}_{2}}$  \\
\hline
\hline
$Y^{\hat{a}I} $ & $-$  & $-$  & $+$  \\
\hline
\hline
$\mathcal{A}_{i}{}^{I}$ & $+$ & $+$ & $+$ \\
\hline
\end{tabular}
\hspace{10mm}
\begin{tabular}{|c|c|}
\hline
Fluxes &   $\Theta$-irrep $\leftrightarrow$ YM Structure Constants/Flux \\
\hline
\hline
$\mathcal{G}$ & $\theta^{IJK8} = \mathcal{G}_{IJ}{}^{K}$ \\
\hline
\hline
$\mathcal{F}$ & $\theta^{ijI8} = \mathcal{F}_{ij}{}^{I}$ \\
\hline
\end{tabular}
\caption{Open-string fields, YM structure constants $\,\mathcal{G}_{IJ}{}^{K}\,$ and flux $\,\mathcal{F}_{ij}{}^{I}\,$ that are parity-even under the O$5$-plane orientifold action $\mathcal{O}_{\mathbb{Z}_{2}} = \Omega_{P} \, \sigma_{\textrm{O}5}$.} 
\label{Table:O5_Fields_open}
\end{center}
\end{table}

\subsubsection*{Fluxes and non-abelian structure constants}

The YM theory living on the stack of D$5$-branes is a non-abelian gauge theory with gauge group $\,\textrm{G}_{\textrm{YM}}$. The vector fields $\,\mathcal{A}_{\bar{\mathsf{M}}}{}^{I}\,$ have non-abelian field strengths
\begin{equation}
\label{F2_def}
\mathcal{F} = d \mathcal{A} + i \mathcal{A} \wedge \mathcal{A} \ .
\end{equation}
Focusing on the internal components of (\ref{F2_def}), we will allow for a background flux of $\,\mathcal{F}_{ij}{}^{I}\,$ in our compactification scheme. In addition, the non-abelian structure of the YM theory will be codified in the structure constants $\,\mathcal{G}_{JK}{}^{I}$. Both quantities are $\mathcal{O}_{\mathbb{Z}_{2}}$-even under the O$5$-plane orientifold action.

Requiring again invariance under the $\,\textrm{SO}(3)\,$ symmetry, we are left with two independent parameters, $\mathfrak{f}\,$ and $\,\mathfrak{g}$, entering
\begin{equation}
\label{open_fluxes_SO(3)}
\mathcal{F}_{i j}{}^{I} =  \mathfrak{f} \, \epsilon_{ijk} \, \delta^{k I} - \omega_{i j}{}^{k} \mathcal{A}_k{}^I + \mathcal{G}_{JK}{}^{I} \mathcal{A}_i{}^J \mathcal{A}_j{}^K  \hspace{7mm} \textrm{ with } \hspace{7mm} \mathcal{G}_{I J}{}^{K} = \mathfrak{g} \, \epsilon_{I J L} \, \delta^{L K} \ .
\end{equation}
While the parameter $\,\mathfrak{f}\,$ corresponds to an open-string background flux contributing to the magnetisation of the D$5$-branes, the parameter $\,\mathfrak{g}\,$ controls the strength of the non-abelian YM interaction on the stack of D$5$-branes. As a matter of terminology, in the rest of this work we will refer to the case $\,\mathfrak{f}=0\,$ ($\,\mathfrak{f}\neq 0\,$) as unmagnetised (magnetised) D$5$-branes, and to the case $\,\mathfrak{g}=0\,$ ($\,\mathfrak{g}\neq 0\,$) as abelian (non-abelian) D$5$-branes.

\subsection{Scherk–Schwarz reduction: bulk \textit{vs.} sources}

The bosonic ten-dimensional type IIB supergravity action consists of two pieces: the closed-string (or bulk) contribution and the one associated with the O$5$/D$5$ sources. Namely,
\begin{equation}
\label{S_IIB}
S_{\textrm{IIB}} = S_{\textrm{bulk}} + S_{\textrm{O$5$/D$5$}} \ .
\end{equation}
In this section, following \cite{Emelin:2021gzx}, we will first present a standard Scherk--Schwarz reduction of the closed-string sector of type IIB supergravity on a seven-dimensional twisted torus $\,\mathbb{T}_{\omega}^{7}$. Then, following \cite{Balaguer:2024cyb,Balaguer:2023jei}, we will incorporate the open-string sector. We will always restrict to the $\,\textrm{SO}(3)$-invariant content of fields and fluxes discussed in the previous sections.

\subsubsection{Scherk–Schwarz ansatz}

Our starting point is a universal ansatz for the ten-dimensional metric in \textit{string frame} of the form
\begin{equation}
\label{10D_metric_ansatz}
ds_{10}^2 = \tau^{-2} \, ds_{3}^2 + \rho^2 \, d\tilde{s}_7^2 \ ,
\end{equation}
where $\,ds_{3}=g_{\mu\nu} \, dx^{\mu} \, dx^{\nu}\,$ and $\,d\tilde{s}_{7}=\tilde{g}_{mn} \, \eta^{m} \, \eta^{n}\,$ are, respectively, the line element of the $D=3$ external spacetime, and the one of a seven-dimensional internal space with unit volume, \textit{i.e.} $\tilde{g} \equiv \textrm{det}[\tilde{g}_{mn}]=1$.

The scalars $\,\rho(x)\,$ and $\,\tau(x)\,$ are universal functions dependent on $x^{\mu}$. The first one accounts for the volume of the internal space (in string frame)
\begin{equation}
\label{string-frame_volume}
\textrm{vol}_7 = \rho^7 \ .
\end{equation}
The second one relates to the ten-dimensional type IIB dilaton $\,\Phi\,$ determining the string coupling constant $\,g_{s}=e^{\Phi}$, and must be chosen as
\begin{equation}
\label{eq:indirect_g_s}
\tau \equiv e^{-2\Phi} \, \textrm{vol}_7 = g_{s}^{-2} \, \textrm{vol}_7  \ ,
\end{equation}
for the dimensional reduction to yield a $D=3$ supergravity in the Einstein frame. It is worth noticing that, due to the $\,\tau^{-2}\,$ factor in front of $ \,ds_{3}^2\,$ in (\ref{10D_metric_ansatz}), a length scale in the external spacetime gets a factor of $\,\tau_{0}^{-1}$, where $\,\tau_{0}\,$ denotes the value of $\,\tau\,$ at a given vacuum solution. This will become relevant when discussing the issue of scale separation between the external spacetime and the internal one.

The unit-volume line element of the internal twisted torus, \textit{i.e.} $d\tilde{s}_{7}\,$ in (\ref{10D_metric_ansatz}), takes the simple form\footnote{As we discussed before, this is the most general seven-dimensional twisted torus metric consistent with the $\,\textrm{SO}(3)\,$ symmetry of \cite{Arboleya:2024vnp}.}
\begin{equation}
\label{7D_metric_ansatz}
d\tilde{s}_{7}^2 = {\ell_{i}}^2 \left[ \left(\eta^2\right)^{2} + \left(\eta^4\right)^{2} + \left(\eta^6\right)^{2} \right]+ {\ell_{a}}^2 \left[ \left(\eta^1\right)^{2} + \left(\eta^3\right)^{2} + \left(\eta^5\right)^{2} \right] +  \ell_7^2 \left(\eta^7\right)^2 \ ,
\end{equation}
in terms of three scalars $\,\ell_{i}(x)\,$, $\,\ell_{a}(x)\,$ and $\,\ell_{7}(x)$, which are subject to the unit-volume condition $\,\ell_7 = \ell_a^{-3} \ell_i^{-3}$. From (\ref{10D_metric_ansatz}) and (\ref{7D_metric_ansatz}), the three dilatons in (\ref{dilatons_def}) associated with the would-be one-cycles on the internal geometry are identified as
\begin{equation}
\label{L_string_def}
L^{s}_{a} = \rho \, \ell_{a}
\hspace{8mm} , \hspace{8mm} 
L^{s}_{i} = \rho \, \ell_{i}
\hspace{8mm} , \hspace{8mm} 
L^{s}_{7} = \rho \, \ell_{7} \ .
\end{equation}
The one-form basis elements $\,\eta^{m}\,$ entering (\ref{7D_metric_ansatz}) are the Maurer--Cartan one-forms on the internal twisted torus. They relate to the original coordinate basis as $\,\eta^{m} = U(y)^{m}{}_{n} \, dy^{n}$, where the \textit{twist matrix} $\,U(y)\in \textrm{G}\,$ is an element of a Lie group $\,\textrm{G}$. The twisted one-forms $\,\eta^{m}\,$ obey a structure equation
\begin{equation}
\label{structure_equation}
d\eta^{p} + \tfrac{1}{2} \, \omega_{mn}{}^{p} \, \eta^{m} \wedge \eta^{n} = 0 \ ,
\end{equation}
where $\,\omega_{mn}{}^{p}=(U^{-1})_{m}{}^{r} (U^{-1})_{n}{}^{s} \left( \partial_{r} U^{p}{}_{s} - \partial_{s} U^{p}{}_{r}  \right)\,$ are the so-called \textit{metric fluxes}. These metric fluxes are simply the structure constants of the algebra of $\,\textrm{G}\,$ associated with the isometry generators, $X_m = (U^{-1})_m{}^n \, \partial_{n}$, on the internal space. Namely,
\begin{equation}
\label{X_isometry_brackets}
[X_m, X_n] = \omega_{mn}{}^{p} \, X_p \ .
\end{equation}
In a standard SS reduction, they satisfy the Jacobi identity
\begin{equation}
\label{ww_constraint}
\omega_{[mn}{}^r \, \omega_{p]r}{}^q = 0  \ ,
\end{equation}
together with a unimodularity (or traceless) condition $\,\omega_{mn}{}^n = 0\,$ that ensures consistency of the reduction at the level of the action \cite{Scherk:1979zr}. As a result, our reduction ansatz contains four independent scalar functions $\,(\tau\,,\,\rho\,,\,\ell_{i}\,,\,\ell_{a})\,$ accounting for the internal space geometry and the ten-dimensional type IIB dilaton $\,\Phi\,$ in agreement with (\ref{dilatons_def}) and (\ref{L_string_def}).

Equipped with the basis of Maurer--Cartan one-forms $\,\eta^{m}\,$ in (\ref{7D_metric_ansatz}), we proceed as in  \cite{Scherk:1979zr} and perform a dimensional reduction of type IIB supergravity \`a la Kaluza--Klein (KK). That is, as if the reduction were carried out on a seven-dimensional torus. For example, the purely internal component of the Neveu--Schwarz--Neveu--Schwarz (NS-NS) three-form field strength $\,H_{(3)}\,$ is expanded as
\begin{equation}
\label{SS_ansatz_H}
H_{(3)} = \frac{1}{3!} \, H_{mnp} \, \eta^{m} \wedge \eta^{n} \wedge \eta^{p} \ ,
\end{equation}
with $\,H_{mnp}\,$ being \textit{constant} flux parameters. The same expansion is performed for the purely internal component of the various $\,F_{(2p+1)}\,$ field strengths in the Ramond--Ramond (R-R) sector of type IIB supergravity. Namely,
\begin{equation}
\label{SS_ansatz_F}
F_{(2p+1)} = \frac{1}{(2p+1)!} \, F_{m_{1} \ldots m_{2p+1}} \, \eta^{m_{1}} \wedge \ldots \wedge \eta^{m_{2p+1}}  \ ,
\end{equation}
with $\,F_{m_{1} \ldots m_{2p+1}}\,$ being \textit{constant} flux parameters. The final outcome of the SS reduction is a $\,D=3\,$ supergravity theory involving the scalar fields (or moduli) $\,(\tau\,,\,\rho\,,\,\ell_{i}\,,\,\ell_{a})\,$ as well as the various flux parameters $\,(\omega_{mn}{}^{p}\,,\,H_{mnp}\,,...)$. The latter induce a non-trivial scalar potential for the moduli $\,(\tau\,,\,\rho\,,\,\ell_{i}\,,\,\ell_{a})$.

In the absence of sources, a standard SS reduction gives rise to a $D=3$ supergravity with maximal $\mathcal{N}=16$ supersymmetry ($32$ supercharges). However, when the O$5$/D$5$ sources in (\ref{O5/D5_location}) are included in the compactification scheme, they break half of the supersymmetries, introduce a source term in the 10D Bianchi identities (see eq.~(\ref{BI_10D}) below), and  induce new terms in the $\,D=3\,$ scalar potential. The resulting $D=3$ theory becomes a half-maximal $\,\mathcal{N}=8\,$ supergravity ($16$ supercharges). Importantly, in the presence of dynamical sources, the ansatz for the RR fluxes in (\ref{SS_ansatz_F}) must be modified for it to comply with the (now sourced) 10D Bianchi identities \cite{Balaguer:2023jei}. The modification is of the form
\begin{equation}
\label{redefinition_F}
F_{m_{1} \ldots m_{2p+1}} \,\, \longrightarrow \,\, \widetilde{F}_{m_{1} \ldots m_{2p+1}} = \underbrace{F_{m_{1} \ldots m_{2p+1}}}_{\textrm{cst}}  \,\, + \,\, \underbrace{\Delta F_{m_{1} \ldots m_{2p+1}}(\psi,\alpha)}_{\textrm{$x$-dependent}} \ ,
\end{equation}
so that the fluxes $\,F_{m_{1} \ldots m_{2p+1}}\,$ in the expansion (\ref{SS_ansatz_F}) are no longer constant and acquire a dependence on the external spacetime coordinates through the open-string moduli (\ref{scalars_open-sector_SO(3)}). This is similar to the modifications due to the Green--Schwarz mechanism for anomaly cancellation \cite{Green:1984sg} in the heterotic string theory: these two instances could be, indeed, heuristically related through duality chains. We will come back to this issue at due time and explicitly show the required modifications $\Delta F_{(2p+1)}$ in the concrete flux compactification scenario we are considering.

\subsubsection{The bulk: closed-string sector}
\label{Closed_Reduction}

In its democratic formulation \cite{Bergshoeff:2001pv} (we follow conventions of \cite{Balaguer:2023jei}), the bosonic field content of type IIB supergravity consists of the universal NS-NS sector $\{ G \,,\, \Phi \,,\, B_{(2)} \,,\, B_{(6)}\}$ and the R-R sector of gauge potentials $\,\lbrace{C_{(2 p)}\rbrace}\,$ with $\,p = 0, \ldots,4\,$. The closed-string sector is governed by the ten-dimensional pseudo-action (in string frame)\footnote{We have not included an $\,|H_{(7)}|^2\,$ term in (\ref{S_bulk}) because a purely internal $\,H_{(7)}\,$ is projected out by the  orientifold action $\,\mathcal{O}_{\mathbb{Z}_{2}} = \Omega_{P} \, \sigma_{\textrm{O}5}\,$ (see Table~\ref{Table:O5_Fluxes_even}). We refer to ref.~\cite{Fernandez-Melgarejo:2023kwk} for a democratic formulation of type IIB supergravity including both $\,H_{(3)}\,$ and $\,H_{(7)}$.}
\begin{equation} 
\label{S_bulk}
S_{\textrm{bulk}} = \frac{1}{2\kappa^{2}} \displaystyle\int \textrm{d}^{10}X \, \sqrt{-G} \, \left[ e^{-2\Phi} \left(R^{(10)} \, + \, 4  (\partial\Phi)^{2} \, - \, \frac{1}{2 \cdot 3!} |H_{(3)}|^2 \right ) \, - \, \frac{1}{2} \sum_{p=0}^4 \frac{|F_{(2p + 1)}|^2}{2 \cdot (2p + 1)!}\right] \ , 
\end{equation}
where $\,X^{\mathsf{M}}=\left(x^{\mu},y^{i},y^{{\hat{a}}}\right)\,$ denotes the ten-dimensional spacetime coordinates, and $\,(2\kappa^{2})^{-1}=(2\pi)(2 \pi \ell_{s})^{-8}\,$ with $\,\ell_{s}=\sqrt{\alpha'}\,$ being the string length scale. The above pseudo-action must be supplemented by the Hodge duality relations
\begin{equation}
\label{Hodge_Duality}
F_{(9)} = \star F_{(1)} 
\hspace{8mm} , \hspace{8mm}
F_{(7)} = - \star F_{(3)}
\hspace{8mm} , \hspace{8mm}
F_{(5)} = \star F_{(5)} \ ,
\end{equation}
to ensure the correct number of degrees of freedom. The NS-NS and RR field strengths satisfy Bianchi identities of the form
\begin{equation}
\label{BI_10D}
d H_{(3)} = 0
\hspace{15mm} \textrm{and} \hspace{15mm}
d  F_{(8-p)} - H_{(3)} \wedge F_{(6-p)} = J_{\textrm{O}p/\textrm{D}p} \ ,
\end{equation}
where $\,J_{\textrm{O}p/\textrm{D}p}\,$ is a $(9-p)$-form that accounts for O$p$/D$p$ sources (in the smeared limit).

\subsubsection*{Reduction of the Ricci scalar}

Let us start by dimensionally reducing the ten-dimensional Ricci scalar $\,R^{(10)}\,$ and dilaton $\,\Phi\,$ contributions in (\ref{S_bulk}). From the ansatz in (\ref{10D_metric_ansatz}), one has that
\begin{equation}
\sqrt{-G} = \tau^{-3} \rho^{7} \ \sqrt{-g} \ ,
\end{equation}
where we have used the unit-volume condition $\,\tilde{g}=1$. The reduction of the ten-dimensional Ricci scalar gives
\begin{equation}
\label{R10}
R^{(10)} = \tau^2 \, R + \rho^{-2} \,  \tilde{R} + \dots ,
\end{equation}
where the Ricci scalars $\,R\,$ and $\,\tilde{R}\,$ are respectively constructed from the external and (unit-volume) internal metrics $\,g_{\mu\nu}\,$ and $\,\tilde{g}_{mn}\,$ in (\ref{10D_metric_ansatz}). In particular,
\begin{equation}
\label{R7}
\begin{array}{rcl}
\tilde{R} &=& - \dfrac{12}{{\ell_i}^2} \partial_\mu {\ell_i} \partial^\mu {\ell_i} - \dfrac{18}{\ell_a \ell_i} \partial_\mu \ell_a \partial^\mu \ell_i - \dfrac{12}{{\ell_a}^2} \partial_\mu {\ell_a} \partial^\mu {\ell_a}  \\[4mm]
&& - \dfrac{1}{4} \, \tilde{g}_{qm} \, \tilde{g}^{nr} \, \tilde{g}^{ps} \omega_{np}{}^q \, \omega_{rs}{}^m - \dfrac{1}{2} \, \tilde{g}^{np} \, \omega_{mn}{}^q \, \omega_{qp}{}^m \ .
\end{array}
\end{equation}
The ellipsis in (\ref{R10}) stands for terms involving spacetime derivatives of $\,\rho\,$ and $\,\tau$. Due to the identification in (\ref{eq:indirect_g_s}), the spacetime derivatives of $\,\rho\,$ and $\,\tau\,$ coming from (\ref{R10}) combine with the dimensional reduction of the kinetic term of $\,\Phi\,$ in (\ref{S_bulk}) to finally yield
\begin{equation}
\frac{1}{2\kappa^{2}} \displaystyle\int \textrm{d}^{10}X \, \sqrt{-G} \, e^{-2\Phi} \, \left(R^{(10)} \, + \, 4  (\partial\Phi)^{2} \right) \,=\, \frac{1}{2\kappa^{2}} \displaystyle\int \textrm{d}^{3}x \, \sqrt{-g} \,  \big( R + L_{\textrm{kin}} - V_{\omega} \big) \ ,
\end{equation}
where
\begin{equation}
\label{L_kin_closed}
L^{\textrm{closed}}_{\textrm{kin}} = - \frac{1}{\tau^2} \partial_\mu \tau \partial^\mu \tau - \frac{7}{\rho^2} \partial_\mu \rho \partial^\mu \rho - \frac{12}{{\ell_i}^2} \partial_\mu {\ell_i} \partial^\mu {\ell_i} - \frac{18}{\ell_a \ell_i} \partial_\mu \ell_a \partial^\mu \ell_i - \frac{12}{{\ell_a}^2} \partial_\mu {\ell_a} \partial^\mu {\ell_a} \ ,
\end{equation}
accounts for the kinetic terms of the $D=3$ closed-string moduli fields, and
\begin{equation}
\begin{aligned} 
\label{V_omega}
V_{\omega} \,=\, \frac{3}{2}\,\tau^{-2} \rho^{-2} \Bigl(\, & {\ell_i}^{2} {\ell_a}^{-4} {\omega_1}^2 + {\ell_i}^{8} {\ell_a}^{4} {\omega_2}^2 \, \, - 2 \ {\ell_i}^6{\ell_a}^6 \omega_2 \, \omega_3 \, + \, {\ell_i}^{4} {\ell_a}^{8} {{\omega}_3}^2 \, + \, \\ 
& + {\ell_i}^{-8} {\ell_a}^{-8} {\omega_4}^2 - 2 {\ell_a}^{-2} \left(\omega_2 \, \omega_4 - 2 \omega_1 \, \omega_5 \right) - {\ell_i}^{-2} \left(2 \omega_3 \, \omega_4 + {\omega_6}^2 \right) \Bigr) \ ,
\end{aligned}
\end{equation}
which originates from the second line in (\ref{R7}) and involves the $\,\textrm{SO}(3)$-invariant metric fluxes in (\ref{metric_fluxes}), contributes to the $\,D=3\,$ scalar potential.

\subsubsection*{Reduction of the gauge fluxes}

We can proceed in the same manner and dimensionally reduce the field strengths in (\ref{S_bulk}). The reduction of the NS-NS three-form flux $\,H_{(3)}\,$ yields a new contribution to the scalar potential
\begin{equation}
\frac{1}{2\kappa^{2}}  \int \textrm{d}^{10}X \, \sqrt{-G^{(10)}} \, \left(-\frac{1}{2 \cdot 3!}\, e^{-2\Phi}\, |H_{(3)}|^2 \right) \,=\, \frac{1}{2\kappa^{2}}  \int \textrm{d}^3 x \sqrt{-g} \, \big(-V_{H} \big) \ ,
\end{equation}
with
\begin{equation}
\label{V_H}
V_{H} \,=\,  \dfrac{1}{2 \cdot 3!} \, \tau^{-2} \, \rho^{-6} \, H_{mnp} \, H^{mnp} \,=\, \dfrac{\tau^{-2}\rho^{-6}}{2} \left( {\ell_a}^{-6} {h_{31}}^2 + 3 \, {\ell_i}^{-4} {\ell_a}^{-2}  {h_{32}}^2   \right)  \ .
\end{equation} 
Note that the internal space indices in (\ref{V_H}) are lowered/raised using the (unit-volume) internal metric in (\ref{7D_metric_ansatz}) and its inverse.

Similarly, the reduction of the RR fluxes $\,F_{(1)}$, $\,F_{(3)}$, $\,F_{(5)}\,$ and $\,F_{(7)}\,$ yields four additional contributions to the scalar potential
\begin{equation}
 \frac{1}{2\kappa^{2}} \displaystyle\int \textrm{d}^{10}X \, \sqrt{-G} \, \left[ - \frac{1}{2} \sum_{p=0}^3 \frac{|F_{(2p + 1)}|^2}{2 \cdot (2p + 1)!}\right] \,=\, \frac{1}{2\kappa^{2}}  \int \textrm{d}^3 x \sqrt{-g} \, \big(-V_{F} \big) \ ,
\end{equation}
with
\begin{equation}
\label{V_F}
\begin{array}{rcl}
V_{F} &=&  \displaystyle\sum_{p=0}^{3} \dfrac{1}{2 \cdot (2p+1)!} \, \tau^{-3} \, \rho^{7-2(2p+1)}  \, F_{m_1\ldots m_{2p+1}} \, F^{m_1\ldots m_{2p+1}} = \\[6mm]
&=& \dfrac{\tau^{-3}\rho^{5}}{2} \, \ell_{i}^{6} \, \ell_{a}^{6} \, f_{1}{}^{2}  + \dfrac{\tau^{-3} \rho}{2} \ \left( \, {\ell_i}^{-6} {f_{31}}^2 + 3 \ {\ell_i}^{4} {\ell_a}^{4}{f_{32}}^2 + 3 \ {\ell_i}^{-2} {\ell_a}^{-4} {f_{33}}^2  \, \right) + \\[4mm]
& & + \, \dfrac{3}{2} \, \tau^{-3} \rho^{-3} {\ell_i}^2 {\ell_a}^2 {f_5}^2 + \dfrac{1}{2} \tau^{-3} \rho^{-7} {f_7}^2  \ .
\end{array}
\end{equation}
In our reduction to three dimensions, $F_{(9)}\,$ will not play any role since we will only consider purely internal flux components and the internal space is seven-dimensional. On the other hand, a RR flux $\,F_{(1)}\,$ is not allowed in a standard SS reduction because the gauge potential $\,C_{(0)}\,$ is a scalar (so it has no legs along the internal space to twist). Therefore, in what follows we will always set 
\begin{equation}
\label{f1=0}
f_{1} = 0 \ .
\end{equation}

The RR contribution to the scalar potential in (\ref{V_F}) is \textit{not} yet the complete story. As already advanced above, the presence of O$5$/D$5$ sources requires a modification of the RR fluxes of the form (\ref{redefinition_F}). This is
\begin{equation}
f_{7} \rightarrow \tilde{f}_{7} = f_{7} + \Delta f_{7}(\psi,\alpha)
\hspace{5mm} , \hspace{5mm} 
f_{5} \rightarrow \tilde{f}_{5} = f_{5} + \Delta f_{5}(\psi,\alpha)
\hspace{5mm} , \hspace{5mm} 
\textrm{etc} \ .
\end{equation}
Importantly, and despite we set $\,f_{1}=0\,$ in our SS reduction, a non-zero $\,\tilde{f}_{1} = \Delta f_{1}(\psi,\alpha)\,$ will be necessary in order to perform a consistent reduction in the presence of sources. In order to determine the precise form of all the modifications $\,\Delta F_{(2p+1)}$ in (\ref{redefinition_F}), with $\,p=0,\ldots, 3$, we will look at the dimensional reduction of the non-abelian extension of the Wess-Zumino term put forward in \cite{Myers:1999ps}.

\subsubsection{The sources: open-string sector}
\label{sec:reduction_open-sector}

The dynamics of D$p$-branes in type II backgrounds with fluxes has been analysed in \cite{Martucci:2005rb}. The effective action for a stack of $N_{\text{D}5}$ coincident D$5$-branes consists of two contributions:  the Dirac--Born--Infeld (DBI) term and the Wess--Zumino (WZ) term. Following the notation of \cite{Myers:1999ps} and \cite{Choi:2018fqw,Balaguer:2023jei}, they are given by
\begin{equation} 
\label{D5_DBI_WZ}
\begin{aligned}
S^{\textrm{DBI}}_{\textrm{D}5} & = -T_{\textrm{D}5} \displaystyle\int_{\textrm{WV}(\textrm{D}5)} \textrm{d}^6\sigma \; \textrm{Tr}\left( e^{-\Phi} \sqrt{-\det[\mathbb{M}_{\mathsf{{\bar M}{\bar N}}}] \det[{\mathbb{Q}^{\hat{a}}}_{\hat{b}}]} \; \right) \ , \\[2mm]
S^{\textrm{WZ}}_{\textrm{D}5} & = \mu_{\textrm{D}5} \displaystyle\int_{\textrm{WV}(\textrm{D}5)}  \textrm{Tr}\left\{ \textrm{P} \left[ e^{i \, \lambda \, \iota_Y \, \iota_Y} \left(\boldsymbol{\hat{C}} \wedge e^{\hat{B}  _{(2)}} \right) \wedge e^{\lambda \, \mathcal{F}}  \right] \right\} \ ,
\end{aligned}
\end{equation}
where $\,\vec{\sigma}\,$ are worldvolume (WV) coordinates on the stack of D5-branes, $\,T_{\textrm{D}5}=(2\pi)(2\pi \ell_{s})^{-6}\,$ is the tension of each D$5$-brane, and $\,\mu_{\textrm{D}5} = T_{\textrm{D}5}>0\,$ is its charge. There is also the parameter $\,\lambda \equiv 2 \pi \alpha' = (2\pi)^{-1}(2 \pi \ell_s)^2\,$ which must be understood as a small expansion parameter. As it will become clear in a moment, all the hatted fields are meant to be Taylor-expanded about the position (in transverse space) of the stack of D$5$-branes.

Before introducing the various objects and symbols entering the non-abelian DBI and WZ actions in (\ref{D5_DBI_WZ}), let us make a pause for notation: the ten-dimensional spacetime index $\,\mathsf{M}\,$ is split as $\,\mathsf{M} = (\mathsf{\bar M}, \hat{a})$, where $\,\mathsf{\bar M} = (\mu, i)\,$ accounts for non-compact and compact directions along the D$5$-branes' worldvolume, and $\,\hat{a}\,$ describes the compact transverse directions. With these splittings in mind, the DBI action involves two matrices
\begin{equation}
\label{M&Q-matrix_def}
\begin{array}{rcl}
\mathbb{M}_{\mathsf{{\bar M}{\bar N}}} &=& \textrm{P} \Big[ \hat{E}_{\mathsf{{\bar M}{\bar N}}} + \hat{E}_{\mathsf{\bar M}\hat{a}} (\mathbb{Q}^{-1} - \delta )^{\hat{a} \hat{b}} \, \hat{E}_{\hat{b} \mathsf{\bar N}} \Big] + \lambda \, \mathcal{F}_{\mathsf{{\bar M}{\bar N}}}  \ , \\[2mm]
{\mathbb{Q}^{\hat{a}}}_{\hat{b}} &=& {\delta^{\hat{a}}}_{\hat{b}} + i \,  \lambda \, [Y^{\hat{a}},Y^{\hat{c}}] \, \hat{E}_{\hat{c} \hat{b}} \ ,
\end{array}
\end{equation}
where
\begin{equation}
\label{Ehat_def}
 \hat{E}_{\mathsf{M} \mathsf{N}} \equiv \hat{G}_{\mathsf{M} \mathsf{N}}+\hat{B}_{\mathsf{M} \mathsf{N}}
\end{equation}
combines the ten-dimensional metric $\,\hat{G}_{\mathsf{MN}}\,$ and the NS-NS two-form potential $\,\hat{B}_{\mathsf{MN}}$, and where $\,\mathcal{F}\,$ -- introduced in (\ref{F2_def}) -- is the non-abelian field strength of the vector field $\,\mathcal{A}\,$ living on the D$5$-branes' worldvolume.

In order to evaluate the DBI action (\ref{D5_DBI_WZ}), we will choose the so-called static gauge and identify the six spacetime coordinates $\,X^{\bar{\mathsf{M}}}=(x^{\mu},y^{i})\,$ with the worldvolume coordinates $\,\vec{\sigma}$. The transverse directions are then re-labelled as

\begin{equation}
y^{\hat{a}}= \lambda \, Y^{\hat{a}} = \lambda \, Y^{\hat{a}I} \, t_{I} 
\hspace{10mm} \textrm{ with } \hspace{10mm}
I=1,\ldots,\dim(\textrm{G}_{\textrm{YM}}) \ ,  
\end{equation}
to emphasise that they are algebra-valued scalars in the worldvolume theory. Without loss of generality, we will choose coordinates such that the stack of D$5$-branes is located at the origin of transverse space, \textit{i.e.}, $\,y^{\hat{a}}=0$. Then, any hatted field is Taylor-expanded about $y^{\hat{a}} = 0$ as
\begin{equation}
\label{Taylor_expansion}
\hat{\Phi}(X^{\mathsf{\bar M}}, y^{\hat{a}})= \sum_{k=0}^{\infty} \frac{\lambda^k}{k!} Y^{\hat{a}_1}\cdots Y^{\hat{a}_k} \partial_{\hat{a}_1} \cdots \partial_{\hat{a}_k} \Phi(X^{\mathsf{\bar M}}, y^{\hat{a}}) \Big|_{y^{\hat{a}}=0} \ ,
\end{equation}
with the $\,k=0\,$ order corresponding to the smeared ($y^{\hat{a}}$-independent) limit. From (\ref{Taylor_expansion}), it follows that the term at order $\,k\,$ in the expansion contains a factor $\,\lambda^{k}\,$ that makes it sub-leading with respect to previous orders. Lastly, the symbol $\,\textrm{P}[\ldots]\,$ in (\ref{D5_DBI_WZ}) denotes the gauge-covariant pullback of the bulk fields to the worldvolume of the D$5$-branes. For example, the pullback of $\,\hat{E}_{\mathsf{ {\bar M}{\bar N}}}\,$ is
\begin{equation} 
\label{PullBack_E}
\textrm{P}[\hat{E}_{\mathsf{ {\bar M} {\bar N}}}]= \hat{E}_{\mathsf{{\bar M}{\bar N}}} + \lambda \, \left(  D_{\mathsf{\bar M}} Y^{\hat{a}} \, \hat{E}_{\hat{a} \mathsf{\bar N}} + D_{\mathsf{\bar N}} Y^{\hat{a}} \, \hat{E}_{\mathsf{\bar M} \hat{a}} \right) + \lambda^2  \, D_{\mathsf{\bar M}} Y^{\hat{a}} D_{\mathsf{\bar N}} Y^{\hat{b}} \hat{E}_{\hat{a} \hat{b}}  \ ,
\end{equation}
and involves the non-abelian covariant derivative
\begin{equation} 
\label{DY_def}
D_{\mathsf{\bar M}} Y^{\hat{a}} = \partial_{\mathsf{\bar M}} Y^{\hat{a}} + i [\mathcal{A}_{\mathsf{\bar M}}, Y^{\hat{a}}] \ .
\end{equation}
For the sake of completeness, the field $\,\boldsymbol{\hat{C}}\,$ in the WZ action (\ref{D5_DBI_WZ}) is a polyform formally containing (the sum of) all the R-R gauge potentials $\,C_{(2p)}$. The symbol $\,\iota_{Y}\,$ stands for the contraction (or interior product) of a differential form with the vector field $\,Y^{\hat{a}}$, namely,
\begin{equation} 
\label{doublecontraction}
\iota_Y \iota_Y \left (\tfrac{1}{2}  \, C_{\hat{a} \hat{b}} \, \eta^{\hat{a}} \wedge \eta^{\hat{b}}  \right ) = - \tfrac{1}{2} \, [Y^{\hat{a}}, Y^{\hat{b}}] \, C_{\hat{a} \hat{b}} \ .
\end{equation}
Lastly, our compactification scheme also incorporates O5-planes. At the perturbative level, orientifold planes are non-dynamical objects, which renders their associated DBI and WZ actions simpler. In particular, 
\begin{equation}
\label{O5_DBI_WZ}
\begin{array}{rcl}
S^{\textrm{DBI}}_{\textrm{O}5} &=& - N_{\textrm{O}5} \, T_{\textrm{O}5} \displaystyle\int_{\textrm{WV}(\textrm{O}5)} \textrm{d}^6\sigma \ e^{- \Phi} \sqrt{- \textrm{det} [G_{\mathsf{\bar{M}\bar{N}}}]} \ ,  \\[2mm]
S^{\textrm{WZ}}_{\textrm{O}5} &=& N_{\textrm{O}5} \, \mu_{\textrm{O}5} \displaystyle\int_{\textrm{WV}(\textrm{O}5)} C_{(6)} \ ,
\end{array}
\end{equation}
where $\,\mu_{\textrm{O}5}=T_{\textrm{O}5}=-2\, T_{\textrm{D}5}<0\,$ since we are considering the variant $\textrm{O}5^{-}$ of O$5$-planes \cite{Gimon:1996rq,Giveon:1998sr,Bergman:2001rp}.

\subsubsection*{Reduction of the DBI}

We now turn to the dimensional reduction of the DBI action in (\ref{D5_DBI_WZ}). As in the reduction of the closed-string sector, we restrict our attention to the $\textrm{SO}(3)$-invariant subset of fields and fluxes identified in equations~(\ref{scalars_open-sector_SO(3)}) and~(\ref{open_fluxes_SO(3)}).
We will expand the action to second order in $\lambda$, which suffices for comparison with the scalar potential of the $D=3$ supergravity, as the latter captures contributions only up to $\mathcal{O}(\lambda^2)$.

According to the definitions in equations~(\ref{Ehat_def}) and~(\ref{Taylor_expansion}), and recalling that 3D vectors and closed-string axions are turned off in our reduction ansatz, the components of $\,\hat{E}_{\mathsf{M}\mathsf{N}}\,$ take the form
\begin{equation}
\label{E_Hat_General}
\hat{E}_{\mu \nu} = \tau^{-2} g_{\mu \nu}
\hspace{5mm} , \hspace{5mm} 
\hat{E}_{\mu m} = \hat{E}_{m \mu} = 0
\hspace{5mm} , \hspace{5mm} 
\hat{E}_{mn} = G_{mn} + \lambda \left(\partial_c B_{m n}\right) Y^c \ ,
\end{equation}
where we only considered the first derivative of $\,B_{m n}\,$ as this will become a $H_{(3)}$-flux. In what follows, we will think of $\,\hat{E}_{\mathsf{M}\mathsf{N}}\,$ as
\begin{equation}
\hat{E}_{\mathsf{M}\mathsf{N}} = \hat{E}_{\mathsf{M}\mathsf{N}}^{(0)} + \lambda \, \hat{E}_{\mathsf{M}\mathsf{N}}^{(1)} \ ,
\end{equation}
which will help us to keep track of $\,\lambda$. The components of the covariant derivative in (\ref{DY_def}) will also be necessary. These are given by
\begin{equation}
\label{cov_der_Y}
\begin{array}{rcl}
D_{\mathsf{\mu}} Y^{\hat{a} I} =  \partial_{\mathsf{\mu}} Y^{\hat{a} I} + \mathcal{G}_{JK}{}^I \mathcal{A}_{\mathsf{\mu}}^J Y^{\hat{a} K}
\hspace{6mm} \textrm{ and } \hspace{6mm}
D_{i} Y^{\hat{a} I} =  \omega_{i\hat{b}}{}^{\hat{a}} \, Y^{\hat{b}I}+ \mathcal{G}_{JK}{}^{I}{\mathcal{A}_i}^J Y^{\hat{a} K} \ .
\end{array}
\end{equation}
Finally, the last building block required for the dimensional reduction of the DBI action is the field strength of the gauge theory on the D$5$-branes. According to (\ref{F2_def}), the relevant components for this computation are
\begin{equation} 
\label{F2_components}
    \mathcal{F}_{\mu i}{}^{I} = 2 \, \partial_{[\mu} \mathcal{A}_{i]}{}^{I} + \mathcal{G}_{JK}{}^{I} \,  \mathcal{A}_{\mu}{}^{J} \,  \mathcal{A}_{i}{}^{K} 
\hspace{3mm} \textrm{ and } \hspace{3mm}
     \mathcal{F}_{i j}{}^{I} = \mathfrak{f} \, \epsilon_{ijk} \, \delta^{k I} - \omega_{i j}{}^{k} \mathcal{A}_k{}^I + \mathcal{G}_{JK}{}^{I} \mathcal{A}_i{}^J \mathcal{A}_j{}^K .
\end{equation}

Equipped with all the above definitions, we now proceed to compute the matrix $\,\mathbb{M}_{\mathsf{{\bar M}{\bar N}}}\,$ in (\ref{M&Q-matrix_def}), which takes the form
\begin{equation}
\label{MMN_expansion}
\mathbb{M}_{\mathsf{{\bar M}{\bar N}}} = \textrm{P}[\hat{E}_{\mathsf{ {\bar M} {\bar N}}}] + \lambda \, \mathcal{F}_{\mathsf{ {\bar M} {\bar N}}} \ . 
\end{equation}
Substituting (\ref{PullBack_E}) into (\ref{MMN_expansion}) yields
\begin{equation}
\mathbb{M}_{\mathsf{{\bar M}{\bar N}}} =  G_{\mathsf{{\bar M}{\bar N}}} + \lambda \, \mathbb{M}_{\mathsf{{\bar M}{\bar N}}}^{(1)} + \lambda^2 \, \mathbb{M}_{\mathsf{{\bar M}{\bar N}}}^{(2)} + \dots \ ,
\end{equation}
with 
\begin{equation}
\mathbb{M}_{\mathsf{{\bar M}{\bar N}}}^{(1)} = \hat{E}_{\mathsf{ {\bar M} {\bar N}}}^{(1)} + \mathcal{F}_{\mathsf{ {\bar M} {\bar N}}}
\hspace{8mm} \textrm{ and } \hspace{8mm}
\mathbb{M}_{\mathsf{{\bar M}{\bar N}}}^{(2)} = D_{\mathsf{\bar M}} Y^{\hat{a}} \, D_{\mathsf{\bar N}} Y^{\hat{b}} \, G_{\hat{a}\hat{b}} \ .
\end{equation}
Let us also recall the relation
\begin{equation}
\label{Det_Trace}
\det \left( {\delta^\mathsf{\bar M}}_{\mathsf{\bar N}} + \lambda {A^\mathsf{\bar M}}_{\mathsf{\bar N}} \right) = 1 + \lambda {A^\mathsf{\bar M}}_{\mathsf{\bar M}} + \frac{\lambda^2}{2} \left( \left( {A^\mathsf{\bar M}}_{\mathsf{\bar M}} \right)^2 - {A^\mathsf{\bar M}}_{\mathsf{\bar N}} {A^\mathsf{\bar N}}_{\mathsf{\bar M}} \right) + \ldots \ ,
\end{equation}
with ${A^\mathsf{\bar M}}_{\mathsf{\bar N}}$ a generic matrix and $\lambda$ a small expansion parameter. Given this result, and using the antisymmetry of the leading-order correction $\hat{E}^{(1)}_{\mathsf{\bar M \bar N}}$, the expansion of the determinant of $\,\mathbb{M}_{\mathsf{\bar M \bar N}}\,$ takes the form
\begin{equation}
\label{eq:Det_M_Expanded}
\begin{array}{rcl}
\sqrt{- \det\left[\mathbb{M}_{\mathsf{{\bar M}{\bar N}}}\right]} & = & \sqrt{- \, \det\left[G_{\mathsf{{\bar M}{\bar R}}}\right]} \, \sqrt{\det \left[{\delta^\mathsf{\bar R}}_{\mathsf{\bar N}} + \lambda \, G^{\mathsf{{\bar R}{\bar S}}} \mathbb{M}_{\mathsf{{\bar S}{\bar N}}}^{(1)} + \lambda^2 \, G^{\mathsf{{\bar R}{\bar S}}} \mathbb{M}_{\mathsf{{\bar S}{\bar N}}}^{(2)} + \ldots \right]} \\[4mm]
& = & \sqrt{- \det\left[G_{\mathsf{{\bar M}{\bar R}}}\right]} \Big( 1 + \frac{\lambda^2}{2} G^{\mathsf{{\bar R}{\bar S}}} \mathbb{M}_{\mathsf{{\bar S}{\bar R}}}^{(2)} - \frac{\lambda^2}{4} G^{\mathsf{{\bar R}{\bar S}}} \mathbb{M}_{\mathsf{{\bar S}{\bar N}}}^{(1)} G^{\mathsf{{\bar N}{\bar P}}} \mathbb{M}_{\mathsf{{\bar P}{\bar R}}}^{(1)} + \ldots  \Big) \ . 
\end{array}
\end{equation}
On the other hand, the $\,\textrm{SO}(3)$-invariant components of the $\,\mathbb{Q}\,$ matrix in (\ref{M&Q-matrix_def}) are given by
\begin{equation}
{\mathbb{Q}^a}_b = {\delta^a}_b + i \lambda \left[Y^a, Y^c \right] \hat{E}_{cb} 
\hspace{5mm} , \hspace{5mm} 
{\mathbb{Q}^a}_7 = 0 
\hspace{5mm} , \hspace{5mm}
{\mathbb{Q}^7}_b = 0 
\hspace{5mm} , \hspace{5mm}
{\mathbb{Q}^7}_7 = 1 \ ,
\end{equation}
so that, making use of (\ref{Det_Trace}) and the fact that $\,Y^{7}=0\,$ in the SO(3)-invariant sector, 
\begin{equation}
\label{Q_expanded}
\sqrt{\det \left[ {\mathbb{Q}^{\hat{a}}}_{\hat{b}} \right]} = 1 + \frac{\lambda^2}{4} \left[ Y^a, Y^b \right] G_{bc} \left[ Y^c, Y^d \right] G_{da} + i \frac{\lambda^2}{2} \left[Y^a, Y^b \right] Y^c \, \partial_c B_{b a} + \ldots \ .
\end{equation}

We can determine the kinetic terms for the open-string scalars and the scalar potential, by inserting (\ref{eq:Det_M_Expanded}) and (\ref{Q_expanded}) in the full expression (\ref{M&Q-matrix_def}) and then using the gauge algebra relations in (\ref{eq:GeneratorRelations}) and the expression (\ref{eq:indirect_g_s}) for the dilaton. 
In particular, the kinetic terms for the open-string axions $\,\alpha\,$ and $\,\psi\,$ arise from 
\begin{equation*}
- T_{\textrm{D}5} \ \textup{Tr} \left[ e^{-\Phi} \sqrt{- \det\left[G_{\mathsf{{\bar M}{\bar R}}}\right]} \left(\frac{\lambda^2}{2} D_\mu Y^a D_\nu Y^b G_{a b} \ G^{\mu \nu} + \frac{\lambda^2}{2} \mathcal{F}_{\mu i} \, G^{i j} \, \mathcal{F}_{j \nu} \, G^{\mu \nu} \right) \right] \ ,
\end{equation*}
and take the form
\begin{equation}
\label{L_kin_open}
L^{\textrm{open}}_{\textup{kin}} = \lambda^2 \, N_{\textrm{D}5} \, T_{\textrm{D}5} \,  \sqrt{- g} \left(- \dfrac{3 \, \ell_a^2 \ell_i^3 \rho^{3/2} }{2 \, \tau^{1/2}} \partial_{\mu} \psi \partial^{\mu} \psi - \dfrac{3 \, \ell_i }{2 \, \rho^{5/2} \tau^{1/2}} \partial_{\mu} \alpha \partial^{\mu} \alpha  \right) + \ldots \ .
\end{equation}
On the other hand, the terms containing only scalars in the DBI action for D$5$-branes (\ref{D5_DBI_WZ}) and O$5$-planes (\ref{O5_DBI_WZ}) produce the open-string contribution to the scalar potential. Recalling that $\,T_{\textrm{O}5}=-2\,T_{\textrm{D}5}$, this takes the form
\begin{equation}
\label{V_DBI}
V_{\textup{DBI}} = -  T_{\textrm{D}5} \, \big(2 N_{\textrm{O}5} - N_{\textrm{D}5} \big) \, V_{\textup{DBI}}^{(0)} + \lambda^2 \, N_{\textrm{D}5} \,  T_{\textrm{D}5} \, V_{\textup{DBI}}^{(2)} \ ,
\end{equation}
with a flux-independent $\lambda^{0}$-contribution
\begin{equation}
\label{V_DBI_0}
V_{\textup{DBI}}^{(0)} = \tau^{-5/2} \rho^{-1/2} {\ell_i}^3 \ ,
\end{equation}
and a $\lambda^{2}$-contribution 
\begin{equation}
\label{V_DBI_2}
\begin{array}{rcl}
V_{\textup{DBI}}^{(2)} &=& \frac{3}{2} \tau ^{-5/2} \rho^{-9/2} {\ell_i}^{-1} \left(\mathfrak{f} - \omega_6  \alpha  + \mathfrak{g} \, \alpha^2 + h_{32} \psi \right)^2  + \frac{3}{2} \tau^{-5/2} \rho^{7/2} {\ell_i}^3 {\ell_a}^4 \mathfrak{g}^2 \psi^4 \\[2mm]
&+&  \frac{1}{2} \tau^{-5/2} \rho^{-1/2} \bigl( 6 \ {\ell_i} {\ell_a}^2 \mathfrak{g}^2 \alpha^2 \psi^2 - 2 \ {\ell_i}^3 h_{31} \,  \mathfrak{g}  \, \psi^3 + 12 \ {\ell_i} {\ell_a}^2 \omega_5 \,  \mathfrak{g} \, \alpha \, \psi^2   \\[2mm]
&+& 3 \ {\ell_i}^{-5} {\ell_a}^{-6}  {\omega_4}^2 \psi^2 + 6 \ \ell_i {\ell_a}^2 {\omega_5}^2 \psi ^2 \bigr) \ .
\end{array}
\end{equation}
Importantly, the flux-independent contribution (\ref{V_DBI_0}) that arises at zeroth order in $\,\lambda\,$ comes along with a coefficient $\,T_{\textrm{D}5} \, \big(2 N_{\textrm{O}5} - N_{\textrm{D}5} \big)\,$ in (\ref{V_DBI}). This is precisely the quantity that enters the Bianchi identity in (\ref{BI_10D}) (with $p=5$) when the O$5$/D$5$ sources in (\ref{O5/D5_location}) are included. More specifically, one finds
\begin{equation}
\label{BI_F3_component}
\left. dF_{(3)} \, \right|_{dy^{7} \wedge \,dy^{a}  \wedge \,dy^{b} \wedge \,dy^{c} } = 3\, \omega_{1} \, f_{32} - 3\,   \omega_{2} \, f_{33} =  \underbrace{(2\pi)^{-1} (2 \pi \ell_s)^8}_{2\kappa^{2}} T_{\textrm{D}5} \, \big(2N_{\textrm{O}5} - N_{\textrm{D}5} \big) \ ,
\end{equation}
so that
\begin{equation}
\label{tadpole_replacement}
2 N_{\textrm{O}5} - N_{\textrm{D}5} = \frac{1}{(2\kappa^{2}) \, T_{\textrm{D}5}}  \left( 3\, \omega_{1} \, f_{32} - 3\,   \omega_{2} \, f_{33} \right) = (2\pi \ell_{s})^{-2} \left( 3\, \omega_{1} \, f_{32} - 3\,   \omega_{2} \, f_{33} \right) \ ,
\end{equation}
and the $F_{(3)}$ flux components are expressed in units of $(2\pi\ell_{s})^{2}$.\footnote{In general, the components of a RR flux $F_{(p+1)}$ are expressed in units of $(2\pi\ell_{s})^{p}$.} The replacement in (\ref{tadpole_replacement}) is essential for matching the scalar potential derived via dimensional reduction with that of a half-maximal $\,D=3\,$ supergravity. This matching also requires to redefine the open-string moduli and flux parameters as
\begin{equation}
\label{open-sector_redefinitions}
\alpha \rightarrow \frac{\alpha}{\epsilon}
\hspace{8mm} , \hspace{8mm}
\psi \rightarrow \frac{\psi}{\epsilon}
\hspace{8mm} , \hspace{8mm}
\mathfrak{g} \rightarrow \epsilon  \, \mathfrak{g}
\hspace{8mm} , \hspace{8mm}
\mathfrak{f} \rightarrow \frac{\mathfrak{f}}{\epsilon} \ ,
\end{equation}
with $\,\epsilon^{2} = (2\kappa^{2}) \, \lambda^{2} \, N_{\textrm{D}5} \,T_{\textrm{D}5}$, so that the prefactor in front of the $\lambda^{2}$-contribution in (\ref{V_DBI}) gets replaced by the $(2 \kappa^2)^{-1}$ that accompanies the 3D supergravity action. Note also that the redefinition (\ref{open-sector_redefinitions}) also removes such a prefactor from the kinetic terms in (\ref{L_kin_open}).

\subsubsection*{Reduction of the WZ}

The WZ term in (\ref{D5_DBI_WZ}) is topological and, therefore, it does not directly contribute to the scalar potential of the reduced $D=3$ theory. However, looking at its dimensional reduction, we can derive the modifications of the RR fluxes in (\ref{redefinition_F}) that are required to consistently incorporate the dynamics of the D$5$-brane sources in the compactification scheme \cite{Balaguer:2023jei, Balaguer:2024cyb}. 

To better explain this point, let us recall that in the democratic formulation of type IIB supergravity, the Bianchi identities (\ref{BI_10D}) for the RR fields can be obtained by performing the variation of the bulk and the WZ action with respect to the $C_{(p + 1)}$ gauge fields\footnote{The DBI action (\ref{D5_DBI_WZ}) does not contain any dependence on the RR gauge fields.} and then applying the Hodge duality relations (\ref{Hodge_Duality}). However, the dynamics of the sources induces corrections of higher order in $\lambda$ in the WZ action and then in the Bianchi identities (\ref{BI_10D}). More concretely, the latter get modified into
\begin{equation}
\label{modified_Bianchi_10D}
d  \tilde{F}_{(8-p)} - H_{(3)} \wedge \tilde{F}_{(6-p)} =  J_{\textrm{O}p/\textrm{D}p} + J_{(9 - p)}^{\textrm{D}5} \ .
\end{equation}
The left-hand side of (\ref{modified_Bianchi_10D}) stems from the variation of the bulk action with respect to $C_{(p + 1)}$, whereas $\,J_{\textrm{O}p/\textrm{D}p}\,$ and $\,J_{(9 - p)}^{\textrm{D}5}\,$ arise from the WZ term respectively as zero and higher-order contributions in $\,\lambda$. Importantly, for the RR fluxes to obey (\ref{modified_Bianchi_10D}), they must be modified as described in (\ref{redefinition_F}): this is why tilded RR field strengths are appearing in (\ref{modified_Bianchi_10D}).

Let us illustrate this issue with an example: the redefinition of the RR flux $\,F_{(1)}$. Both in the modified field-strengths and in $ J_{(9 - p)}^{\textrm{D}5}$ we will stick to corrections up to $\mathcal{O} (\lambda^2)$, which are the only ones that the three-dimensional gauged supergravity will be sensitive to. The modified flux $\,{\tilde F}_{(1)} = F_{(1)} + \Delta F_{(1)}\,$ satisfies the Bianchi identity 
\begin{equation}
\label{F1tilde_Bianchi}
d {\tilde F}_{(1)} = J^{\textrm{D}5}_{(2)} \hspace{10mm} \Longleftrightarrow \hspace{10mm}
\left\{ \begin{matrix} 
d F_{(1)} = 0 \hspace{5mm} & \textrm{ at } \,\,\,\, \mathcal{O}(\lambda^0) \\[4mm]
d \Delta F_{(1)} = J^{\textrm{D}5}_{(2)} \hspace{5mm} & \textrm{ at } \,\,\,\, \mathcal{O}(\lambda^2)
\end{matrix}\right.  \ .
\end{equation}
In the democratic formulation of type IIB supergravity \cite{Bergshoeff:2001pv}, $\,C_{(0)}\,$ is dual to an eight-form potential $\,C_{(8)}$. We will then get (\ref{F1tilde_Bianchi}) by varying the full action (\ref{S_IIB}) with respect to $\,C_{(8)}$. The relevant kinetic term in the bulk action (\ref{S_bulk}) is
\begin{equation}
S_{\textrm{bulk}} \supset - \frac{1}{2 \kappa^2} \int \frac{1}{4} \,  F_{(9)} \wedge \star F_{(9)} \, ,
\end{equation}
whose variation with respect to $\,C_{(8)}\,$ yields\footnote{In our conventions, $\varepsilon^{ aibjck7 \mu_0 \mu_1 \mu_2 } = + 1 \,$.} (up to a standard boundary term)
\begin{equation}
\label{delta_bulk_C8}
\delta S_{\textrm{bulk}} = \frac{1}{2 \kappa^2} \int \frac{1}{8!} \, \delta C_{\mathsf{M}_1 \dots \mathsf{M}_8 } \, \partial_{\mathsf{M}_9} 
( \star F_{(9)} )_{\mathsf{M}_{10}} \,  \varepsilon^{\mathsf{M}_1 \dots \mathsf{M}_{10}} \ .
\end{equation}
As an example, the variation with respect to the $\textrm{SO}(3)$-invariant component $\,C_{\mu_0\mu_1\mu_2 ij ab 7}\,$ yields 
\begin{equation}
\label{Der_F1}
\frac{1}{2 \kappa^2} \, \partial_{[c|} \tilde{F}_{(1) |k] } = \frac{1}{2 \kappa^2} \, \omega_{k c}{}^7 \tilde{f}_1  \ ,
\end{equation}
where we have used the Hodge duality relations (\ref{Hodge_Duality}) and imposed the redefinition (\ref{redefinition_F}) of the RR fluxes.

%

As far as the WZ action (\ref{D5_DBI_WZ}) is concerned, the contribution to the polyform in the integrand that depends on $\,C_{(8)}\,$ is given by
\begin{equation}
\label{WZ_Delta_f1}
S_{\textrm{D}5}^{\textrm{WZ}} \supset  \mu_{\textrm{D}5} \displaystyle\int_{\textrm{WV}(\textrm{D}5)}  \textrm{Tr} \left\{ \textrm{P} \left[i \, \lambda \, \iota_Y \, \iota_Y \big(\hat{C}_{(8)} \big) \right] \right\}   \ .
\end{equation}
Using the expansion in (\ref{Taylor_expansion}), and keeping at most $\mathcal{O}(\lambda^2)$ terms, the integrand in (\ref{WZ_Delta_f1}) contains specific contributions of the form
\begin{equation}
\begin{array}{rcl}
\textrm{P} \left[i \, \lambda \, \iota_Y \, \iota_Y \big(\hat{C}_{(8)} \big) \right]_{\mu_0 \mu_1 \mu_2 ijk} & = & 
\textrm{P} \left[ -\frac{\lambda}2 \,  \mathcal{G}_{IJ}{}^K Y^{\hat{a }I} Y^{\hat{b} J} {\hat C}_{\mu_0 \mu_1 \mu_2 ijk\hat{a}\hat{b}} \, t_K \right] =\\[2mm]
&-& \frac{\lambda^2}2 \, \mathcal{G}_{IJ}{}^K Y^{\hat{a} I} Y^{\hat{b} J} Y^{\hat{c}} \, \partial_{\hat{c}} {C}_{\mu_0 \mu_1 \mu_2 ijk\hat{a}\hat{b}} \, t_K \\[2mm] 
& - &  \frac{\lambda^2}2 \,  \left( \mathcal{G}_{IJ}{}^K Y^{\hat{a} I} Y^{\hat{b} J} D_{\mu_0} Y^{\hat{c}} {C}_{\hat{c} \mu_1 \mu_2 i jk\hat{a}\hat{b}} \, t_K + \ldots \right) \\[2mm]
& - & \frac{\lambda^2}2 \, \left( \mathcal{G}_{IJ}{}^K Y^{\hat{a} I} Y^{\hat{b} J} D_i Y^{\hat{c}} {C}_{\mu_0 \mu_1 \mu_2 \hat{c} jk\hat{a}\hat{b}} \, t_K + \ldots \right) \ ,
\end{array}
\end{equation}
where the ellipsis in the two last brackets stand for two analogous terms in which the world-volume index of the gauge covariant derivative $D$ is $(\mu_1,\mu_2)$ or $(j,k)$.  Focusing again on the $\textrm{SO}(3)$-invariant component $\,C_{\mu_0\mu_1\mu_2 ij ab 7}$, only the last term is of relevance. In particular, using (\ref{cov_der_Y}), it contributes to the variation of the WZ term as
\begin{equation}
\label{delta_WZ_C8}
\delta S_{\textrm{D}5}^{\textrm{WZ}} = -  \lambda^2 \, N_{\textrm{D}5} \, \mu_{\textrm{D}5} \, \mathfrak{g} \, \psi^3 \, \omega_{k c}{}^{7} \, \delta C_{\mu_0 \mu_1 \mu_2 i j a b 7} \ .
\end{equation}
Bringing together (\ref{delta_bulk_C8})-(\ref{Der_F1}) and (\ref{delta_WZ_C8}), the variation of the full action (\ref{S_IIB}) yields
\begin{equation}
\label{full_variation_C8}
\delta S_{\textrm{IIB}} = \omega_{k c}{}^7 \left ( \frac{1}{2 \kappa^2} \tilde{f}_1 \, - \lambda^2  N_{\textrm{D}5} \,\mu_{\textrm{D}5} \, \mathfrak{g} \psi^3 \right ) \, \delta C_{\mu_0 \mu_1 \mu_2 i j a b 7}  \, .
\end{equation}
Demanding (\ref{full_variation_C8}) to vanish, one can read off the modification for the flux $\,F_{(1)} = f_1 \eta^7\,$ at $\,\mathcal{O}(\lambda^2)$, which by virtue of (\ref{f1=0}) and the redefinitions in (\ref{open-sector_redefinitions}), turns out to be
\begin{equation}
\label{ftilde_1_redef}
\tilde{f}_{1} = \Delta f_1 = \mathfrak{g} \, \psi^3  \ .
\end{equation}

The modifications of the RR fluxes $\,F_{(3)}$, $\,F_{(5)}\,$ and $\,F_{(7)}\,$ are obtained following the same procedure as above (see \cite{Balaguer:2023jei,Balaguer:2024cyb}). For a given RR potential $\,F_{(p+2)}$, the relevant contribution to the polyform (\ref{D5_DBI_WZ}) is identified with
\begin{equation}
\label{D5_DBI_WZ_relevant_contribution}
S_{\textrm{D}5}^{\textrm{WZ}} \, \supset \mu_{\textrm{D}5} \displaystyle\int_{\textrm{WV}(\textrm{D}5)}  \textrm{Tr} \left\{ \textrm{P} \big[ e^{i \lambda \iota_Y \iota_Y} \, \hat{C}_{(7-p)} \wedge e^{\hat{B}_{(2)}} \wedge e^{\lambda \mathcal{F}} \big] \right\}  \ .
\end{equation}
Then the integrand in (\ref{D5_DBI_WZ_relevant_contribution}) is expanded up to $\,\mathcal{O}(\lambda^2)\,$ terms, while neglecting the $\,\mathcal{O}(\lambda)\,$ contributions since any single generator $\,t_I\,$ has vanishing trace. Some long (and tedious) computations yield
\begin{equation}
\label{ftilde_3_redef}
\tilde{f}_{31} = f_{31} 
\hspace{7mm} , \hspace{7mm}
\tilde{f}_{32} = f_{32} - \psi^2 \left( \omega_5 + \mathfrak{g} \, \alpha \right) \hspace{7mm} , \hspace{7mm}
\tilde{f}_{33} = f_{33} - \tfrac{1
}{2} \psi^2 \omega_4 \ ,
\end{equation}
together with
\begin{equation}
\label{ftilde_5_redef}
\tilde{f}_{5} = f_{5} + \tfrac{1}{2} \psi \left[ \alpha \left( 2\omega_5 - \omega_6 \right) + \psi \, h_{32} + 2 \, \mathfrak{f} + 2 \, \mathfrak{g} \, \alpha^2 \right] \ ,
\end{equation}
and
\begin{equation}
\label{ftilde_7_redef}
\tilde{f}_{7} =  f_{7}  + \tfrac{1}{2} \alpha \left[ 3 \, \alpha \, \omega_6 - 3 \, \psi \, h_{32} - 6 \, \mathfrak{f} - 2 \, \mathfrak{g} \, \alpha^2 \right] \ .
\end{equation}

\subsubsection{Summary of the reduction}

The reduction of the type IIB closed-string and open-string sectors down to three dimensions has given rise to an Einstein-scalar theory for the closed-string moduli $\,(\tau\,,\,\rho\,,\,\ell_{i}\,,\,\ell_{a})\,$ and the open-string ones $\,(\psi\,,\,\alpha)$. The 3D action reads
\begin{equation}
\label{3D_effective_action}
S_{\textrm{3D}} = \frac{1}{2\kappa^{2}} \displaystyle\int \textrm{d}^{3}x \, \sqrt{-g} \,  \big( R + L_{\textrm{kin}} - V \big) \ ,
\end{equation}
with kinetic terms in (\ref{L_kin_closed}) and (\ref{L_kin_open}) (after the redefinitions in (\ref{open-sector_redefinitions})) combining into
\begin{equation}
\label{L_kin_3D}
\begin{array}{rcl}
L_{\textrm{kin}} &=& - \dfrac{1}{\tau^2} \partial_\mu \tau \partial^\mu \tau - \dfrac{7}{\rho^2} \partial_\mu \rho \partial^\mu \rho - \dfrac{12}{{\ell_i}^2} \partial_\mu {\ell_i} \partial^\mu {\ell_i} - \dfrac{18}{\ell_a \ell_i} \partial_\mu \ell_a \partial^\mu \ell_i - \dfrac{12}{{\ell_a}^2} \partial_\mu {\ell_a} \partial^\mu {\ell_a} \\[4mm]
&& - \dfrac{3 \, \ell_i }{2 \, \rho^{5/2} \tau^{1/2}} \partial_{\mu} \alpha \partial^{\mu} \alpha - \dfrac{3 \, \ell_a^2 \ell_i^3 \rho^{3/2} }{2 \, \tau^{1/2}} \partial_{\mu} \psi \partial^{\mu} \psi   \ ,
\end{array}
\end{equation}
and a scalar potential $\,V\,$ induced by the various types of fluxes and the YM structure constants given by
\begin{equation}
\label{V_3D}
V = V_{\omega} \, (\ref{V_omega}) + V_{H} \, (\ref{V_H}) + V_{\widetilde{F}} \, (\ref{V_F}) + \widetilde{V}_{\textrm{DBI}} \, (\ref{V_DBI}) \ .
\end{equation}
It is worth emphasising that the contributions $\,V_{\widetilde{F}}\,$ and $\,\widetilde{V}_{\textrm{DBI}}\,$ in (\ref{V_3D}) are nothing but the ones in (\ref{V_F}) and (\ref{V_DBI}) with the RR fluxes in there being replaced by the tilded counterparts in (\ref{ftilde_1_redef}) and (\ref{ftilde_3_redef})-(\ref{ftilde_7_redef}).

\subsection{Jacobi and Bianchi identities}
\label{sec:Jacobi&Bianchi_id}

The flux parameters and Yang–Mills structure constants are required to satisfy a set of algebraic consistency conditions. The first one is the Jacobi identity $\,\omega \, \omega = 0\,$ given in (\ref{ww_constraint}), which yields
\begin{equation}
\label{QC_ww}
\begin{array}{rclcrclcrcl}
\omega_{3} \, \omega_{4} + \omega_{5} \,(\omega_{5} +\omega_{6}) &=& 0 
& \hspace{3mm} , & \hspace{3mm}    
\omega_{4} \,(2\, \omega_{5} +\omega_{6}) &=& 0  \\[2mm]
\omega_{1} \, \omega_{3} - \omega_{2} \,(\omega_{5} +\omega_{6}) &=& 0 
& \hspace{3mm} , & \hspace{3mm}    
\omega_{3} \,(2\, \omega_{5} +\omega_{6}) &=& 0  
& \hspace{3mm} , & \hspace{3mm}    
\omega_{1} \, \omega_{4} &=& 0 \ . \ \\[2mm]
\omega_{2} \, \omega_{4} - \omega_{1} \,(\omega_{5} +\omega_{6}) &=& 0
& \hspace{3mm} , & \hspace{3mm}    
\omega_{1} \, \omega_{3} - 2\, \omega_{2} \, \omega_{5} &=& 0  \ .
\end{array}
\end{equation}
The Jacobi identity can be reinterpreted as a nilpotency condition, $\,d^{2}=0$, on a \textit{twisted} exterior derivative defined on the internal space. A violation of the Jacobi identity has been associated with the presence of Kaluza--Klein monopoles in the compactification scheme \cite{Villadoro:2007yq}.

In addition to the Jacobi identity (\ref{ww_constraint}), one must also impose the Bianchi identities (\ref{BI_10D}) for the NS-NS and R-R gauge potentials of type IIB supergravity:
\begin{itemize}

\item There is the Bianchi identity $\,d H_{(3)} = 0\,$ which yields
\begin{equation}
\label{QC_DH3}
\omega_{3} \, h_{32} \,\,=\,\, 0    
\hspace{8mm} , \hspace{8mm} 
2\,\omega_{2} \, h_{32} - \omega_{3} \, h_{31} \,\,= \,\,0   \ , 
\end{equation}
and signals the absence of spacetime filling NS$5$-branes.

\item There is the Bianchi identity $\,d F_{(3)} = J_{\textrm{O$5$/D$5$}}\,$ where $\,J_{\textrm{O$5$/D$5$}}\,$ accounts for the O$5$-planes and D$5$-branes in the compactification (in the smeared limit). For the O5/D5 sources in (\ref{O5/D5_location}) one finds
\begin{equation}
\label{O5/D5_unrestrited}
3\, \omega_{1} \, f_{32} - 3\,   \omega_{2} \, f_{33} =  (2\pi \ell_{s})^{2}  \, \big(2N_{\textrm{O}5} - N_{\textrm{D}5} \big)  \ .
\end{equation}
In addition, one also finds 
\begin{equation}
\label{QC_DF3}
\begin{array}{rcl}
\omega_{2} \, f_{31} + (2\, \omega_{5} + \omega_{6}) \, f_{32} + 2 \, \omega_{3} \, f_{33} & = & 0 \ , \\[2mm]
\omega_{1} \, f_{31} - (2\, \omega_{5} + \omega_{6}) \, f_{33} - 2 \, \omega_{4} \, f_{32}  & = &  0  \ , 
\end{array}
\end{equation}
which signal the absence of spacetime filling O$5$/D$5$-sources different from the ones in (\ref{O5/D5_location}).

\item There is the Bianchi identity $\,d F_{(5)} - H_{(3)} \wedge F_{(3)} = 0\,$ which yields
\begin{equation}
\label{QC_DF5}
3 \, \omega_{4} \, f_{5} - h_{31} \, f_{31} + 3 \, h_{32} \, f_{33} \,\, = \,\, 0 \ , 
\end{equation}
and signals the absence of spacetime filling D$3$-branes (and O$3$-planes)

\item There is the Jacobi identity of the YM gauge algebra in (\ref{eq:GeneratorRelations}), \textit{i.e.} $\,\mathcal{G}_{[I J}{}^K \, \mathcal{G}_{P]K}{}^Q = 0$, which is automatically satisfied for our choice of $\,\textrm{G}_{\textrm{YM}\,}$ in (\ref{G_YM_SO(3)}).

\item There is the (non-abelian) Bianchi identity in the YM sector $\,(d + \mathcal{A} \, \wedge) \mathcal{F}  = 0\,$ which yields the constraint $\,\mathcal{G}_{IJ}{}^{K} \, \mathcal{F}_{jk}{}^{I}=0\,$ for a background YM flux.\footnote{Note that $\,\omega_{[i j}{}^l \, \mathcal{F}_{k]l }{}^I = 0\,$ in the $\,\textrm{SO}(3)$-invariant sector.} Plugging the SO(3)-invariant components in (\ref{open_fluxes_SO(3)}), this constraint reduces to
\begin{equation}
\label{QC_open_string}
\mathfrak{f} \, \mathfrak{g} = 0 \ .
\end{equation}

\end{itemize}

\section{Three-dimensional flux vacua with open strings}
\label{sec:3D_supergravity}

As shown in \cite{Arboleya:2024vnp}, the closed-string sector of the type IIB compactifications with O$5$/D$5$ sources that we have presented in the previous section describes a (half-maximal) $\,\mathcal{N}=8$, $\,D=3$ gauged supergravity \cite{Nicolai:2001ac,deWit:2003ja} (see also \cite{Deger:2019tem}). This theory contains $\,64\,$ scalar fields (see Table~\ref{Table:O5_Fields}), which describe a coset space 
\begin{equation}
\label{M_N=8_closed}
\mathcal{M}^{\textrm{closed}}_{\mathcal{N}=8} = \frac{\textrm{SO}(8,8)}{\textrm{SO}(8) \times \textrm{SO}(8)} \ .
\end{equation}
However, the $\,\textrm{SO}(3)$-invariant field content -- which consists of the four dilatons in (\ref{dilatons_def}) and the four axions in (\ref{axions_def}) -- describes itself an $\,\mathcal{N}=2\,$ subsector of half-maximal supergravity. This $\,\mathcal{N}=2\,$ subsector is better described in terms of four complex fields, $\,R$, $\,S$, $\,T$ and $\,U$, yielding a so-called RSTU-model \cite{Arboleya:2024vnp}. The four complex fields in the RSTU-models describe a coset geometry that consists of four copies of the Poincar\'e (upper) half-plane, namely, 
\begin{equation}
\label{M_N=2_closed}
\mathcal{M}^{\textrm{closed}}_{\mathcal{N}=2} = \left[ \frac{\textrm{SL}(2)}{\textrm{SO}(2)} \right]_{R} \times\,\,\, \left[ \frac{\textrm{SL}(2)}{\textrm{SO}(2)} \right]_{S} \times\,\,\, \left[ \frac{\textrm{SL}(2)}{\textrm{SO}(2)} \right]_{T} \times\,\,\, \left[ \frac{\textrm{SL}(2)}{\textrm{SO}(2)} \right]_{U} \subset \mathcal{M}^{\textrm{closed}}_{\mathcal{N}=8} \ .
\end{equation}
Finally, when turning off the axions in (\ref{axions_def}), as we do in this work, one is left with an $\,\mathcal{N}=1\,$ supergravity model of the type investigated in \cite{Emelin:2021gzx}. The scalar geometry described by the four dilatons in (\ref{dilatons_def}) is just
\begin{equation}
\label{M_N=1_closed}
\mathcal{M}^{\textrm{closed}}_{\mathcal{N}=1} = \mathbb{R}_{R} \times\,\,\, \mathbb{R}_{S} \times\,\,\, \mathbb{R}_{T} \times\,\,\, \mathbb{R}_{U} \ .
\end{equation}

In this section we will extend the setup of \cite{Arboleya:2024vnp} to accommodate the open-string sector which, when restricted to $\,\textrm{SO}(3)$-invariant fields, includes the two additional fields $\,(\psi,\alpha)\,$ in (\ref{scalars_open-sector_SO(3)}). We will explicitly show that the 3D action in (\ref{3D_effective_action})-(\ref{V_3D}) can be recast as a half-maximal gauged supergravity coupled to $\,\mathfrak{N}=3\,$ vector multiplets.

\subsection{Matter-coupled RSTU-models}

In order to accommodate a YM theory with gauge group $\,\textrm{G}_{\textrm{YM}}$, let us consider half-maximal $\,D=3\,$ supergravity coupled to a number
\begin{equation}
\label{N=dimG}
\mathfrak{N}=\textrm{dim}(\textrm{G}_{\textrm{YM}})
\end{equation}
of vector multiplets \cite{Nicolai:2001ac,deWit:2003ja}. The theory now contains $\,64+8\mathfrak{N}\,$ scalar fields, which describe a coset space 
\begin{equation}
\label{M_N=8_open}
\mathcal{M}^{\textrm{closed/open}}_{\mathcal{N}=8} = \frac{\textrm{SO}(8,8+\mathfrak{N})}{\textrm{SO}(8) \times \textrm{SO}(8+\mathfrak{N})} \ .
\end{equation}
A useful parameterisation of the $\,64+8\mathfrak{N}\,$ scalars is given in terms of a so-called coset representative \cite{Maharana:1992my}
\begin{equation}
\label{Vielb_N=2}
\mathcal{V} = \left( \begin{matrix}
     \boldsymbol{e} & 0 & 0 \\[1mm] 
     - \boldsymbol{c}^T \, \boldsymbol{e} & \boldsymbol{e}^{-T} & \boldsymbol{a}^T \\[1mm]
     - \boldsymbol{a} \, \boldsymbol{e} & 0 & \mathbb{I}_\mathfrak{N}
     \end{matrix} \right) \in \frac{\textrm{SO}(8,8+\mathfrak{N})}{\textrm{SO}(8) \times \textrm{SO}(8+\mathfrak{N})}  \ ,
\end{equation}
which involves an $\,\mathfrak{N} \times 8\,$ matrix $\,\boldsymbol{a}$, and two $\,8 \times 8\,$ matrices $\,\boldsymbol{e} \in \textrm{GL}(8)/\textrm{SO}(8)\,$ and $\,\boldsymbol{c} = \boldsymbol{b} + \frac{1}{2} \boldsymbol{a}^T \boldsymbol{a}\,$ with $\,\boldsymbol{b}=-\boldsymbol{b}^{T}$. The matrices $\,\boldsymbol{a}$, $\,\boldsymbol{e}\,$ and $\,\boldsymbol{b}\,$ contain $\,3\mathfrak{N}$, $\,36\,$ and $\,28\,$ scalars, respectively. From the coset representative in (\ref{Vielb_N=2}), we introduce a scalar-dependent matrix
\begin{equation}
\label{M_scalar_matrix}
M = \mathcal{V} \, \mathcal{V}^{T}  =     
\left( \begin{matrix}
\boldsymbol{g} & - \boldsymbol{g} \, \boldsymbol{c} & -\boldsymbol{g} \, \boldsymbol{a}^{T} \\[1mm]
- \boldsymbol{c}^T \, \boldsymbol{g} & \boldsymbol{g}^{-1}+\boldsymbol{c}^{T}\,\boldsymbol{g}\,\boldsymbol{c}+\boldsymbol{a}^{T}\,\boldsymbol{a} & \boldsymbol{a}^{T}+\boldsymbol{c}^{T}\,\boldsymbol{g}\,\boldsymbol{a}^{T} \\[1mm]
- \boldsymbol{a} \, \boldsymbol{g} & \boldsymbol{a}+\boldsymbol{a}\,\boldsymbol{g}\,\boldsymbol{c}  & \mathbb{I}_\mathfrak{N}+\boldsymbol{a}\,\boldsymbol{g}\,\boldsymbol{a}^{T}
\end{matrix} \right) \in \textrm{SO}(8,8+\mathfrak{N}) \ ,
\end{equation}
with $\,\boldsymbol{g}=\boldsymbol{e}\,\boldsymbol{e}^{T}$, which will be the one appearing in the Lagrangian of the $D=3$ half-maximal supergravity (see (\ref{3D_sugra_action})).

\subsubsection{$\textrm{SO}(3)$ symmetry}

In order to recover the dimensional reduction performed in Section~\ref{sec:IIB_reduction}, we must restrict to the subset of scalars which are invariant under a specific $\,\textrm{SO}(3)\,$ subgroup embedded inside the duality group $\,\textrm{SO}(8,8+\mathfrak{N})\,$ of half-maximal supergravity as
\begin{equation}
\label{SO(3)_embedding}
\textrm{SO}(3) \subset \textrm{SO}(3) \times \textrm{SO}(2,2) \times \textrm{SO}(2,2+\frac{\mathfrak{N}}{3}) \subset \textrm{SO}(6,6) \times \textrm{SO}(2,2+\frac{\mathfrak{N}}{3}) \subset \textrm{SO}(8,8+\mathfrak{N}) \ .
\end{equation}
As discussed before, $\,\textrm{SO}(3)\,$ invariance requires a YM gauge group $\,\textrm{G}_{\textrm{YM}}\,$ consisting of $\,k\,$ factors $\,\textrm{SO}(3)_{(k)}$. In other words, the half-maximal $\,D=3\,$ supergravity must be coupled to $\,\mathfrak{N} = 3k\,$ vector multiplets with $\,k \in \mathbb{N}$. From (\ref{SO(3)_embedding}), the commutant of $\,\textrm{SO}(3)\,$ inside $\,\textrm{SO}(8,8+\mathfrak{N})\,$ is $\,\textrm{SO}(2,2) \times \textrm{SO}(2,2+\frac{\mathfrak{N}}{3})$. Using the fact that $\,\textrm{SO}(2,2) \sim \textrm{SL}(2) \times \textrm{SL}(2)$, the scalar geometry of the $\,\textrm{SO}(3)$-invariant sector of half-maximal supergravity is identified with the coset space
\begin{equation}
\label{M_N=2_open}
\mathcal{M}^{\textrm{closed/open}}_{\mathcal{N}=2} = \left[ \frac{\textrm{SL}(2)}{\textrm{SO}(2)} \right]_{R} \times \left[ \frac{\textrm{SL}(2)}{\textrm{SO}(2)} \right]_{S} \times \,\, \frac{\textrm{SO}(2,2+\frac{\mathfrak{N}}{3})}{\textrm{SO}(2) \times \textrm{SO}(2+\frac{\mathfrak{N}}{3})}  \subset  \mathcal{M}^{\textrm{closed/open}}_{\mathcal{N}=8}\ .
\end{equation}
This is nothing but the open-string extension of the RSTU-models of \cite{Arboleya:2024vnp} with an arbitrary number $\,\mathfrak{N} = 3k\,$ of vector multiplets. Note that the open-string sector combines with the complex fields $\,T\,$ and $\,U\,$ to get an enhancement to $\,\textrm{SO}(2,2+\frac{\mathfrak{N}}{3})$.

Committing with our choice of gauge group $\,\textrm{G}_{\textrm{YM}}\,$ in (\ref{G_YM_SO(3)}), \textit{i.e.} $\,\mathfrak{N}=3$, the coset space (\ref{M_N=2_open}) accounts for the dilatons (\ref{dilatons_def}) and axions (\ref{axions_def}) coming from the closed-string sector, as well as the two new axion-like fields (\ref{scalars_open-sector_SO(3)}) originating from the open-string sector.  Therefore, there are the four complex closed-string scalars $R$, $S$, $T$ and $U$ of \cite{Arboleya:2024vnp},
%
%
and the two real open-string axions $\,\alpha\,$ and $\,\psi\,$ in (\ref{scalars_open-sector_SO(3)}). They enter the coset representative $\,\mathcal{V}\,$ in (\ref{Vielb_N=2}) through the building blocks
\begin{equation}
\label{coset_blocks}
\boldsymbol{e} \,\, = \,\, \left( 
\begin{matrix}  
\boldsymbol{e}_{TU} \otimes \mathbb{I}_{3}  & 0 \\ 
0 & \boldsymbol{e}_{RS}
\end{matrix} 
\right)
\hspace{10mm} \textrm{ and } \hspace{10mm}
\boldsymbol{b} \,\, = \,\,
\left( 
\begin{matrix}  
\boldsymbol{b}_{T} \otimes \mathbb{I}_{3} & 0 \\ 
0 & \boldsymbol{b}_{R}
\end{matrix}
\right)  \ ,
\end{equation}
with
\begin{equation}
\label{coset_block_e}
\boldsymbol{e}_{TU} \,\, = \,\, \frac{1}{\sqrt{\textrm{Im}T \, \textrm{Im}U}} \left( 
\begin{matrix}  
1 & 0 \\ 
\textrm{Re}U & \textrm{Im}U 
\end{matrix} 
\right)
\hspace{10mm} , \hspace{10mm}
\boldsymbol{e}_{RS} \,\, = \,\, \frac{1}{\sqrt{\textrm{Im}R \,\textrm{Im}S}} \left( 
\begin{matrix}  
1 & 0 \\ 
\textrm{Re}S & \textrm{Im}S 
\end{matrix} 
\right) \ ,
\end{equation}
and
\begin{equation}
\label{coset_block_b}
\boldsymbol{b}_{T} \,\, = \,\,
\left( 
\begin{matrix}  
0 & \textrm{Re}T  \\ 
-\textrm{Re}T & 0 
\end{matrix}
\right)  
\hspace{10mm} , \hspace{10mm}
\boldsymbol{b}_{R} \,\, = \,\,
\left( 
\begin{matrix}  
0 & \textrm{Re}R \\ 
-\textrm{Re}R & 0 
\end{matrix}
\right)  \ ,
\end{equation}
as well as
\begin{equation}
\label{coset_block_a}
\boldsymbol{a} \,\, = \,\, \left( \begin{matrix}
    \psi & \alpha  & 0 & 0  & 0 & 0  & 0 & 0 \\
    0 & 0  & \psi & \alpha  & 0 & 0  & 0 & 0 \\
    0 & 0  & 0 & 0 & \psi & \alpha  & 0 & 0 
\end{matrix} \right) \ .
\end{equation}

\subsubsection{RSTU-models with open strings: the Lagrangian}

The bosonic part of the Lagrangian of half-maximal $\,D=3\,$ gauged supergravity takes the form \cite{Nicolai:2001ac,Deger:2019tem} (see Appendix~\ref{Appendix:half-maximal_sugra} for conventions, etc.)
\begin{equation}
\label{3D_sugra_action}
S_{\textrm{\textrm{sugra}}} = \frac{1}{2\kappa^{2}} \displaystyle\int \textrm{d}^{3}x \, \sqrt{-g} \,  \big( R + L_{\textrm{kin}} - V \big) \ .
\end{equation}
When particularised to the extended RSTU-models presented in the previous section, which include the open-string fields $\,(\psi,\alpha)\,$ in (\ref{scalars_open-sector_SO(3)}), the scalar-dependent matrix $\,M\,$ in (\ref{M_scalar_matrix}) with building blocks (\ref{coset_blocks})-(\ref{coset_block_a}) yields kinetic terms of the form
\begin{equation}
\label{Lkin_half-max}
\begin{array}{rcl}
L_{\textrm{kin}} &=& - \dfrac{1}{8} \, \textrm{Tr} \left[ \partial_\mu M\, \partial^{\mu} M^{-1} \right] = \\[4mm]
&=&  - \dfrac{1}{2} \left[ \, 
\dfrac{\partial R \, \partial \bar{R}}{(\textrm{Im}R)^2} 
+ \dfrac{\partial S \, \partial \bar{S}}{(\textrm{Im}S)^2} \, \right] +\\[6mm]
& & - \dfrac{3}{2} \left[ \, 
\dfrac{ \Big( \partial\textrm{Re}T + \frac{1}{2} (\psi \, \partial \alpha - \alpha \, \partial \psi )\Big)^{2} + \left( \partial\textrm{Im}T\right)^{2}} {(\textrm{Im}T)^2} 
+ \dfrac{\partial U \, \partial \bar{U}}{(\textrm{Im}U)^2}  \right. +\\[6mm]
& & \left. \quad\quad \,\, + \, \dfrac{\textrm{Im}U}{\textrm{Im}T}  \left( \partial \alpha  \right)^2 
+ \dfrac{ \left( \partial \psi  + \textrm{Re}U \, \partial \alpha\right)^2 }{\textrm{Im}T \, \textrm{Im}U} \, \right] \ .
\end{array}
\end{equation}

There is also the scalar potential $\,V\,$ in (\ref{3D_sugra_action}). In $\,\mathcal{N}=8$, $\,D=3\,$ gauged supergravity, the scalar potential is fully determined by the so-called \textit{embedding tensor} $\,\Theta$. This is an object that encodes all the couplings in the $D=3$ supergravity which are consistent with half-maximal supersymmetry. From a string-theoretic viewpoint, the embedding tensor $\,\Theta\,$ encodes all the fluxes in the closed- and open-string sectors. Moreover, it admits a group-theoretic decomposition into irreducible representations of $\,\textrm{SO}(8,8+\mathfrak{N})\,$ of the form
\begin{equation}
\label{Theta_def}
\Theta_{\mathcal{M}\mathcal{N}|\mathcal{P}\mathcal{Q}} = \theta_{\mathcal{MNPQ}} + 2 \left( \eta_{\mathcal{M}[\mathcal{P}} \theta_{\mathcal{Q}]\mathcal{N}} - \eta_{\mathcal{N}[\mathcal{P}} \theta_{\mathcal{Q}]\mathcal{M}} \right) + 2 \eta_{\mathcal{M}[\mathcal{P}} \eta_{\mathcal{Q}]\mathcal{N}} \theta \ ,
\end{equation}
where $\mathcal{M}$ is a fundamental index of  $\,\textrm{SO}(8,8+\mathfrak{N})$, in our case $\,\mathfrak{N}=3$, and the individual components transform as
\begin{equation}
\label{Theta_irreps}
\theta_{\mathcal{MNPQ}} = \theta_{[\mathcal{MNPQ}]} \in \mathbf{3876}, \qquad
\theta_{\mathcal{MN}} = \theta_{(\mathcal{MN})} \in \mathbf{189} \ (\text{traceless}), \qquad
\theta \in \mathbf{1} \ .
\end{equation}
In (\ref{Theta_def}), the matrix $\,\eta_{\mathcal{\mathcal{MN}}}$\, denotes the non-degenerate $\textrm{SO}(8,8+\mathfrak{N})$-invariant metric (see Appendix~\ref{Appendix:half-maximal_sugra} for more details). Very importantly, consistency of the half-maximal $\,D=3\,$ supergravity requires that the embedding tensor (\ref{Theta_def}) satisfies a set of \textit{quadratic constraints}, which takes the form (see \textit{e.g.} \cite{Eloy:2024lwn})
\begin{equation}
\label{QC_half-max}
\Theta_{\mathcal{KL}|[\mathcal{M}}{}^{\mathcal{R}} \, \Theta_{\mathcal{N}]\mathcal{R}|\mathcal{P}\mathcal{Q}} + \Theta_{\mathcal{K}\mathcal{L}|[\mathcal{P}}{}^{\mathcal{R}} \, \Theta_{\mathcal{Q}]\mathcal{R}|\mathcal{M}\mathcal{N}} = 0 \ .
\end{equation}
Any embedding tensor $\,\Theta\,$ in (\ref{Theta_def}) specified by the various irreps in (\ref{Theta_irreps}) that satisfies the quadratic constraint (\ref{QC_half-max}) defines a half-maximal $\,D=3\,$ supergravity. Then, the scalar potential takes the schematic form
\begin{equation}
\label{V_half-max}
V(M) = g^{2} \, \Theta  \, \Theta \left( M^{4} + \eta \, M^{3} + \dots \right) \ ,
\end{equation}
where $\,g\,$ is the gauge coupling in the $D=3$ supergravity. We refer again the reader to the Appendix~\ref{Appendix:half-maximal_sugra} for more details on the exact form of the scalar potential. The gauge coupling $\,g\,$ can be eliminated from (\ref{V_half-max}) through a redefinition of the embedding tensor $\,\Theta \rightarrow g^{-1} \, \Theta\,$, so we will set $\,g=1\,$ in the rest of the work without loss of generality.

\subsubsection{Matching with the type IIB reduction with O$5$/D$5$ sources}

In order to match the dimensional reduction of Section~\ref{sec:IIB_reduction}, we must first impose the condition
\begin{equation}
\label{scalar_current}
\partial_{\mu}(\textrm{Re}T) + \frac{1}{2} (\psi \, \partial_{\mu} \alpha - \alpha \, \partial_{\mu} \psi ) = 0 \ ,
\end{equation}
in the kinetic terms (\ref{Lkin_half-max}) and, subsequently, turn off the closed-string axions.\footnote{This may initially appear awkward. However, the condition (\ref{scalar_current}) arises from the fact that we have turned off the three-dimensional vector fields originating from the dimensional reduction, in order to match the Einstein-scalar Lagrangian in (\ref{3D_sugra_action}). One such vectors, we denote it $A_{\mu}$, descends from $\,C_{(4)}\,$ and has precisely the scalar current (\ref{scalar_current}) as a source. An explicit computation of the covariant derivative of $\textrm{Re}T$ following (\ref{cov_der_3D}) gives $\,D_{\mu}\textrm{Re}T = \partial_{\mu} \textrm{Re}T + A_{\mu}$. Therefore, $\,A_{\mu}\,$ is massive and, when it is turned off, its equation of motion forces the scalar current (\ref{scalar_current}) to vanish. This scalar current is extracted from the kinetic terms in (\ref{Lkin_half-max}) upon gauge covariantisation of the derivatives, \textit{i.e.}, $\,\partial \rightarrow D$.} In our parameterisation of the Poincar\'e (upper) half-plane, this amounts to set to zero the real parts of the four closed-string complex fields, namely,
\begin{equation}
\label{no-axions_condition}
\textrm{Re}R=0
\hspace{5mm} , \hspace{5mm} 
\textrm{Re}S=0
\hspace{5mm} , \hspace{5mm} 
\textrm{Re}T=0
\hspace{5mm} , \hspace{5mm} 
\textrm{Re}U=0 \ .
\end{equation}
Then, the kinetic terms in (\ref{Lkin_half-max}) and the scalar potential in (\ref{V_half-max}) computed using the framework of half-maximal $\,D=3\,$ supergravity, precisely match the kinetic terms in (\ref{L_kin_3D}) and scalar potential in (\ref{V_3D}) obtained from the dimensional reduction of type IIB supergravity down to $\,D=3$. This matching requires the correspondence between embedding tensor components and type IIB fluxes displayed in Tables~\ref{Table:O5_Fluxes_even} and \ref{Table:O5_Fields_open}, as well as a dictionary between closed-string dilatons of the form
\begin{equation} 
\label{dictionary_dilatons}
\textrm{Im}R = \frac{\tau \, {\ell_i}^3 \, {\ell_a}^3}{\rho} 
\hspace{5mm} , \hspace{5mm} 
\textrm{Im}S = \frac{{\ell_a}^{3}}{\tau^{\frac{1}{2}} \, \rho^{\frac{1}{2}}}
\hspace{5mm} , \hspace{5mm} 
\textrm{Im}T = \frac{\tau^{\frac{1}{2}} \, \rho^{\frac{1}{2}}}{{\ell_i}^{2} \, {\ell_a}}
\hspace{5mm} , \hspace{5mm} 
\textrm{Im}U = \frac{1}{{\rho}^{2} \, {\ell_i} \, {\ell_a}} \ .
\end{equation}
We have explicitly verified that, under the flux/embedding tensor correspondence provided in Tables~\ref{Table:O5_Fluxes_even} and \ref{Table:O5_Fields_open}, the quadratic constraint in (\ref{QC_half-max}) yields exactly (and exclusively) the Jacobi and Bianchi identities discussed in Section~\ref{sec:Jacobi&Bianchi_id}, with the exception of (\ref{O5/D5_unrestrited}). The underlying reason is that half-maximal ($\mathcal{N}=8$) supersymmetry is fully compatible with the presence of O$5$/D$5$ sources of the type specified in (\ref{O5/D5_location}). Consequently, their number -- given by $\,J_{\textrm{O5/D5}}\,$ in (\ref{O5/D5_unrestrited}) -- remains unconstrained from the perspective of half-maximal $\,D=3\,$ supergravity.

\subsubsection{Symmetries of the scalar potential}
\label{sec:scaling_symmetry_general}

The scalar potential of the RSTU-models coupled to the open-string fields $\,(\psi,\alpha)\,$ features an interesting scaling symmetry. This symmetry rescales the closed- and open-string moduli as
\begin{equation}
\label{scaling_RSTU}
\begin{array}{cc}
R' = \lambda_{R} \, R 
\hspace{8mm} , \hspace{8mm}
S' = \lambda_{S} \, S
\hspace{8mm} , \hspace{8mm}
T' = \lambda_{T} \, T
\hspace{8mm} , \hspace{8mm}
U' = \lambda_{U} \, U 
\hspace{8mm} , \hspace{8mm}
 \\[3mm]
 \alpha' = \left(\frac{\lambda_{T}}{\lambda_{U}}\right)^{\frac12}\,\alpha
\hspace{5mm} , \hspace{5mm} 
\psi' = \left(\lambda_{T}\lambda_{U}\right)^{\frac12}\, \psi \ , 
\end{array}
\end{equation}
with the constant $\lambda$'s being real parameters. These transformations must be accompanied by appropriate compensating transformations of the metric fluxes
\begin{equation}
\label{scaling_metric}
\begin{array}{c}
\omega_1' = \left( \frac{\lambda_R \lambda_S \lambda_T^3}{\lambda_U}\right)^{\frac12} \omega_1 
\hspace{8mm} , \hspace{8mm}
\omega_2' = \lambda_T^2 \,  \omega_2
\hspace{8mm} , \hspace{8mm}
\omega_3' = \frac{\lambda_T}{\lambda_S} \, \omega_3 \ , \\[3mm]
\omega_4' = \frac{\lambda_R}{\lambda_U} \, \omega_4
\hspace{8mm} , \hspace{8mm}
\omega_5' = \left( \frac{\lambda_R \lambda_T}{\lambda_S \lambda_U}\right)^{\frac12} \omega_5
\hspace{8mm} , \hspace{8mm}
\omega_6' = \left( \frac{\lambda_R \lambda_T}{\lambda_S \lambda_U} \right)^{\frac12} \omega_6 \ ,
\end{array}
\end{equation}
of the gauge fluxes
\begin{equation}
\label{scaling_gauge}
\begin{array}{c}
h_{31}' = \left( \frac{\lambda_R \lambda_S \lambda_T^3}{\lambda_U^3}\right)^{\frac12} h_{31} 
\hspace{8mm} , \hspace{8mm}
h_{32}' = \left( \frac{\lambda_R \lambda_T}{\lambda_S \lambda_U^3}\right)^{\frac12} h_{32} \ , \\[3mm]
f_{31}' = \frac{\lambda_R}{\lambda_S} \, f_{31}
\hspace{8mm} , \hspace{8mm}
f_{32}' = \left( \frac{\lambda_R \lambda_U \lambda_T^3}{\lambda_S}\right)^{\frac12} f_{32}
\hspace{8mm} , \hspace{8mm}
f_{33}' = \lambda_R \, \lambda_T \, f_{33}  \ , \\[3mm]
f_5' = \left( \frac{\lambda_R \lambda_T}{\lambda_S \lambda_U} \right)^{\frac12} \lambda_T \, f_5
\hspace{8mm} , \hspace{8mm}
f_7' = \left( \frac{\lambda_R \lambda_T}{\lambda_S \lambda_U^3} \right)^{\frac12} \, \lambda_T \, f_7  \ ,
\end{array}
\end{equation}
and of the open-string fluxes
\begin{equation}
\label{scaling_open_fluxes}
\begin{array}{c}
\mathfrak{f}' = \frac{\lambda_{T}}{\lambda_{U}}\,\left(\frac{\lambda_{R}}{\lambda_{S}}\right)^{\frac12} \, \mathfrak{f}
\hspace{8mm} , \hspace{8mm}
\mathfrak{g}' = \left(\frac{\lambda_{R}}{\lambda_{S}}\right)^{\frac12} \, \mathfrak{g} \ .
\end{array}
\end{equation}
This symmetry implies that, when closed-string axions are turned off as in our type IIB compactifications -- see (\ref{axions_def_null}) or equivalently (\ref{no-axions_condition}) --, any flux vacuum can be brought to closed-string vacuum expectation values (VEV's) of the form
\begin{equation}
\label{GTTO_trick}
\left\langle  R \, \right\rangle = \left\langle  S \right\rangle = \left\langle  T \right\rangle = \left\langle  U \right\rangle = i  \ ,
\end{equation}
without loss of generality. Once this procedure has been carried out, a non-zero VEV for any of the open-string axions, $\,\left\langle \alpha \right\rangle\,$ or $\,\left\langle \psi \right\rangle$, can no longer be reabsorbed through a redefinition of the fluxes. Nonetheless, there are two special situations in which the scaling symmetry just discussed gets enhanced with an additional shift symmetry of (some of) the open-string axions. When this happens, (some of) the open-string axions can also be set to zero while maintaining generality.

\subsubsection*{Shift symmetries of $\,\alpha\,$ and $\,\psi\,$}

When $\mathfrak{g} = 0$, the scaling symmetry in (\ref{scaling_RSTU}) gets enhanced by an additional real parameter $\,c_{\alpha}\,$ that accounts for an independent shift of the open-string modulus $\,\alpha$. More concretely, the transformation rules for the closed- and open-string moduli become
\begin{equation}
\label{scaling_metric_g=0}
\begin{array}{cc}
    R' = \lambda_{R} \, R 
\hspace{5mm} , \hspace{5mm} 
S' = \lambda_{S} \, S
\hspace{5mm} , \hspace{5mm} 
T' = \lambda_{T} \, T +\frac{c_{\alpha}}{2} \sqrt{\lambda_{T}\lambda_{U}} \, \psi
\hspace{5mm} , \hspace{5mm} 
U' = \lambda_{U} \, U  \ ,
 \\[3mm]
 \alpha' = \left(\frac{\lambda_{T}}{\lambda_{U}}\right)^{\frac12}\,\alpha + c_{\alpha}
\hspace{5mm} , \hspace{5mm} 
\psi' = \left(\lambda_{T}\lambda_{U}\right)^{\frac12}\, \psi \ , 
\end{array}
\end{equation}
and only the compensating transformations for the fluxes $\,f_{7}\,$ and $\,\mathfrak{f}\,$ must be modified with respect to the expressions in (\ref{scaling_gauge}) and (\ref{scaling_open_fluxes}). In particular, the new compensating transformations read
\begin{equation}
\label{scaling_gauge_open_g=0}
\begin{array}{c}
f_7' = \left( \frac{\lambda_R \lambda_T}{\lambda_S \lambda_U^3} \right)^{\frac12} \left[ \lambda_T \, f_7 +3 \, c_{\alpha}\left(\lambda_{T}\lambda_{U}\right)^{\frac12} \, \mathfrak{f} + \frac{3}{2} \, c_{\alpha}^{2} \, \omega_{6} \right] \, ,
\\[3mm]
\mathfrak{f}' = \left( \frac{\lambda_R \lambda_T}{\lambda_S \lambda_U} \right)^{\frac12}\left[ \left(\frac{\lambda_{T}}{\lambda_{U}}\right)^{\frac12} \,  \mathfrak{f}+c_{\alpha} \, \omega_{6} \right] \, .
\end{array}
\end{equation}
This shift symmetry allows us to set $\,\left\langle \alpha \right\rangle=0\,$ without loss of generality when searching for flux vacua with $\,\mathfrak{g}=0$. Summarising,
\begin{equation}
\label{summary_shift_alpha}
\mathfrak{g} = 0 \quad \Rightarrow \quad \textrm{Flux vacua with $\,\left\langle \alpha \right\rangle = 0\,$ are general.}
\end{equation}

Finally, if $\,\mathfrak{g}=0\,$ \textit{and} $\,\omega_{4}=\omega_{5}=0$, then the scaling symmetry in (\ref{scaling_RSTU}) gets enhanced by yet another real parameter $\,c_{\psi}\,$ that accounts for an independent shift of the open-string modulus $\,\psi$. The scalar potential then becomes invariant under the field redefinitions
\begin{equation}
\label{scaling_metric_g=0_w4=w5=0}
\begin{array}{cc}
R' = \lambda_{R} \, R
\hspace{5mm} , \hspace{5mm} 
S' = \lambda_{S} \, S
\hspace{5mm} , \hspace{5mm} 
T' = \lambda_{T} \, T + \frac{1}{2} \left(\frac{\lambda_{T}}{\lambda_{U}}\right)^{\frac12} \left[ c_{\alpha}\lambda_{U} \, \psi - c_{\psi} \, \alpha \right] \, 
\hspace{5mm} , \hspace{5mm}
U' = \lambda_{U} \, U \ ,\\[3mm]
\alpha' = \left(\frac{\lambda_{T}}{\lambda_{U}}\right)^{\frac12}\,\alpha + c_{\alpha}
\hspace{5mm} , \hspace{5mm} 
\psi' = \left(\lambda_{T}\lambda_{U}\right)^{\frac12}\, \psi + c_{\psi} \ , 
\end{array}
\end{equation}
and the corresponding flux rescalings in (\ref{scaling_gauge}), with appropriate modifications for the parameters $\,f_{5}$, $\,f_{7}\,$ and $\,\mathfrak{f}\,$ given by
\begin{equation}
\label{scaling_gauge_open_g=0_w4=w5=0}
\begin{array}{c}
f_5' = \left( \frac{\lambda_R \lambda_T}{\lambda_S \lambda_U} \right)^{\frac12} \left[ \lambda_T \, f_5 + \, c_{\psi}\left(\frac{\lambda_{T}}{\lambda_{U}}\right)^{\frac12} \, \mathfrak{f} - \frac{1}{2} \, c_{\alpha} c_{\psi} \, \omega_{6} + \frac{1}{2} \, \frac{c_{\psi}^2}{\lambda_{U}} \, h_{32}\right] \ ,
\\[4mm]
f_7' = \left( \frac{\lambda_R \lambda_T}{\lambda_S \lambda_U^3} \right)^{\frac12} \left[ \lambda_T \, f_7 +3 \, c_{\alpha}\left(\lambda_{T}\lambda_{U}\right)^{\frac12} \, \mathfrak{f} + \frac{3}{2} \, c_{\alpha}^{2} \, \omega_{6} \right] \ ,
\\[4mm]
\mathfrak{f}' = \left( \frac{\lambda_R \lambda_T}{\lambda_S \lambda_U} \right)^{\frac12}\left[ \left(\frac{\lambda_{T}}{\lambda_{U}}\right)^{\frac12} \,  \mathfrak{f}+c_{\alpha} \, \omega_{6} - \frac{c_{\psi}^2}{\lambda_{U}} \, h_{32}\right] \ .
\end{array}
\end{equation}
These two shift symmetries allows us to set $\,\left\langle \alpha \right\rangle=0\,$ and $\,\left\langle \psi \right\rangle=0\,$ without loss of generality when searching for flux vacua with $\,\mathfrak{g}=0\,$ and $\,\omega_{4}=\omega_{5}=0$. Namely,
\begin{equation}
\label{summary_shift_alpha_psi}
\mathfrak{g} = 0 \,\,\,\, \textrm{ \textit{and} } \,\,\,\, \omega_{4}=\omega_{5} = 0 \quad \Rightarrow \quad \textrm{Flux vacua with $\,\left\langle \alpha \right\rangle = \left\langle \psi \right\rangle = 0\,$ are general.}
\end{equation}
The above shift symmetries of the open-string axions will be exploited in the next section when charting the landscape of RSTU-models with open strings.

\subsection{Flux vacua with open strings}

The algebraic condition in (\ref{QC_open_string}) imposes stringent restrictions on the possible open-string sectors. In particular, it allows for three open-string scenarios:

\begin{itemize}

\item[$i)$] Having unmagnetised D$5$-branes supporting an abelian YM theory, \textit{i.e.},
\begin{equation}
\mathfrak{f} = 0 
\hspace{10mm} \textrm{ and } \hspace{10mm}
\mathfrak{g}= 0 \ .
\end{equation}

\item[$ii)$] Having unmagnetised D$5$-branes supporting a non-abelian YM theory, \textit{i.e.},
\begin{equation}
\mathfrak{f} = 0 
\hspace{10mm} \textrm{ and } \hspace{10mm}
\mathfrak{g} \neq 0 \ .
\end{equation}

\item[$iii)$] Having abelian YM background fluxes magnetising the D$5$-branes, \textit{i.e.},
\begin{equation}
\mathfrak{f} \neq 0 
\hspace{10mm} \textrm{ and } \hspace{10mm}
\mathfrak{g} = 0 \ .
\end{equation}
\end{itemize}

In the following we will consider the three scenarios, and will exhaustively chart the landscape of flux vacua within RSTU-models coupled to $\,\mathfrak{N}=3\,$ matter multiplets. In order to be exhaustive when extremising the scalar potential  subject to the Jacobi and Bianchi identities in Section~\ref{sec:Jacobi&Bianchi_id}, we will make use of the algebraic geometry software \textsc{Singular} \cite{DGPS}. Also, in order to make contact with the type IIB compactifications on co-calibrated $\textrm{G}_{2}$ spaces of \cite{Emelin:2021gzx,Arboleya:2024vnp}, we will set
\begin{equation}
\label{G2_co-calibrated_cond}
H_{(3)} = F_{(5)} = 0 \ ,
\end{equation}
and only allow for metric fluxes $\,\omega,$ gauge fluxes $\,F_{(3)}\,$ and $\,F_{(7)}$, together with open-string fluxes and YM structure constants $\,\mathcal{F}\,$ and $\,\mathcal{G}$. 


For the flux vacua we will find, we will present their associated flux configurations, the number $\,\mathcal{N}\,$ of supersymmetries preserved based on the number of massless gravitini in the spectrum (\textit{i.e.} gravitini with normalised mass $\,m_{3/2}L=1\,$ in units of the AdS$_3$ radius $\,L$), and the spectrum of scalar fluctuations, which will be used to establish their perturbative stability within half-maximal $\,D=3\,$ supergravity. In the case of AdS$_{3}$ vacua, we will also present the conformal dimension for the would-be dual CFT$_{2}$ operators, $\,\Delta$, which corresponds to the larger root of
\begin{equation}
\label{Delta_d=2}
m^2 L^2 = \Delta (\Delta - d) \hspace{8mm} \textrm{ with } \hspace{8mm} d=2 \ .
\end{equation}

\subsubsection{Flux vacua with unmagnetised and abelian D$5$-branes}

\begin{table}[t!]
\begin{center}
\scalebox{0.9}{
\renewcommand{\arraystretch}{1.5}
\begin{tabular}{!{\vrule width 1.5pt}l!{\vrule width 1pt}c!{\vrule width 1pt}c!{\vrule width 1pt}cccccc!{\vrule width 1pt}ccc!{\vrule width 1pt}c!{\vrule width 1pt}c!{\vrule width 1.5pt}}
\Xcline{4-14}{1.5pt}
\multicolumn{3}{c!{\vrule width 1pt}}{}& \multicolumn{6}{c!{\vrule width 1pt}}{$\omega$} & \multicolumn{3}{c!{\vrule width 1pt}}{$F_{(3)}$} & \multicolumn{1}{c!{\vrule width 1pt}}{$F_{(7)}$}& \multicolumn{1}{c!{\vrule width 1pt}}{$\,\,\, \mathcal{G} \,\,\, $}\\ 
\noalign{\hrule height 1.5pt}
     \hspace{3mm} ID & Type &  SUSY & $\omega_{1}$ & $ \omega_{2}$ & $  \omega_{3}$ & $\omega_{4}$ & $ \omega_{5}$ & $ \omega_{6}$ & $f_{31}$ & $f_{32}$ & $f_{33}$ & $  f_{7}$ & $  \mathfrak{g}$  \\ 
\noalign{\hrule height 1pt}
     $ \textbf{vac~1} $ & \multirow{3}{*}{$\textrm{Mkw}_{3}$} & $\mathcal{N}=0,4$ & $ \kappa $ & $ \xi$ & $  0 $  & $ 0 $ & $ 0 $ & $  0 $ & $0$ & $ \kappa$ & $ -\xi$ & $0 $ &  \multirow{3}{*}{$\mathfrak{g}$} \\    
     \cline{1-1}\cline{3-13} 
     $ \textbf{vac~2} $ &  & $\mathcal{N}=0$ &  $ 0 $ & $ \kappa$ & $  0 $ & $0$ & $ 0$ & $  0$ & $0$ & $ 0$ & $ -\kappa$ & $0$ &  \\ 
     \cline{1-1}\cline{3-13}
     $ \textbf{vac~3} $ &  & $\mathcal{N}=0$ & $ 0$ & $ \kappa $ & $  \kappa$ & $0$ & $ 0$ & $  0$ & $0$ & $ 0$ & $ 0$ & $0$ &  \\ 
\noalign{\hrule height 1pt}
     $ \textbf{vac~4}$ & \multirow{2}{*}{${\textrm{AdS}_{3}}$} & $\mathcal{N}=4$ & $ 0$ & $ 0$ & $ 0$ & $0$ & $ 0$ & $  \kappa$ & $\pm\kappa$ & $ 0$ & $ 0$ & $  -\kappa$ & \multirow{2}{*}{$\mathfrak{g}$}
     \\ 
     \cline{1-1}\cline{3-13}
     $ \textbf{vac~5}$ &  & $\mathcal{N}=0$ & $ 0$ & $0$ & $ 0$ & $0$ & $ 0$ & $\kappa$ & $\pm\kappa$ & $ 0$ & $ 0$ & $\kappa$ & 
     \\
\noalign{\hrule height 1pt}
     $ \textbf{vac~6}$ & \multirow{2}{*}{$\textrm{AdS}_{3}$} & $\mathcal{N}=3$ & $ \kappa$ & $ 0$ & $  0$ & $0$ & $ -\kappa$ & $ \kappa$ & $\pm\kappa $ & $ 0$ & $ \mp\kappa$ & $-2\kappa$ & \multirow{2}{*}{$\mathfrak{g}$} \\ 
     \cline{1-1}\cline{3-13}
     $ \textbf{vac~7}$ &  & $\mathcal{N}=1$ & $ \kappa$ & $ 0$ & $  0$ & $0$ & $ -\kappa$ & $ \kappa$ & $\mp\kappa $ & $ 0$ & $ \pm\kappa$ & $2\kappa$  &  \\
\noalign{\hrule height 1pt}
     $ \textbf{vac~8}$ & \multirow{2}{*}{${\textrm{AdS}_{3}}$} & $\mathcal{N}=1$ & $ 0$ & $ 0$ & $  0$ & $0$ & $ -\kappa$ & $ \kappa$ & $\pm\kappa$ & $ 0$ & $ 0$ & $\kappa$ & \multirow{2}{*}{$\mathfrak{g}$} \\ 
     \cline{1-1}\cline{3-13}
     $ \textbf{vac~9}$ &  & $\mathcal{N}=0$ & $ 0$ & $ 0$ & $  0$ & $0$ & $ -\kappa$ & $ \kappa$ & $\pm\kappa$ & $ 0$ & $ 0$ & $-\kappa$ &  \\ 
\noalign{\hrule height 1pt}
      \cellcolor[HTML]{A3C1AD}$ \textbf{vac~10}$ & $\textrm{AdS}_{3}$ & $\mathcal{N}=0$ & $ 0$ & $ 2\kappa$ & $  \kappa$ & $0$ & $ 0$ & $ 0$ & $\kappa$ & $ \pm\kappa$ & $ -\kappa$ & $\pm\kappa$ & $\mathfrak{g}$ \\
     \hline 
      \cellcolor[HTML]{A3C1AD}$ \textbf{vac~11}$ & $\textrm{AdS}_{3}$ & $\mathcal{N}=0$ & $ 0$ & $ 2\kappa$ & $  \kappa$ & $0$ & $ 0$ & $ 0$ & $\kappa$ & $ \pm\kappa$ & $ -\kappa$ & $\mp\kappa$ & $\mathfrak{g}$ \\
\noalign{\hrule height 1pt} 
     $ \textbf{vac~12}$ & \multirow{4}{*}{${\textrm{AdS}_{3}}$} & $\mathcal{N}=4$ & $ 0$ & $ 0$ & $  \mp\kappa$ & $\mp\kappa$ & $ \kappa$ & $ -2\kappa$ & $\mp 2\kappa$ & $ 0$ & $ 0$ & $  2\kappa$ &  \multirow{4}{*}{$\mathfrak{g}$} \\
      \cline{1-1}\cline{3-13}
     $ \textbf{vac~13}$ &  & $\mathcal{N}=1$ & $ 0$ & $ 0$ & $  \mp\kappa$ & $\mp\kappa$ & $ \kappa$ & $ -2\kappa$ & $\mp 2\kappa$ & $ 0$ & $ 0$ & $  -2\kappa$ &  \\
     \cline{1-1}\cline{3-13}
     $ \textbf{vac~14}$ &  & $\mathcal{N}=0$ & $ 0$ & $ 0$ & $  \pm\kappa$ & $\pm\kappa$ & $ \kappa$ & $ -2\kappa$ & $\mp 2\kappa$ & $ 0$ & $ 0$ & $  2\kappa$ &  \\
     \cline{1-1}\cline{3-13}
     $ \textbf{vac~15}$ &  & $\mathcal{N}=0$ & $ 0$ & $ 0$ & $  \pm\kappa$ & $\pm\kappa$ & $ \kappa$ & $ -2\kappa$ & $\mp 2\kappa$ & $ 0$ & $ 0$ & $  -2\kappa$ &  \\
\noalign{\hrule height 1.5pt}
\end{tabular}}
\caption{Fluxes producing the unmagnetised ($\mathcal{F}=0$) flux vacua with $\,\left\langle\alpha\right\rangle = \left\langle\psi\right\rangle = 0$ of \cite{Arboleya:2024vnp}. They all are compatible with turning on an arbitrary non-abelian YM parameter $\,\mathfrak{g}$. The \textbf{vac~1} is Mkw$_{3}$ and generically non-supersymmetric, but becomes $\mathcal{N}=4$ when $\kappa = \pm \xi$. For the AdS$_{3}$ supersymmetric vacua, the $\,\mathcal{N}=p\,$ supersymmetry is realised as $\,\mathcal{N}=(p,0)\,$ or $\,\mathcal{N}=(0,p)\,$  depending on the upper/lower sign choice of the fluxes. The green-marked vacua correspond to solutions in which the closed-string moduli are stabilised.}
\label{Table:flux_vacua_paper}
\end{center}
\end{table}

\begin{table}[h!]
\begin{center}
\scalebox{0.9}{
\renewcommand{\arraystretch}{1.8}
\begin{tabular}{!{\vrule width 1.5pt}c!{\vrule width 1pt}c!{\vrule width 1pt}c!{\vrule width 1pt}cccccc!{\vrule width 1pt}cc!{\vrule width 1pt}ccc!{\vrule width 1pt}c!{\vrule width 1pt}c!{\vrule width 1.5pt}}
\noalign{\hrule height 1.5pt}
     ID & Scalar spectrum  \\ 
\noalign{\hrule height 1pt}
     $ \textbf{vac~1} $ & $ g^{-2} \, m^2 =  0_{(30+\textcolor{BrickRed}{8\mathfrak{N}})}, \left(\frac{\kappa^2}{16} \right)_{(9)}, \left(\frac{\kappa^2}4\right)_{(9)} ,
     \left(\frac{\xi^2}4\right)_{(9)}  , 
     \left(\frac{9\kappa^2}{16} \right)_{(1)}, \left[ \frac{\left( \kappa \, \pm \,  2\xi \right)^2}{16} \right]_{(3)}  $  \\[2mm] 
     \hline 
     $ \textbf{vac~2} $ & \multirow{2}{*}{$g^{-2} \, m^2 =  \left( \frac{\kappa^2}{4} \right)_{(15)} , 0_{(49+\textcolor{BrickRed}{8\mathfrak{N}})} $}  \\ 
     \cline{1-1} 
     $ \textbf{vac~3} $ &  \\ 
\noalign{\hrule height 1pt}
     $ \textbf{vac~4} $ & $\begin{array}{ccl}
        m^2 L^2 &=&  8_{(19)}, \ 0_{(45+\textcolor{BrickRed}{8\mathfrak{N}})}  \\[-2mm]
         \Delta &=&  4_{(19)}, \ 2_{(45+\textcolor{BrickRed}{8\mathfrak{N}})} 
     \end{array}$
     \\ 
    \hline 
     $ \textbf{vac~5} $ &  $\begin{array}{ccl}
        m^2 L^2 &=&  8_{(19+\textcolor{BrickRed}{3\mathfrak{N}})}, \ 0_{(45+\textcolor{BrickRed}{5\mathfrak{N}})}  \\[-2mm]
         \Delta &=&  4_{(19+\textcolor{BrickRed}{3\mathfrak{N}})}, \ 2_{(45+\textcolor{BrickRed}{5\mathfrak{N}})} 
     \end{array}$
     \\
\noalign{\hrule height 1pt}
     $ \textbf{vac~6} $ & $\begin{array}{ccl}
     m^2 L^2 &=&  8_{(10)}, \ 4_{(18)} , \ 0_{(36+\textcolor{BrickRed}{5\mathfrak{N}})} , \ \textcolor{BrickRed}{-1}_{(\textcolor{BrickRed}{3\mathfrak{N}})} \\[-2mm]
      \Delta &=&  4_{(10)}, \ (1+\sqrt{5})_{(18)} , \ 2_{(36+\textcolor{BrickRed}{5\mathfrak{N}})} , \ \textcolor{BrickRed}{1}_{(\textcolor{BrickRed}{3\mathfrak{N}})}
     \end{array}$  \\ 
     \hline
     $ \textbf{vac~7} $ &  $\begin{array}{ccl}
     m^2 L^2 &=&  8_{(10)}, \ 4_{(18)} , \ \textcolor{BrickRed}{3}_{(\textcolor{BrickRed}{3\mathfrak{N}})} , \ 0_{(36+\textcolor{BrickRed}{5\mathfrak{N}})}  \\[-2mm]
      \Delta &=&  4_{(10)}, \ (1+\sqrt{5})_{(18)} , \ \textcolor{BrickRed}{3}_{(\textcolor{BrickRed}{3\mathfrak{N}})}, \ 2_{(36+\textcolor{BrickRed}{5\mathfrak{N}})} 
     \end{array}$  \\
\noalign{\hrule height 1pt}
     $ \textbf{vac~8} $ & $\begin{array}{ccl}
     m^2 L^2 &=& 24_{(10)}, \ 8_{(25+\textcolor{BrickRed}{6\mathfrak{N}})}, \ 0_{(29+\textcolor{BrickRed}{2\mathfrak{N}})} \\[-2mm]
      \Delta &=& 6_{(10)}, \ 4_{(25+\textcolor{BrickRed}{6\mathfrak{N}})}, \ 2_{(29+\textcolor{BrickRed}{2\mathfrak{N}})} \end{array}$ \\ 
     \hline 
     $ \textbf{vac~9} $ & $\begin{array}{ccl}
     m^2 L^2 &=& 24_{(10)}, \ 8_{(25+\textcolor{BrickRed}{3\mathfrak{N}})}, \ 0_{(29+\textcolor{BrickRed}{5\mathfrak{N}})} \\[-2mm]
      \Delta &=& 6_{(10)}, \ 4_{(25+\textcolor{BrickRed}{3\mathfrak{N}})}, \ 2_{(29+\textcolor{BrickRed}{5\mathfrak{N}})} \end{array}$  \\ 
\noalign{\hrule height 1pt}
      \cellcolor[HTML]{A3C1AD}$ \textbf{vac~10} $ & $\begin{array}{ccl}
     m^2 L^2 &=&  80_{(3)}, \ 48_{(9)}, \ 24_{(4)}, \ 8_{(7)}, \ 0_{(41+\textcolor{BrickRed}{8\mathfrak{N}})} \\[-2mm]
     \Delta &=&  10_{(3)}, \ 8_{(9)}, \ 6_{(4)}, \ 4_{(7)}, \ 2_{(41+\textcolor{BrickRed}{8\mathfrak{N}})}
     \end{array}$   \\[1mm]
     \hline 
      \cellcolor[HTML]{A3C1AD}$ \textbf{vac~11} $ & $\begin{array}{ccl} 
     m^2 L^2 &=& 48_{(15)}, \ 8_{(13)}, \ 0_{(36+\textcolor{BrickRed}{8\mathfrak{N}})} \\[-2mm]
     \Delta &=& 8_{(15)}, \ 4_{(13)}, \ 2_{(36+\textcolor{BrickRed}{8\mathfrak{N}})}
     \end{array}$   \\[1mm]
\noalign{\hrule height 1pt} 
     $ \textbf{vac~12} $ & \multirow{2}{*}{$\begin{array}{ccl}
     m^2 L^2 &=&  15_{(8)}, \ 8_{(19)}, \ 3_{(8+\textcolor{BrickRed}{4\mathfrak{N}})}, \ 0_{(29+\textcolor{BrickRed}{4\mathfrak{N}})} \\[-2mm]
     \Delta &=&  5_{(8)}, \ 4_{(19)}, \ 3_{(8+\textcolor{BrickRed}{4\mathfrak{N}})}, \ 2_{(29+\textcolor{BrickRed}{4\mathfrak{N}})}
     \end{array}$}  \\
     \cline{1-1} 
     $ \textbf{vac~14} $ &    \\
     \hline
     $ \textbf{vac~13} $ &  \multirow{2}{*}{$\begin{array}{ccl}
     m^2 L^2 &=&  15_{(8)}, \ 8_{(19+\textcolor{BrickRed}{3\mathfrak{N}})}, \ 3_{(8+\textcolor{BrickRed}{4\mathfrak{N}})}, \ 0_{(29+\textcolor{BrickRed}{\mathfrak{N}})} \\[-2mm]
     \Delta &=&  5_{(8)}, \ 4_{(19+\textcolor{BrickRed}{3\mathfrak{N}})}, \ 3_{(8+\textcolor{BrickRed}{4\mathfrak{N}})}, \ 2_{(29+\textcolor{BrickRed}{\mathfrak{N}})}
     \end{array}$}  \\
     \cline{1-1}
     $ \textbf{vac~15} $ &  \\
\noalign{\hrule height 1.5pt}
\end{tabular}}
\caption{Scalar masses at the flux vacua of Table~\ref{Table:flux_vacua_paper}. The subscript in $\#_{(s)}$ denotes the multiplicity of the mass $\,\#\,$ in the spectrum. For the AdS$_{3}$ vacua we have normalised the scalar spectrum using the AdS$_{3}$ radius $L^2=-2/V_{0}$. In addition, for each AdS$_3$ vacuum, the conformal dimension $\Delta$ of the would-be CFT$_{2}$ dual operators are also indicated.}
\label{Table:flux_vacua_paper_scalars}
\end{center}
\end{table}

The first scenario allowed by the Bianchi identity (\ref{QC_open_string}) corresponds to the choice
\begin{equation}
\mathfrak{f} = 0 
\hspace{10mm} \textrm{ and } \hspace{10mm}
\mathfrak{g} = 0 \ ,
\end{equation}
so no open-string fluxes or YM structure constants are present in the compactification. Still, the open-string fields couple to the closed-string ones in the Lagrangian via the closed-string fluxes.

The closed-string flux vacua identified in \cite{Arboleya:2024vnp} are recovered by extremising the scalar potential under the condition
\begin{equation}
\left\langle \psi \right\rangle = \left\langle \alpha \right\rangle = 0 \ ,
\end{equation}
while simultaneously enforcing the Jacobi and Bianchi identities presented in Section~\ref{sec:Jacobi&Bianchi_id}. These flux vacua are summarised in Table~\ref{Table:flux_vacua_paper} (where $\,\mathfrak{g}=0\,$ is understood).
For all of them, a non-zero VEV for $\,\left\langle \alpha \right\rangle\,$ can be freely turned on due to the shift symmetry in (\ref{summary_shift_alpha}). In contrast, a non-vanishing $\,\left\langle \psi \right\rangle\,$ is generally not allowed, with the exception of the Minkowski solution Mkw$_{3}$ and the AdS$_3$ solutions labeled \textbf{vac~4,5} and \textbf{vac~10,11} in Table~\ref{Table:flux_vacua_paper}, where the shift symmetry (\ref{summary_shift_alpha_psi}) permits an arbitrary $\,\left\langle \psi \right\rangle\,$ as well. 

Even if $\,\left\langle \psi \right\rangle = \left\langle \alpha \right\rangle = 0$, the open-string sector contributes to the spectrum of scalar fluctuations. The scalar mass spectra associated with the flux vacua of Table~\ref{Table:flux_vacua_paper} are collected in Table~\ref{Table:flux_vacua_paper_scalars}, where an arbitrary number of vector multiplets $\,\mathfrak{N}=3 k$, with $\,k \in \mathbb{N}$, has been considered. The masses associated with open-string modes are highlighted in red in Table~\ref{Table:flux_vacua_paper_scalars}, from which we observe that additional mass values, originally absent in the closed-string setup of \cite{Arboleya:2024vnp}, occur exclusively at \textbf{vac~6,7}. For the rest of flux vacua, the open-string sector simply changes the multiplicity of the mass values already encountered in \cite{Arboleya:2024vnp}. However, certain closed-string vacua in \cite{Arboleya:2024vnp} that shared identical mass spectra become non-degenerate once the masses from the open-string sector are included. This is the case for the groups of solutions \textbf{vac~4,5}, \textbf{vac~6,7}, \textbf{vac~8,9}, and \textbf{vac~12,13,14,15} in Table~\ref{Table:flux_vacua_paper_scalars}. Lastly, for the sake of completeness, the gravitini mass spectra associated with the flux vacua of Table~\ref{Table:flux_vacua_paper} are presented in Table~\ref{Table:flux_vacua_paper_gravitini} of Appendix~\ref{app_gravitini_masses}.

Contrary to expectations \cite{Danielsson:2016mtx}, the dynamics of the open-string sector does not introduce any perturbative instability in the spectrum of scalar fluctuations about the closed-string flux vacua of \cite{Arboleya:2024vnp}, as all scalar masses in Table~\ref{Table:flux_vacua_paper_scalars} respect the BF bound for stability in AdS$_3$, \textit{i.e.} $\,m^{2}L^{2} \ge -1$. This differs from the case of AdS$_7$ and AdS$_4$ flux vacua, where open strings were found to develop modes with masses below the BF bound \cite{Danielsson:2017max}. We also observe that the $\Delta$'s associated with the $\,8\mathfrak{N}\,$ scalars in the open-string sector turn out to be integer-valued for all the AdS$_{3}$ flux vacua in Table~\ref{Table:flux_vacua_paper_scalars}. As a curiosity, \textbf{vac~6} therein is the only AdS$_{3}$ vacuum whose would-be dual CFT$_{2}$ contains relevant operators, and these precisely stem from the open-string sector. However, it is important to make clear that the existence of a holographic CFT$_{2}$ dual to these AdS$_{3}$ flux vacua is not guaranteed at all.

To close this section, let us present the spectrum of scalar fluctuations within half-maximal supergravity about the supersymmetric vacuum of \cite{VanHemelryck:2025qok} when $\,\mathfrak{N}\,$ vector multiplets arising from open strings are included. This supersymmetric vacuum is a cousin of our \textbf{vac~10}. The scalar spectrum reads
\begin{equation}
\label{spectrum_susy}
\begin{array}{ccl}
m^2 L^2 &=&  80_{(4)} \, , \, 48_{(7)} \, , \,  24_{(4)} \, , \, 8_{(9)} \, , \, 0_{(40+\textcolor{BrickRed}{8\mathfrak{N}})} \ , \\[2mm]
\Delta &=&  10_{(4)} \, , \, 8_{(7)} \, , \, 6_{(4)} \, , \, 4_{(9)} \, , \, 2_{(40+\textcolor{BrickRed}{8\mathfrak{N}})} \ ,
\end{array} 
\end{equation}
so the values of the masses are the same as in the non-supersymmetric \textbf{vac~10} in Table~\ref{Table:flux_vacua_paper_scalars}, but the multiplicities are slightly different. The gravitini masses are the same as for \textbf{vac~10}.

\subsubsection{Flux vacua with unmagnetised and non-abelian D$5$-branes}
\label{sec:vacua_unmagnetised}

\begin{table}[t!]
\begin{center}
\scalebox{0.66}{
\renewcommand{\arraystretch}{1.5}
\hspace{-3mm}\begin{tabular}{!{\vrule width 1.5pt}c!{\vrule width 1pt}c!{\vrule width 1pt}c!{\vrule width 1pt}cccccc!{\vrule width 1pt}ccc!{\vrule width 1pt}c!{\vrule width 1pt}c!{\vrule width 1.5pt}}
\Xcline{4-13}{1.5pt}
\multicolumn{3}{c!{\vrule width 1pt}}{}& \multicolumn{6}{c!{\vrule width 1pt}}{$\omega$} & \multicolumn{3}{c!{\vrule width 1pt}}{$F_{(3)}$} & \multicolumn{1}{c!{\vrule width 1pt}}{$F_{(7)}$} \\ 
\noalign{\hrule height 1.5pt}
     ID & Type & SUSY & $\omega_{1}$ & $ \omega_{2}$ & $  \omega_{3}$ & $\omega_{4}$ & $ \omega_{5}$ & $ \omega_{6}$ & $f_{31}$ & $f_{32}$ & $f_{33}$ & $  f_{7}$ \\ 
     \noalign{\hrule height 1pt}
     $ \textbf{vac~16} $ & \multirow{2}{*}{$\textrm{AdS}_3$} & $\mathcal{N}= 3 $ & $\mathfrak{g}\,\alpha$ & $ 0 $ & $ 0 $  & $ 0 $ & $ -\mathfrak{g}\,\alpha $ & $ \mathfrak{g}\,\alpha  $ & $ \pm \mathfrak{g}\,\alpha $ & $ 0 $ & $ \mp \mathfrak{g}\,\alpha$ & $ -\frac{\mathfrak{g}\,\alpha}{2} (4+\alpha^2) $  \\    
     \cline{1-1}\cline{3-13} 
     $ \textbf{vac~17} $ &  & $\mathcal{N}= 1$ & $  \mathfrak{g}\,\alpha $ & $ 0 $ & $ 0 $  & $ 0 $ & $ -\mathfrak{g}\,\alpha $ & $ \mathfrak{g}\,\alpha  $ & $ \pm \mathfrak{g}\,\alpha $ & $ 0 $ & $ \mp \mathfrak{g}\,\alpha$ & $ \frac{\mathfrak{g}\,\alpha}{2} (4-\alpha^2) $ \\ 
     \noalign{\hrule height 1pt}
     $ \textbf{vac~18} $ & \multirow{2}{*}{$\textrm{AdS}_3$} & $\mathcal{N}= 1$ & $ 0  $ & $ 0 $ & $ 0 $  & $ 0 $ & $ -\mathfrak{g}\,\alpha $ & $ \mathfrak{g}\,\alpha  $ & $ \pm \mathfrak{g}\,\alpha $ & $ 0 $ & $ 0 $ & $ \frac{\mathfrak{g}\,\alpha}2 (2-\alpha^2)$  \\    
     \cline{1-1}\cline{3-13} 
     $ \textbf{vac~19} $ &  & $\mathcal{N}= 0$ & $ 0  $ & $ 0 $ & $ 0 $  & $ 0 $ & $ -\mathfrak{g}\,\alpha $ & $ \mathfrak{g}\,\alpha  $ & $ \pm \mathfrak{g}\,\alpha $ & $ 0 $ & $ 0 $ & $ -\frac{\mathfrak{g}\,\alpha}2 (2+\alpha^2)$  \\ 
     \noalign{\hrule height 1pt}
     $ \textbf{vac~20} $ & \multirow{4}{*}{$\textrm{AdS}_3$} & $\mathcal{N}= 4$ & $ 0 $ & $ 0 $ & $ \pm \frac{\mathfrak{g}\,\alpha}2 $  & $ \pm \frac{\mathfrak{g}\,\alpha}2 $ & $ -\frac{\mathfrak{g}\,\alpha}2 $ & $  \mathfrak{g}\,\alpha $ & $ \pm \mathfrak{g}\,\alpha $ & $ 0 $ & $ 0 $ & $ -\frac{\mathfrak{g}\,\alpha}2 (2+\alpha^2) $ \\  
     \cline{1-1}\cline{3-13} 
     $ \textbf{vac~21} $ &  & $\mathcal{N}=1$ & $ 0 $ & $ 0 $ & $ \pm \frac{\mathfrak{g}\,\alpha}2 $  & $ \pm \frac{\mathfrak{g}\,\alpha}2 $ & $ -\frac{\mathfrak{g}\,\alpha}2 $ & $  \mathfrak{g}\,\alpha $ & $ \pm \mathfrak{g}\,\alpha $ & $ 0 $ & $ 0 $ & $ \frac{\mathfrak{g}\,\alpha}2 (2-\alpha^2) $   \\ 
     \cline{1-1}\cline{3-13}
     $ \textbf{vac~22} $ &  & $\mathcal{N}= 0$ & $ 0 $ & $ 0 $ & $ \pm \frac{\mathfrak{g}\,\alpha}2 $  & $ \pm \frac{\mathfrak{g}\,\alpha}2 $ & $ -\frac{\mathfrak{g}\,\alpha}2 $ & $  \mathfrak{g}\,\alpha $ & $ \mp \mathfrak{g}\,\alpha $ & $ 0 $ & $ 0 $ & $- \frac{\mathfrak{g}\,\alpha}2 (2+\alpha^2) $  \\
     \cline{1-1}\cline{3-13}
     $ \textbf{vac~23}$ &  & $\mathcal{N}= 0$ & $ 0 $ & $ 0 $ & $ \pm \frac{\mathfrak{g}\,\alpha}2 $  & $ \pm \frac{\mathfrak{g}\,\alpha}2 $ & $ -\frac{\mathfrak{g}\,\alpha}2 $ & $  \mathfrak{g}\,\alpha $ & $ \mp \mathfrak{g}\,\alpha $ & $ 0 $ & $ 0 $ & $ \frac{\mathfrak{g}\,\alpha}2 (2-\alpha^2) $  \\
      \noalign{\hrule height 1pt}
     $ \textbf{vac~24} $ & \multirow{2}{*}{$\textrm{AdS}_3$} & $\mathcal{N}= 4$ & $ 0  $ & $ 0 $ & $ 0 $  & $ 0 $ & $0 $ & $ \mathfrak{g}\,\alpha  $  & $  \pm\mathfrak{g}\,\alpha $ & $0$ & $0$ & $ -\frac{\mathfrak{g}\,\alpha}{2} (2+\alpha^2) $  \\    
     \cline{1-1}\cline{3-13}
     $ \textbf{vac~25} $ &  & $\mathcal{N}= 0$ & $ 0  $ & $ 0 $ & $ 0 $  & $ 0 $ & $0$ & $ \mathfrak{g}\,\alpha  $  & $  \pm\mathfrak{g}\,\alpha $ & $0$ & $0$ & $ \frac{\mathfrak{g}\,\alpha}{2} (2-\alpha^2) $  \\
     \noalign{\hrule height 1pt}
     $ \textbf{vac~26} $ & $\textrm{AdS}_3$ & $\mathcal{N}= 0$ & $ 0  $ & $ 0 $ & $ 0 $  & $ 0 $ & $0 $ & $ \mathfrak{g} \, \frac{(2+\alpha^2)}{\alpha}  $  & $  \pm \mathfrak{g} \, \frac{\sqrt{4+2\alpha^2+\alpha^4}}{\alpha} $ & $0$ & $0$ & $ -\mathfrak{g} \, \frac{(2+\alpha^2)^2}{2\alpha} $  \\    
\noalign{\hrule height 1pt}
     $ \textbf{vac~27} $ & $\textrm{AdS}_3$ & $\mathcal{N}= 0$ & $ 0  $ & $ 0 $ & $ 0 $  & $ 0 $ & $-\mathfrak{g}\,\frac{(2+\alpha^2)}{\alpha} $ & $ \mathfrak{g}\,\frac{(2+\alpha^2)}{\alpha}  $  & $  \pm\mathfrak{g}\frac{\sqrt{4+2\alpha^2+\alpha^4}}{\alpha} $ & $0$ & $0$ & $ -\mathfrak{g}\,\frac{(2+\alpha^2)^2}{2\alpha} $  \\ 
\noalign{\hrule height 1pt}
    $ \textbf{vac~28} $ & \multirow{2}{*}{$\textrm{AdS}_3$} & $\mathcal{N}= 0$ & $ 0  $ & $ 0 $ & $ -\mathfrak{g}\,\frac{(2+\alpha^2)}{2\alpha}  $  & $ -\mathfrak{g}\frac{(2+\alpha^2)}{2\alpha}  $ & $-\mathfrak{g}\,\frac{(2+\alpha^2)}{2\alpha} $ & $ \mathfrak{g}\,\frac{(2+\alpha^2)}{\alpha}  $  & $  \pm\mathfrak{g}\frac{\sqrt{4+2\alpha^2+\alpha^4}}{\alpha} $ & $0$ & $0$ & $ -\mathfrak{g}\,\frac{(2+\alpha^2)^2}{2\alpha} $  \\  
     \cline{1-1}\cline{3-13} 
     $ \textbf{vac~29} $ & & $\mathcal{N}= 0$ & $ 0  $ & $ 0 $ & $ \mathfrak{g}\,\frac{(2+\alpha^2)}{2\alpha}  $  & $ \mathfrak{g}\frac{(2+\alpha^2)}{2\alpha}  $ & $-\mathfrak{g}\,\frac{(2+\alpha^2)}{2\alpha} $ & $ \mathfrak{g}\,\frac{(2+\alpha^2)}{\alpha}  $  & $  \pm\mathfrak{g}\frac{\sqrt{4+2\alpha^2+\alpha^4}}{\alpha} $ & $0$ & $0$ & $ -\mathfrak{g}\,\frac{(2+\alpha^2)^2}{2\alpha} $   \\ 
     \noalign{\hrule height 1pt}
      \cellcolor[HTML]{89CFF0}$ \textbf{vac~30} $ & \multirow{2}{*}{$\textrm{AdS}_3$} & $\mathcal{N}= 0$ & $ 0  $&  $ \frac{\mathfrak{g}\,\alpha}{x_1\,x_2} $  & $ \frac{\mathfrak{g}\,\alpha\,x_2}{2x_1\,x_3} $ & $ 0 $  & $0 $ & $ 0  $  & $  \mathfrak{g}\,\alpha\,x_2 $ & $\frac{\mathfrak{g}\,\alpha}{\sqrt{2x_1}}$ & $-\frac{\mathfrak{g}\,\alpha\,x_3}{x_2}$ & $\frac{\mathfrak{g}\,\alpha}{4x_1} $ \\    
     \cline{1-1}\cline{3-13} 
      \cellcolor[HTML]{89CFF0}$ \textbf{vac~31} $ &  & $\mathcal{N}= 0$ & $ 0  $&  $ -\frac{\mathfrak{g}\,\alpha}{x_1\,x_2} $  & $ -\frac{\mathfrak{g}\,\alpha\,x_2}{2x_1\,x_3} $ & $ 0 $  & $0 $ & $ 0  $  & $  -\mathfrak{g}\,\alpha\,x_2 $ & $\frac{\mathfrak{g}\,\alpha}{\sqrt{2x_1}}$ & $\frac{\mathfrak{g}\,\alpha\,x_3}{x_2}$ & $\frac{\mathfrak{g}\,\alpha}{4x_1} $  \\ 
     \hline
      \cellcolor[HTML]{89CFF0}$\textbf{vac~32} $ & \multirow{2}{*}{$\textrm{AdS}_3$} & $\mathcal{N}= 0$ & $ 0  $&  $ \frac{\mathfrak{g}\,\alpha}{x_1\,x_2} $  & $ \frac{\mathfrak{g}\,\alpha\,x_2}{2x_1\,x_3} $ & $ 0 $  & $0 $ & $ 0  $  & $  \mathfrak{g}\,\alpha\,x_2 $ & $-\frac{\mathfrak{g}\,\alpha}{\sqrt{2x_1}}$ & $-\frac{\mathfrak{g}\,\alpha\,x_3}{x_2}$ & $\frac{\mathfrak{g}\,\alpha}{4x_1} $ \\ 
     \cline{1-1}\cline{3-13} 
      \cellcolor[HTML]{89CFF0}$ \textbf{vac~33} $ &  & $\mathcal{N}= 0$ & $ 0  $&  $ -\frac{\mathfrak{g}\,\alpha}{x_1\,x_2} $  & $ -\frac{\mathfrak{g}\,\alpha\,x_2}{2x_1\,x_3} $ & $ 0 $  & $0 $ & $ 0  $  & $  -\mathfrak{g}\,\alpha\,x_2 $ & $-\frac{\mathfrak{g}\,\alpha}{\sqrt{2x_1}}$ & $\frac{\mathfrak{g}\,\alpha\,x_3}{x_2}$ & $\frac{\mathfrak{g}\,\alpha}{4x_1} $  \\ 
\noalign{\hrule height 1pt}
     $ \textbf{vac~34} $ & $\textrm{AdS}_3$ & $\mathcal{N}=0$ & $ \mathfrak{g}\,\alpha\,y_{1} $ &  $ 0 $  & $ 0 $ & $ 0 $  & $\mathfrak{g}\,\alpha\,y_{2} $ & $ -\mathfrak{g}\,\alpha\,x  $  & $ \pm\mathfrak{g}\,\alpha\,\frac{y_{2}\,y_{3}}{y_{1}} $ & $ 0 $ & $ \pm\mathfrak{g}\,\alpha\,y_{3} $ & $ \mathfrak{g}\,\alpha\,y_{4} $  \\ 
\noalign{\hrule height 1.5pt}
\end{tabular}}
\caption{Fluxes producing unmagnetised ($\mathcal{F}=0$) flux vacua with arbitrary $\,\mathfrak{g}\,$ and $\,\alpha$. The blue-marked vacua correspond to solutions in which the closed-string moduli are stabilised.}
\label{Table:flux_vacua_unmagnetised}
\end{center}
\end{table}

\begin{table}[h!]
\begin{center}
\scalebox{0.80}{
\renewcommand{\arraystretch}{1.8}
\begin{tabular}{!{\vrule width 1.5pt}c!{\vrule width 1pt}c!{\vrule width 1pt}c!{\vrule width 1pt}cccccc!{\vrule width 1pt}cc!{\vrule width 1pt}ccc!{\vrule width 1pt}c!{\vrule width 1pt}c!{\vrule width 1.5pt}}
\noalign{\hrule height 1.5pt}
     ID & Scalar spectrum  \\ 
\noalign{\hrule height 1pt}
     $ \textbf{vac~16} $ & $\begin{array}{ccl}
        m^2 L^2 &=&  8_{(10)}, 4_{(18)} , 3_{(9)} , \ 0_{(51)}  \\[-2mm]
         \Delta &=&  4_{(10)}, \ (1+\sqrt{5})_{(18)}, \ 3_{(9)}, \ 2_{(51)} 
     \end{array}$ \\ 
     \hline
     $ \textbf{vac~17} $ &  $\begin{array}{ccl}
        m^2 L^2 &=&  8_{(15)}, 4_{(18)} , 3_{(8)} , \ 0_{(46)} , \ -1_{(1)}  \\[-2mm]
         \Delta &=&  4_{(15)}, \ (1+\sqrt{5})_{(18)}, \ 3_{(8)}, \ 2_{(46)} , \ 1_{(1)}
     \end{array}$ \\
\noalign{\hrule height 1pt}
      $ \textbf{vac~18} $ & $\begin{array}{ccl}
        m^2 L^2 &=&  24_{(20)} , \ 8_{(31)} , \ 0_{(37)} \\[-2mm]
         \Delta &=&  6_{(20)} , \ 4_{(31)} , \ 2_{(37)}
     \end{array}$ \\ 
     \hline
     $ \textbf{vac~19} $ & $\begin{array}{ccl}
        m^2 L^2 &=&  24_{(15)} , \ 8_{(37)} , \ 0_{(36)} \\[-2mm]
         \Delta &=&  6_{(15)} , \ 4_{(37)} , \ 2_{(36)}
     \end{array}$ \\
\noalign{\hrule height 1pt}
     $ \textbf{vac~20} $ & \multirow{2}{*}{$\begin{array}{ccl}
        m^2 L^2 &=&  15_{(16)}, 8_{(25)} , 3_{(12)} , \ 0_{(35)}  \\[-2mm]
         \Delta &=&  5_{(16)}, \ 4_{(25)}, \ 3_{(12)}, \ 2_{(35)} 
     \end{array}$ }\\ 
     \cline{1-1}
      $ \textbf{vac~22} $ & \\
      \hline
     $ \textbf{vac~21} $ &  \multirow{2}{*}{$\begin{array}{ccl}
        m^2 L^2 &=&  24_{(5)} , \ 15_{(16)} , \ 8_{(19)} , \ 3_{(12)} , 0_{(36)}  \\[-2mm]
         \Delta &=& 6_{(5)} , \ 5_{(16)} , \ 4_{(19)} , \ 3_{(12)} , \ 2_{(36)}
     \end{array}$ } \\ 
     \cline{1-1} 
     $ \textbf{vac~23} $ &  \\
      \noalign{\hrule height 1pt}
    $ \textbf{vac~24} $ & $ \begin{array}{ccl}
        m^2 L^2 &=&   8_{(37)} , \ 0_{(51)}  \\[-2mm]
         \Delta &=&  4_{(37)} , \ 2_{(51)}
     \end{array} $
     \\
    \hline
     $ \textbf{vac~25} $ &  $\begin{array}{ccl}
        m^2 L^2 &=&  24_{(5)} , \ 8_{(31)} , \ 0_{(52)}  \\[-2mm]
         \Delta &=& 6_{(5)} , \ 4_{(31)} , \ 2_{(52)}
     \end{array}$ \\
\noalign{\hrule height 1pt}
    $ \textbf{vac~26} $ & $ \begin{array}{ccl}
        m^2 L^2 &=&  \left[ \frac{8 (4+8\alpha^2+3\alpha^4)}{4+2\alpha^2+\alpha^4} \right]_{(5)} , \ \left[ \frac{8 (2+\alpha^2)^2}{4+2\alpha^2+\alpha^4} \right]_{(12)} , \ \left[ \frac{8 \alpha^2 (2+\alpha^2)}{4+2\alpha^2+\alpha^4} \right]_{(18)} , \left[ \frac{16 (2+\alpha^2)}{4+2\alpha^2+\alpha^4} \right]_{(1)} , \  8_{(1)} , \ 0_{(51)} 
   \end{array}$
     \\[2.5mm]
 \noalign{\hrule height 1pt}
     $ \textbf{vac~27} $ &  $ \begin{array}{ccl}
        m^2 L^2 &=&  \left[ \frac{8 (4+8\alpha^2+3\alpha^4)}{4+2\alpha^2+\alpha^4} \right]_{(10)} , \ \left[ \frac{24 (2+\alpha^2)^2}{4+2\alpha^2+\alpha^4} \right]_{(10)} , \ \left[ \frac{8 (2+\alpha^2)^2}{4+2\alpha^2+\alpha^4} \right]_{(21)},  \\[1mm] & &  \left[ \frac{8 \alpha^2 (2+\alpha^2)}{4+2\alpha^2+\alpha^4} \right]_{(9)} , \ \left[ \frac{16 (2+\alpha^2)}{4+2\alpha^2+\alpha^4} \right]_{(2)} , \ 8_{(1)} , \ 0_{(35)} \\
   \end{array} $ \\[7mm]
 \noalign{\hrule height 1pt}
     $ \textbf{vac~28} $ &  \multirow{2}{*}{$\begin{array}{ccl}
        m^2 L^2 &=&  \left[ \frac{8 (4+8\alpha^2+3\alpha^4)}{4+2\alpha^2+\alpha^4} \right]_{(5)} , \ \left[ \frac{3 (4+12\alpha^2+5\alpha^4)}{4+2\alpha^2+\alpha^4} \right]_{(8)} , \ \left[ \frac{15 (2+\alpha^2)^2}{4+2\alpha^2+\alpha^4} \right]_{(8)} ,\left[ \frac{8 (2+\alpha^2)^2}{4+2\alpha^2+\alpha^4} \right]_{(12)} , \\[1mm] & &   \ \left[ \frac{3 (2+\alpha^2)^2}{4+2\alpha^2+\alpha^4} \right]_{(12)} , \ \left[ \frac{8 \alpha^2 (2+\alpha^2)}{4+2\alpha^2+\alpha^4} \right]_{(6)} ,  \left[ \frac{16 (2+\alpha^2)}{4+2\alpha^2+\alpha^4} \right]_{(1)} , 8_{(1)} , \ 0_{(35)} 
   \end{array}$ } \\[2mm]
     \cline{1-1} 
     $ \textbf{vac~29} $ & \\
\noalign{\hrule height 1pt}
 \cellcolor[HTML]{89CFF0}$ \textbf{vac~30} $ &  \multirow{2}{*}{$\begin{array}{ccl}
        m^2 L^2 &=&   0_{(46)} \, , \, \textrm{ see Figure}~\ref{Fig:scalar_masses_plot}\\
        \end{array}$} \\
     \cline{1-1} 
      \cellcolor[HTML]{89CFF0}$ \textbf{vac~31} $ & \\
     \noalign{\hrule height 1pt}
 \cellcolor[HTML]{89CFF0}$ \textbf{vac~32} $ &  \multirow{2}{*}{$\begin{array}{ccl}
        m^2 L^2 &=&   0_{(46)}\, , \, \textrm{ see Figure}~\ref{Fig:scalar_masses_plot}\\
        \end{array}$} \\
     \cline{1-1} 
      \cellcolor[HTML]{89CFF0}$ \textbf{vac~33} $ & \\
\noalign{\hrule height 1pt}
$ \textbf{vac~34} $ &  $\begin{array}{ccl}
        m^2 L^2 &=& 0_{(23)} \, , \, \textrm{ see Figure}~\ref{Fig:scalar_masses_plot}\\ 
        \end{array}$ \\
\noalign{\hrule height 1.5pt}
\end{tabular}}
\caption{Scalar normalised masses at the AdS$_{3}$ flux vacua of Table~\ref{Table:flux_vacua_unmagnetised}. The subscript in $\#_{(s)}$ denotes the multiplicity of the mass $\,\#\,$ in the spectrum. In addition, the conformal dimension $\Delta$ of the would-be CFT$_{2}$ dual operators are also indicated for those AdS$_3$ vacua whose spectrum does not depend on $\,\alpha$.}
\label{Table:flux_vacua_unmagnetised_scalars}
\end{center}
\end{table}

\begin{figure}[t] 
  \begin{minipage}[b]{0.5\linewidth}
    \centering
      \hspace{7mm} \textbf{vac~30,31}\\
      \vspace{2mm}
    \includegraphics[width=0.95\linewidth]{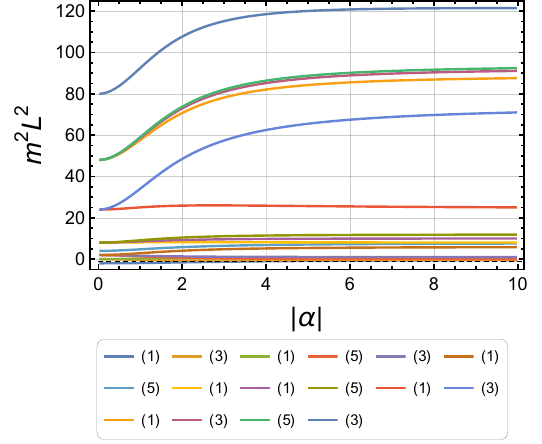} 
    \vspace{3ex}
  \end{minipage}
  \begin{minipage}[b]{0.5\linewidth}
    \centering
     \hspace{7mm} \textbf{vac~32,33}\\
      \vspace{2mm}
    \includegraphics[width=0.95\linewidth]{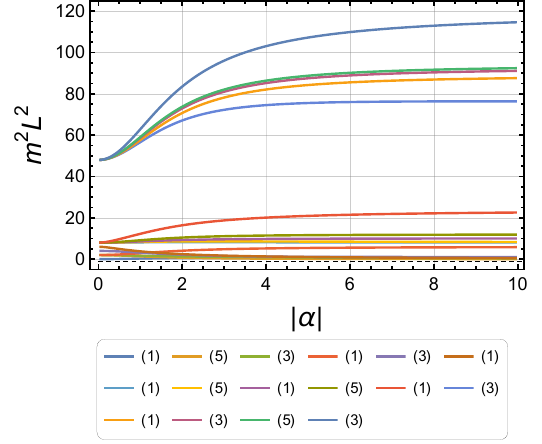}   
    \vspace{3ex}
  \end{minipage}  
  \begin{minipage}{\linewidth}
\hspace{26mm}\textbf{vac 34} \,\,\textrm{(branch $1$)}
\hspace{41mm}\textbf{vac 34} \,\,\textrm{(branch $2$)}
    \vspace{2mm} 
  \end{minipage}
  
  \begin{minipage}[b]{0.5\linewidth}
    \centering
    \includegraphics[width=0.95\linewidth]{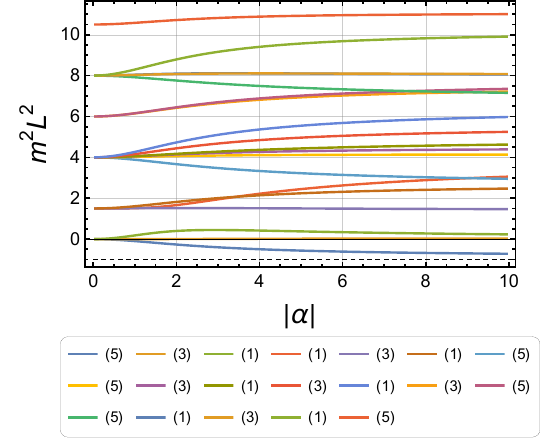}  
  \end{minipage}
  \begin{minipage}[b]{0.5\linewidth}
    \centering
    \includegraphics[width=0.95\linewidth]{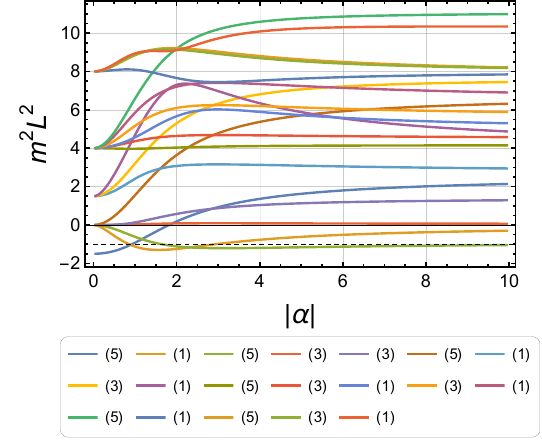}  
  \end{minipage} 
  \caption{Non-zero and $\alpha$-dependent normalised scalar masses at \textbf{vac~30,31}, \textbf{vac~32,33} and the two-fold \textbf{vac~34} of Table~\ref{Table:flux_vacua_unmagnetised}. The multiplicities are displayed in the legends below the plots. The black dashed line corresponds to the BF bound $m^{2}L^{2}=-1$. A normalised scalar mass below this value signals a perturbative instability of the corresponding AdS$_{3}$ vacuum.}
  \label{Fig:scalar_masses_plot} 
\end{figure}

The second scenario allowed by the Bianchi identity (\ref{QC_open_string}) corresponds to the choice
\begin{equation}
\mathfrak{f} = 0 
\hspace{10mm} \textrm{ and } \hspace{10mm}
\mathfrak{g} \neq 0 \ .
\end{equation}
This amounts to turning on a non-abelian YM theory on the D$5$-branes without allowing for YM background fluxes (\textit{i.e.} unmagnetised D$5$-branes).

The extremisation of the scalar potential, subject to the Jacobi and Bianchi identities in Section~\ref{sec:Jacobi&Bianchi_id}, enforces the condition
\begin{equation}
\left\langle \psi \right\rangle = 0 \ .
\end{equation}
One is then left with the open-string VEV $\,\left\langle \alpha \right\rangle\,$ which, this time, cannot be reabsorbed in a redefinition of the fluxes. Using \textsc{Singular} \cite{DGPS} to solve this algebraic problem exhaustively, one finds, first of all, the set of flux vacua in Table~\ref{Table:flux_vacua_paper} with arbitrary $\,\mathfrak{g} \neq 0$. In addition, there are also the AdS$_{3}$ vacua in Table~\ref{Table:flux_vacua_unmagnetised} which are genuine solutions with $\,\left\langle \alpha \right\rangle \neq 0$: they become either trivial or singular in the $\,\alpha \rightarrow 0\,$ limit.

In order to present the fluxes associated with \textbf{vac~30} to \textbf{vac~33} in Table~\ref{Table:flux_vacua_unmagnetised}, we have introduced three $\alpha$-dependent quantities $\,x_{1,2,3}\,$ of the form
\begin{equation}
x_{1}=\frac{1}{4(2+\alpha^2)}
\hspace{8mm} , \hspace{8mm} 
x_{2}=-\sqrt{\frac{8+3\alpha^2}{2}}
\hspace{8mm} , \hspace{8mm} 
x_{3}=4+2\alpha^2\pm\alpha\,\sqrt{2+\alpha^2} \ .
\end{equation}
Similarly, when presenting the fluxes associated with \textbf{vac~34}, we have introduced four $\alpha$-dependent quantities $\,y_{1,2,3,4}\,$ of the form
\begin{equation}
\label{vac_34_algebraic_1}
y_{3}=\sqrt{-y_{1}\,(y_{1}+2y_{2})}
\hspace{8mm} , \hspace{8mm} 
y_{4}=(2+\alpha^2)+y_{2}(1+\frac32\alpha^2) \ ,
\end{equation}
where $\,y_{1,2}\,$ are implicitly given in terms of $\,\alpha\,$ through the relations
\begin{equation}
\label{vac_34_algebraic_2}
\begin{array}{rcl}
2y_2^3+\frac{9 y_1^3}{4}+3 y_1^2\,y_2+\left(1+y_2+2 y_2^2\right) y_1&=& 0 \ , \\[2mm]
y_1^2+2 \alpha ^2 \left(1+y_2\right){}^2+4 \left(1+y_2+ y_1\,y_2\right)&=& 0 \ .
\end{array}
\end{equation}
The set of algebraic equations (\ref{vac_34_algebraic_1})-(\ref{vac_34_algebraic_2}) admits two distinct $y$-configurations or branches for a given value of $\alpha$. This implies that \textbf{vac~34} in Table~\ref{Table:flux_vacua_unmagnetised} is actually two-fold.

The scalar mass spectra associated with the unmagnetised AdS$_{3}$ flux vacua of Table~\ref{Table:flux_vacua_unmagnetised} are collected in Table~\ref{Table:flux_vacua_unmagnetised_scalars}. For the sake of presentation, the normalised scalar masses for \textbf{vac~30,31}, \textbf{vac~32,33} and the two-fold  \textbf{vac~34} are plotted as a function of $\,\alpha\,$ in Figure~\ref{Fig:scalar_masses_plot}. It turns out that, for small values of $\,\alpha$, \textbf{vac~30,31} and \textbf{vac~34} (branch $2$) feature perturbative instabilities (or tachyons), namely, scalars with a negative mass $\,m^2 L^2 < -1\,$ below the BF bound. However, these tachyons disappear when $\,\alpha\,$ increases. The rest of the unmagnetised AdS$_{3}$ vacua in Table~\ref{Table:flux_vacua_unmagnetised} are perturbatively stable within half-maximal $\,D=3\,$ supergravity. Note also that, with the exception of \textbf{vac~16,17}, only integer-valued $\Delta$'s appear for those AdS$_{3}$ flux vacua whose scalar mass spectrum is $\alpha$-independent. Finally, the gravitino masses associated with the unmagnetised AdS$_{3}$ flux vacua of Table~\ref{Table:flux_vacua_unmagnetised} are presented in Table~\ref{Table:flux_vacua_unmagnetised_gravitini} of Appendix~\ref{app_gravitini_masses}.

\subsubsection{Flux vacua with abelian magnetic fluxes on the D$5$-branes}

\begin{table}[t]
\begin{center}
\scalebox{0.57}{
\renewcommand{\arraystretch}{1.5}
\hspace{-2mm}\begin{tabular}{!{\vrule width 1.5pt}c!{\vrule width 1pt}c!{\vrule width 1pt}c!{\vrule width 1pt}cccccc!{\vrule width 1pt}ccc!{\vrule width 1pt}c!{\vrule width 1.5pt}}
\Xcline{4-13}{1.5pt}
\multicolumn{3}{c!{\vrule width 1pt}}{}& \multicolumn{6}{c!{\vrule width 1pt}}{$\omega$} & \multicolumn{3}{c!{\vrule width 1pt}}{$F_{(3)}$} & \multicolumn{1}{c!{\vrule width 1pt}}{$F_{(7)}$} \\ 
\noalign{\hrule height 1.5pt}
     ID & Type & SUSY & $\omega_{1}$ & $ \omega_{2}$ & $  \omega_{3}$ & $\omega_{4}$ & $ \omega_{5}$ & $ \omega_{6}$ & $f_{31}$ & $f_{32}$ & $f_{33}$ & $  f_{7}$  \\ 
\noalign{\hrule height 1pt}
      \cellcolor[HTML]{E25822}$ \textbf{vac~35} $ & \multirow{2}{*}{$\textrm{AdS}_3$} & $\mathcal{N}= 0$ & $ 0 $ & $ \pm 4 \sqrt{\frac23} \mathfrak{f}$ & $ \pm 3 \sqrt{\frac23} \mathfrak{f} $  & $ 0 $ & $ 0 $ & $  0 $ & $\pm  \sqrt{\frac32} \mathfrak{f}$ & $ \mp \sqrt{2} \mathfrak{f}$ & $ \mp \sqrt{\frac23} \mathfrak{f}$ & $0 $  \\  
     \cline{1-1}\cline{3-13} 
      \cellcolor[HTML]{E25822}$ \textbf{vac~36} $ &  & $\mathcal{N}=0$ & $ 0 $ & $\pm 4 \sqrt{\frac23} \mathfrak{f}$ & $ \pm 3 \sqrt{\frac23} \mathfrak{f} $  & $ 0 $ & $ 0 $ & $  0 $ & $\pm  \sqrt{\frac32} \mathfrak{f}$ & $ \pm \sqrt{2} \mathfrak{f}$ & $ \mp \sqrt{\frac23} \mathfrak{f}$ & $0 $  \\ 
     \hline
      \cellcolor[HTML]{E25822}$ \textbf{vac~37} $ & \multirow{2}{*}{$\textrm{AdS}_3$} & $\mathcal{N}=0$ & $ 0 $ & $ \pm 4 \sqrt{\frac23} \mathfrak{f}$ & $ \pm \sqrt{\frac23} \mathfrak{f} $  & $ 0 $ & $ 0 $ & $  0 $ & $\pm \sqrt{\frac32} \mathfrak{f}$ & $ \mp \sqrt{2}\mathfrak{f}$ & $ \mp 3 \sqrt{\frac23} \mathfrak{f}$ & $0 $ \\
     \cline{1-1}\cline{3-13}
      \cellcolor[HTML]{E25822}$ \textbf{vac~38}$ &  & $\mathcal{N}=0$ & $ 0 $ & $ \pm 4 \sqrt{\frac23} \mathfrak{f}$ & $ \pm \sqrt{\frac23} \mathfrak{f} $  & $ 0 $ & $ 0 $ & $  0 $ & $\pm \sqrt{\frac32} \mathfrak{f}$ & $ \pm \sqrt{2} \mathfrak{f}$ & $ \mp 3 \sqrt{\frac23} \mathfrak{f}$ & $0 $  \\
     \noalign{\hrule height 1pt}
      \cellcolor[HTML]{E25822}$ \textbf{vac~39} $ & \multirow{2}{*}{$\textrm{AdS}_3$} & $\mathcal{N}=0$ & $ 0 $ & $ \pm \frac{\sqrt{2}}{c_2} \mathfrak{f} $ & $\pm \frac{\sqrt{2} c_1}{c_2} \mathfrak{f} $  & $ 0 $ & $ 0 $ & $ 0 $ & $\pm  \frac{2\sqrt{2} c_1^2}{c_2} \mathfrak{f}$ & $ \pm \sqrt{2} \mathfrak{f} $ & $\mp \frac{\sqrt{2}c_1}{c_2} \mathfrak{f}$ & $0$  \\    
     \cline{1-1}\cline{3-13} 
       \cellcolor[HTML]{E25822}$\textbf{vac~40} $ &  & $\mathcal{N}=0$ & $ 0 $ & $ \pm \frac{\sqrt{2}}{c_2} \mathfrak{f} $ & $ \pm \frac{\sqrt{2}c_1}{c_2} \mathfrak{f} $  & $ 0 $ & $ 0 $ & $ 0 $ & $\pm \frac{2\sqrt{2}c_1^2}{c_2} \mathfrak{f}$ & $ \mp \sqrt{2}\mathfrak{f}$ & $\mp \frac{\sqrt{2}c_1}{c_2} \mathfrak{f}$ & $0$  \\ 
\noalign{\hrule height 1pt}
     $ \textbf{vac~41} $ & \multirow{2}{*}{$\textrm{AdS}_3$} & $\mathcal{N}=0$ & $\pm \sqrt{\frac{2}{5+4c_3}} \mathfrak{f} $ & $ 0 $ & $ 0 $  & $ 0 $ & $ \mp \frac{3+2c_3}{2} \sqrt{\frac{2}{5+4c_3}} \mathfrak{f} $ & $ \pm \frac{3+2c_3}{2} \sqrt{\frac{2}{5+4c_3}} \mathfrak{f} $ & $\pm c_3 \sqrt{\frac{2}{5+4c_3}} \mathfrak{f}$ & $ 0 $ & $\mp \frac{2c_3}{3+2c_3} \sqrt{\frac{2}{5+4c_3}} \mathfrak{f}$ & $\mp \frac{3+2c_3}{2} \sqrt{\frac{2}{5+4c_3}} \mathfrak{f}$ \\   
     \cline{1-1}\cline{3-13} 
     $ \textbf{vac~42} $ &  & $\mathcal{N}=0$ & $ \pm \sqrt{\frac{2}{5+4c_3}} \mathfrak{f} $ & $ 0 $ & $ 0 $  & $ 0 $ & $ \mp \frac{3+2c_3}{2} \sqrt{\frac{2}{5+4c_3}} \mathfrak{f} $ & $\pm \frac{3+2c_3}{2} \sqrt{\frac{2}{5+4c_3}} \mathfrak{f} $ & $\mp c_3 \sqrt{\frac{2}{5+4c_3}} \mathfrak{f}$ & $ 0 $ & $\pm \frac{2c_3}{3+2c_3} \sqrt{\frac{2}{5+4c_3}} \mathfrak{f}$ & $\mp \frac{3+2c_3}{2} \sqrt{\frac{2}{5+4c_3}} \mathfrak{f}$  \\ 
\noalign{\hrule height 1.5pt}
\end{tabular}}
\caption{Fluxes producing abelian ($\mathcal{G}=0$) and magnetised AdS$_{3}$ flux vacua with arbitrary~$\mathfrak{f}$. The red-marked vacua correspond to solutions in which the closed-string moduli are stabilised.}
\label{Table:flux_vacua_magnetised}
\end{center}
\end{table}

The third and last scenario allowed by the Bianchi identity (\ref{QC_open_string}) corresponds to the choice
\begin{equation}
\mathfrak{f} \neq 0 
\hspace{10mm} \textrm{ and } \hspace{10mm}
\mathfrak{g} = 0 \ .
\end{equation}
Equivalently, turning on three abelian YM fluxes of the form $\,\mathcal{F}_{ij}{}^K = \mathfrak{f} \, \epsilon_{ij}{}^{K}\,$ on the D$5$-branes.

Since $\,\mathfrak{g}=0$, we can exploit the shift symmetry (\ref{summary_shift_alpha}) to set $\,\left\langle \alpha \right\rangle =0\,$ without loss of generality. Then, the extremisation of the scalar potential, combined with the Jacobi and Bianchi identities in Section~\ref{sec:Jacobi&Bianchi_id}, enforces again the condition
\begin{equation}
\left\langle \psi \right\rangle = 0 \ .
\end{equation}
Using again \textsc{Singular} \cite{DGPS} to solve the extremisation problem exhaustively, one encounters the set of magnetised AdS$_{3}$ flux vacua displayed in Table~\ref{Table:flux_vacua_magnetised}. In order to present the fluxes associated with \textbf{vac~39,40}, we have introduced the (exact) constants
\begin{equation}
\label{const_vac_39,40_1}
c_2=\sqrt{3c_1-c_1^{2}-1}
\hspace{5mm} \textrm{ and } \hspace{5mm}
c_1 =\frac{1}{4\sqrt{2}}\sqrt{\sqrt{15}\left(c_4- \sqrt{-c_4^2+ 32 \sqrt{15}c_4^{-1}+39}\right)-3} \ ,
\end{equation}
with
\begin{equation}
\label{const_vac_39,40_2}
c_4=\sqrt{(64 \sqrt{3505}+4741)^{\frac13}+201(64 \sqrt{3505}+4741)^{-\frac13}+13} \ .
\end{equation}
Also, in order to present the fluxes associated with \textbf{vac~41,42}, we have introduced the (exact) constant
\begin{equation}
\label{const_vac_41,42}
c_3 = -\tfrac{1}{12} \left[ 14 + 28 \left( 2c_5^{-1} \right)^{\frac13} - \left( 4 c_5\right)^{\frac13} \right]
\hspace{8mm} \textrm{ with } \hspace{8mm}
c_5 = 151 + 9\sqrt{417} \ .
\end{equation}
The scalar and gravitini mass spectra\footnote{In order to lighten Table~\ref{Table:flux_vacua_magnetised_scalars}, we display the numerical value of (most of) the scalar masses instead of the exact expressions which, for example, involve the constants in (\ref{const_vac_39,40_1})-(\ref{const_vac_39,40_2}) and (\ref{const_vac_41,42}).} are collected in Table~\ref{Table:flux_vacua_magnetised_scalars} and Table~\ref{Table:flux_vacua_magnetised_gravitini} of Appendix~\ref{app_gravitini_masses}, respectively. From the former, it follows that \textbf{vac~41,42} are perturbatively unstable as they contain a tachyon with mass $\,m^2 L^2 \big{|}_{\textrm{tachyon}} \sim -1.268\,$ below the BF bound ($m^2 L^2 \ge -1$) for stability in AdS$_{3}$ \cite{Breitenlohner:1982bm}. The rest of the magnetised AdS$_{3}$ vacua in Table~\ref{Table:flux_vacua_magnetised} are perturbatively stable within half-maximal $\,D=3\,$ supergravity. Finally, all the spectra in Table~\ref{Table:flux_vacua_magnetised_scalars} contain irrational masses and so happens when looking at the associated $\Delta$'s.

\begin{table}[]
\begin{center}
\scalebox{0.9}{
\renewcommand{\arraystretch}{1.5}
\begin{tabular}{!{\vrule width 1.5pt}c!{\vrule width 1pt}c!{\vrule width 1pt}c!{\vrule width 1pt}cccccc!{\vrule width 1pt}cc!{\vrule width 1pt}ccc!{\vrule width 1pt}c!{\vrule width 1pt}c!{\vrule width 1.5pt}}
\noalign{\hrule height 1.5pt}
     ID & Scalar spectrum  \\
\noalign{\hrule height 1pt}
      \cellcolor[HTML]{E25822}$ \textbf{vac~35} $ & \multirow{4}{*}{$\begin{array}{ccl} m^2 L^{2} &=& 119.702_{(3)} , \, \left[\frac{4 \left( 119 + \sqrt{8473}\right)}{9}\right]_{(5)} , \, \left( \frac{832}{9} \right)_{(3)} , \, 88.848_{(1)} , \, 74.847_{(3)} , \, 24_{(1)} , \\ & &  \left[\frac{4 \left( 119 - \sqrt{8473}\right)}{9}\right]_{(5)}, \, 10.067_{(1)} , \, 8_{(6)} , \, 5.951_{(1)} , \, 1.007_{(3)} , \, 0.912_{(1)} , \, 0_{(55)} \end{array}$} \\
     \cline{1-1}
      \cellcolor[HTML]{E25822}$ \textbf{vac~36} $ & \\
     \cline{1-1}
      \cellcolor[HTML]{E25822}$ \textbf{vac~37} $ &   
     \\
     \cline{1-1}
      \cellcolor[HTML]{E25822}$ \textbf{vac~38} $ &  \\
     \noalign{\hrule height 1pt}
      \cellcolor[HTML]{E25822}$ \textbf{vac~39} $ & \multirow{2}{*}{$ \begin{array}{ccl} m^2 L^2 &=& 92.7707_{(3)}, \, 69.8938_{(5)} , \, 69.2676_{(1)} , \, 68.5297_{(3)} , \, 54.0276_{(3)} , \, 24_{(1)} , \, 11.9693_{(5)} ,\\[1mm] & &  10.0417_{(1)} ,  8_{(6)} , \, 5.9118_{(1)} , \, 0.2667_{(3)} , \, 0_{(55)} , \, -0.6611_{(1)}  \end{array}$} \\[2mm]
     \cline{1-1} 
      \cellcolor[HTML]{E25822}$ \textbf{vac~40} $ &
     \\
\noalign{\hrule height 1pt}
     $ \textbf{vac~41} $ & \multirow{2}{*}{$ \begin{array}{ccl} m^2 L^2 &=& 9.073_{(5)}, \, 9.058_{(1)} , \, 9.056_{(3)} , \, 8_{(1)} , \, 7.578_{(5)} , \, 6.451_{(1)} , \, 5.828_{(1)} , \, 5.571_{(3)} , \\[1mm] & &  4.912_{(1)} , \, 4.486_{(3)} , \, 3.985_{(5)} , \, 3.946_{(3)} , \, 2.725_{(1)} , \, 2.267_{(5)} , \, 0.262_{(3)} , \, 0.049_{(3)} , \\[1mm] & & 0_{(33)} , \, -0.583_{(5)} , \, -0.829_{(5)}, \, {\color{red}{-1.268_{(1)}}}  \end{array}$} \\[5mm]
     \cline{1-1} 
     $ \textbf{vac~42} $ & 
     \\[5mm]
\noalign{\hrule height 1.5pt}
\end{tabular}}
\caption{Scalar normalised masses at the AdS$_{3}$ flux vacua of Table~\ref{Table:flux_vacua_magnetised}. The subscript in $\#_{(s)}$ denotes the multiplicity of the mass $\,\#$. The presence of a tachyon ($m^{2} L^{2}<-1$) in the spectrum of \textbf{vac~41} and \textbf{vac~42} has been highlighted in red.}
\label{Table:flux_vacua_magnetised_scalars}
\end{center}
\end{table}

\section{On scale separation in AdS$_{3}$ flux vacua with open strings}
\label{sec:flux_vacua_phenomenology}

In this section we will address the question of the existence of scale-separated AdS$_{3}$ vacua with dynamical open strings. To this end, it proves convenient to undo the ``go to the origin" trick performed in the previous section (see eq.(\ref{GTTO_trick})), and express the moduli VEV's in terms of the fluxes by exploiting the symmetries of the scalar potential discussed in Section~\ref{sec:scaling_symmetry_general}. Once this is done, the flux-dependent VEV's of the four dilatons of the closed-string sector determine the string coupling $\,g_{s}=e^{\Phi}\,$ and the internal volume $\,\text{vol}_7\,$ (in string frame) as
\begin{equation}
\label{gs&vol7}
g_{s}^2 = \frac{1}{(\textrm{Im}U)^{3} \, \textrm{Im}R}
\hspace{10mm} \textrm{ and } \hspace{10mm}
\left(\textrm{vol}_{7}\right)^{4} = \frac{(\textrm{Im}T)^{3}}{(\textrm{Im}R)^{2} \, (\textrm{Im}S) \, (\textrm{Im}U)^{12}} \ ,
\end{equation}
where we have used the definitions (\ref{string-frame_volume})-(\ref{eq:indirect_g_s}) and the closed-string moduli dictionary (\ref{dictionary_dilatons}). Importantly, moduli stabilisation is achieved only for the following AdS$_{3}$ vacua:
\begin{itemize}
\item[$i)$] Unmagnetised and abelian D$5$-branes: \textbf{vac 10} and \textbf{vac 11} in Table~\ref{Table:flux_vacua_paper}.

\item[$ii)$] Unmagnetised and non-abelian D$5$-branes: \textbf{vac 30} to \textbf{vac 33} in Table~\ref{Table:flux_vacua_unmagnetised}.

\item[$iii)$] Magnetised and abelian D$5$-branes: \textbf{vac 35} to \textbf{vac 40} in Table~\ref{Table:flux_vacua_magnetised}.

\end{itemize}

\subsubsection*{Fluxes, Bianchi identities and internal geometry}

In all the above AdS$_{3}$ vacua, closed-string moduli stabilisation requires non-trivial metric fluxes $\,(\omega_{2},\omega_{3})\,$ as well as gauge fluxes $\,(f_{31},f_{32},f_{33})\,$ and (possibly) $\,f_{7}$. With only these fluxes activated, the first Bianchi identity in (\ref{QC_DF3}) yields the only non-trivial restriction
\begin{equation}
\label{tadpole_restricted}
\omega_2 \, f_{31} + 2 \, \omega_3 \, f_{33} = 0  \ ,
\end{equation}
which is consistently satisfied in all our AdS$_{3}$ vacua. Imposing (\ref{tadpole_restricted}), the unrestricted Bianchi identity component in (\ref{O5/D5_unrestrited}) for the allowed O$5$/D$5$ sources in (\ref{O5/D5_location}) becomes
\begin{equation}
\label{Bianchi_unrestricted_general}
6\,  f_{33}^{2} \, \frac{\omega_3}{f_{31}}  =  (2\pi \ell_{s})^{2}  \, \big(2 N_{\textrm{O}5} - N_{\textrm{D}5} \big)   \ .
\end{equation}
Therefore, the sign of the flux ratio $\,\omega_{3}/f_{31}\,$ determines whether there is an excess of O$5$-planes or D$5$-branes in the compactification scheme.

Before moving to discuss specific flux examples, let us comment on the general class of internal geometries that are consistent with just having metric fluxes $(\omega_{2},\omega_{3})$ in (\ref{metric_fluxes}). When only these two flux parameters are turned on, the algebra (\ref{X_isometry_brackets}) spanned by the isometry generators $\,(X_{a},X_{i})$\, and $\,X_{7}\,$ on the internal twisted torus (group manifold) simplifies to
\begin{equation}
\label{algebra_isometry_example_CORFU}
\begin{array}{cccc}
\left[ X_{2}, X_{7} \right]  = \omega_{3}  \, X_{1} & \hspace{8mm} , & \hspace{8mm}  \left[ X_{1},X_{7} \right] = - \omega_{2} \, X_{2}  & , \\[2mm] 
\left[ X_{4}, X_{7} \right]  = \omega_{3} \, X_{3} & \hspace{8mm} , & \hspace{8mm}  \left[ X_{3},X_{7} \right] = - \omega_{2} \, X_{4}  & , \\[2mm] 
\left[ X_{6}, X_{7} \right]  = \omega_{3} \, X_{5} & \hspace{8mm} , & \hspace{8mm}  \left[ X_{5},X_{7} \right] = - \omega_{2} \, X_{6}  & . 
\end{array}
\end{equation}
The brackets in (\ref{algebra_isometry_example_CORFU}) describe a $2$-step solvable algebra with degenerate Killing--Cartan metric. More concretely, (\ref{algebra_isometry_example_CORFU}) defines a specific example of a seven-dimensional solvmanifold for which the Maurer–Cartan equation (\ref{structure_equation}) admits an explicit integration. The resulting geometry can be interpreted as three copies of the three-dimensional solvmanifold E$_2$, all sharing a common seventh direction (see \cite{VanHemelryck:2025qok} for more details).

\subsection{Flux vacua with unmagnetised and abelian D$5$-branes}

Let us first consider \textbf{vac 10} and \textbf{vac 11} in Table~\ref{Table:flux_vacua_paper} which stabilise the closed-string moduli and are compatible with $\,\mathfrak{f}=\mathfrak{g}=0$. Changing to the more standard picture in which moduli VEV's are expressed in terms of free fluxes, one finds
\begin{equation}
\label{vac10&11_closed_VEVs}
\textrm{Im}R =  -\frac{f_{31} \left( \pm f_{32}^3 \, f_7\right)^{\frac12}}{\omega_3^2 \, f_{33}}
\hspace{1mm} , \hspace{1mm} 
\textrm{Im}S = \frac{\omega_3 \, f_{33}^2}{f_{31} \left( \pm f_{32}^3 \, f_7\right)^{\frac12}} 
\hspace{1mm} , \hspace{1mm} 
\textrm{Im}T = -\frac{\left( \pm f_{32}^3 \, f_7\right)^{\frac12}}{\omega_3 \, f_{33}}
\hspace{1mm} , \hspace{1mm} 
\textrm{Im}U = \left( \pm \frac{f_{32}}{f_7} \right)^{\frac12} \ ,
\end{equation}
together with
\begin{equation}
\label{vac10&11_open_VEVs}
\alpha = \textrm{free}
\hspace{10mm} \textrm{ and } \hspace{10mm}
\psi = \textrm{free} \ ,
\end{equation}
where the upper (lower) sign choice in (\ref{vac10&11_closed_VEVs}) corresponds to \textbf{vac~10} (\textbf{vac~11}). The value $\,V_{0}\,$ of the scalar potential at the vacua sets the AdS$_{3}$ radius $\,L\,$ as
\begin{equation}
L^{2} = -\frac{2}{V_{0}} = 64\frac{(f_{32}^3 f_7)^2}{(\omega_3 f_{33})^6} \, f_{31}^2      \ .
\end{equation}

An example of a flux choice that achieves scale separation is the one discussed in \cite{Arboleya:2025ocb}, namely,
\begin{equation}
\label{flux_example_CORFU}
\omega_{2} = a
\hspace{3mm} , \hspace{3mm}
\omega_{3} = b^{2} 
\hspace{3mm} , \hspace{3mm}
f_{31} = 2 \, a \, b^{2}
\hspace{3mm} , \hspace{3mm}
f_{32} = a \, b \, N^{5}
\hspace{3mm} , \hspace{3mm}
f_{33} = - a^{2}
\hspace{3mm} , \hspace{3mm}
f_{7}= 2 \, a^{2} \, b \, N^{11} ,
\end{equation}
with $\,a,b,N \in \mathbb{N}\,$ so that all the fluxes are integer-valued. These fluxes yield
\begin{equation}
\label{scales_example_CORFU}
g_{s} \sim N^{-2}
\hspace{7mm} , \hspace{7mm}
\textrm{vol}_{7} \sim  N^{\frac{31}2}
\hspace{7mm} , \hspace{7mm}
\tau_{0}^{-1} L \sim N^{\frac{13}2}
\ ,
\end{equation}
where $\,\tau_{0}^{-1} L\,$ characterises the size of the external spacetime using the 10D metric in (\ref{10D_metric_ansatz}). The parameter $\,N\,$ plays the role of a scaling parameter which must be taken arbitrarily large (formally $N\rightarrow \infty$) in order to achieve parametrically-controlled scale separation and weak coupling, \textit{i.e.}, $\,g_{s} \ll 1\,$ and $\,\tau_{0}^{-1} L \gg \left(\textrm{vol}_{7}\right)^{\frac{1}{7}}$. Importantly, the scaling parameter $\,N\,$ decouples from the unrestricted Bianchi identity (\ref{Bianchi_unrestricted_general}) since
\begin{equation}
(2\pi \ell_{s})^{2} \, \big(2 N_{\textrm{O}5} - N_{\textrm{D}5} \big)  = 3 \, a^{3}  > 0 \ .
\end{equation}
There must then be an excess of O$5$-planes over D$5$-branes in the compactification scheme. A similar analysis shows that the would-be one-cycles in \eqref{L_string_def} scale as
\begin{equation}
\label{scalings_one-forms_example_CORFU}
L^{s}_{a} = \rho \, \ell_{a} \sim N^{\frac{3}{2}}
\hspace{8mm} , \hspace{8mm}
L^{s}_{i} = \rho \, \ell_{i} \sim N^{\frac{3}{2}}
\hspace{8mm} \textrm{ and } \hspace{8mm}
L^{s}_{7} = \rho \, \ell_{7} \sim N^{\frac{13}{2}}  \ ,
\end{equation}
so that $\,L^{s}_{7}\,$ becomes of the same order as $\,\tau_{0}^{-1} L\,$ in (\ref{scales_example_CORFU}), as originally noticed in \cite{VanHemelryck:2025qok}. It is also worth emphasising the distinct scaling behaviour of the would-be one-cycles in (\ref{scalings_one-forms_example_CORFU}), which reveals an anisotropic internal geometry, with the seventh direction playing a special role. This anisotropy aligns with the obstruction to achieving parametric scale separation in isotropic compactifications, as recently discussed in \cite{Tringas:2025uyg}.

Finally, it was left as an open question in \cite{Arboleya:2025ocb} whether the dynamics of the open-string sector could introduce perturbative instabilities in the spectrum of scalar fluctuations about the non-supersymmetric and scale-separated \textbf{vac~10} and \textbf{vac~11} of \cite{Arboleya:2024vnp}. We can now answer the question. From Table~\ref{Table:flux_vacua_paper_scalars}, we observe that the open-string moduli turn out to be flat directions of the scalar potential, as can be inferred from the appearance of $\,8\mathfrak{N}\,$ zero masses in the spectrum of scalar fluctuations about \textbf{vac 10} and \textbf{vac~11}.\footnote{This is \textit{not} a generic feature of the flux vacua listed in Table~\ref{Table:flux_vacua_paper_scalars}.} Therefore, the open-string sector does not introduce any perturbative instability in this class of scale-separated type IIB AdS$_{3}$ flux vacua.

\subsection{Flux vacua with unmagnetised and non-abelian D$5$-branes}

\begin{table}[]
\begin{center}
\scalebox{0.68}{
\renewcommand{\arraystretch}{1.7}
\begin{tabular}{!{\vrule width 1pt}c!{\vrule width 1pt}c!{\vrule width 1pt}c!{\vrule width 1pt}c!{\vrule width 1pt}c!{\vrule width 1pt}c!{\vrule width 1pt}c!{\vrule width 1pt}c!{\vrule width 1pt}}
\noalign{\hrule height 1.5pt}
     ID & $V_{0} \equiv \left\langle V \right\rangle$ &  $  \text{Im} R$ & $\text{Im} S$ & $\text{Im} T$ & $\text{Im} U$ & $\alpha$ & $\psi$ \\ 
\noalign{\hrule height 1pt}
     \cellcolor[HTML]{89CFF0}$ \textbf{vac~30} $ & \multirow{2}{*}{$-\frac{g^2\alpha_0^4\left(8+3\alpha_0^2\right)^2 \mathfrak{g}^4}{128 f_{31}^2}$} & \multirow{2}{*}{$\left|\frac{2f_{33}\left(\alpha_0\pm2\sqrt{2+\alpha_0^2}\right)(f_{31})^{\frac12}}{\mathfrak{g}\alpha_0\left(8+3\alpha_0^2\right)(\omega_2)^{\frac12}}\right|$} & \multirow{2}{*}{$\left|\frac{f_{33}\mathfrak{g}\alpha_0\left(\alpha_0\pm2\sqrt{2+\alpha_0^2}\right)}{\left(f_{31}^3\omega_2\right)^{\frac12}}\right|$} &\multirow{2}{*}{ $ \left|\frac{\left(f_{31}\omega_2\right)^{\frac12}}{2\mathfrak{g}\alpha_0\left(2+\alpha_0^2\right)^{\frac12}}\right|$} & \multirow{2}{*}{$\left|\frac{4f_{32}^2\mathfrak{g}\alpha_0\left(2+\alpha_0^2\right)^{\frac12}}{\left(f_{31}\omega_2\right)^{\frac32}}\right|$} & \multirow{2}{*}{$\frac{f_{31}\omega_2}{2^{\frac32}f_{32}\mathfrak{g}\sqrt{2+\alpha_0^2}}$}  & \multirow{2}{*}{$0$}
     \\
     \cline{1-1} 
     \cellcolor[HTML]{89CFF0}$ \textbf{vac~31} $  & & & &  & &  &
     \\
     \hline
      \cellcolor[HTML]{89CFF0}$ \textbf{vac~32} $ & \multirow{2}{*}{$-\frac{g^2\alpha_0^4\left(8+3\alpha_0^2\right)^2 \mathfrak{g}^4}{128 f_{31}^2}$} & \multirow{2}{*}{$\left|\frac{2f_{33}\left(\alpha_0\pm2\sqrt{2+\alpha_0^2}\right)(f_{31})^{\frac12}}{\mathfrak{g}\alpha_0\left(8+3\alpha_0^2\right)(\omega_2)^{\frac12}}\right|$} & \multirow{2}{*}{$\left|\frac{f_{33}\mathfrak{g}\alpha_0\left(\alpha_0\pm2\sqrt{2+\alpha_0^2}\right)}{\left(f_{31}^3\omega_2\right)^{\frac12}}\right|$}  &\multirow{2}{*}{ $ \left|\frac{\left(f_{31}\omega_2\right)^{\frac12}}{2\mathfrak{g}\alpha_0\left(2+\alpha_0^2\right)^{\frac12}}\right|$} & \multirow{2}{*}{$\left|\frac{4f_{32}^2\mathfrak{g}\alpha_0\left(2+\alpha_0^2\right)^{\frac12}}{\left(f_{31}\omega_2\right)^{\frac32}}\right|$} & \multirow{2}{*}{$-\frac{f_{31}\omega_2}{2^{\frac32}f_{32}\mathfrak{g}\sqrt{2+\alpha_0^2}}$}  & \multirow{2}{*}{$0$}
     \\
      \cline{1-1}
      \cellcolor[HTML]{89CFF0}$ \textbf{vac~33} $ &  &  & &  &   & &
     \\
\noalign{\hrule height 1.5pt}
\end{tabular}}
\caption{Moduli VEV's for the unmagnetised \textbf{vac~30} to \textbf{vac~33} of Table \ref{Table:flux_vacua_unmagnetised} after undoing the ``go to the origin" trick.}
\label{Table:fields_vacua_unmagnetised}
\end{center}
\end{table}

Let us move to study \textbf{vac 30} to \textbf{vac 33} in Table~\ref{Table:flux_vacua_unmagnetised} which stabilise the closed-string moduli and are compatible with $\,\mathfrak{f}=0\,$ and $\mathfrak{g}\neq 0$. Switching again to the picture in which moduli VEV's are expressed in terms of free fluxes, one finds the vacuum configurations displayed in Table~\ref{Table:fields_vacua_unmagnetised}. There we have introduced the flux-dependent quantity $\,\alpha_{0}\,$ which is implicitly defined as the solution of the algebraic equation
\begin{equation}
\alpha_0^4\left(2+\alpha_0^2\right)=2^{-9} \left(\frac{f_{31} \, \omega_2}{f_{32}}\right)^6\frac{1}{f_7^2 \, \mathfrak{g}^4} \ .
\end{equation}
The non-zero metric flux $\,\omega_{3}\,$ is determined by the relation (\ref{tadpole_restricted}).

As a proof of concepts, we will present an example achieving scale separation within this class of AdS$_{3}$ vacua. Let us make a choice of fluxes of the form
\begin{equation}
\label{example:flux_scalings_unmagnetised}
\mathfrak{g} = a N^2
\hspace{2mm} , \hspace{2mm} 
\omega_2 =  b  
\hspace{2mm} , \hspace{2mm} 
\omega_3 = b 
\hspace{2mm} , \hspace{2mm} 
f_{31} = 2b 
\hspace{2mm} , \hspace{2mm} 
f_{32} = c N^2 
\hspace{2mm} , \hspace{2mm} 
f_{33} = - b
\hspace{2mm}, \hspace{2mm} 
f_7 = d N^{4} \ ,
\end{equation}
where $a$, $b$, $c$, $d$, $N \in \mathbb{N}$ so that the closed-string fluxes and the non-abelian structure constants are integer-valued. Then the string coupling, the internal volume (in string frame), and the size of the external spacetime scale with $\,N\,$ as
\begin{equation}
\label{example:flux_scalings_unmagnetised_scalings}
g_s \sim N^{-1} 
\hspace{7mm} , \hspace{7mm} 
\text{vol}_7 \sim N^{\frac{11}2}  \hspace{7mm} , \hspace{7mm}
\tau_{0}^{-1} L \sim N^{\frac52} \ ,
\end{equation}
whereas the net number of O$5$/D$5$ sources, $(2\pi \ell_{s})^{2} \, \big(2 N_{\textrm{O}5} - N_{\textrm{D}5} \big)  = 3  b^2$, does not scale with $\,N$. The scaling parameter $\,N\,$ must then be taken arbitrarily large in order to achieve scale separation, $(\tau_{0}^{-1} L)^{7} \gg \textrm{vol}_{7}$, at small string coupling, $g_{s} \ll 1$. The characteristic sizes of the would-be one-cycles in \eqref{L_string_def} become
\begin{equation}
\label{scalings_one-forms_unmagnetised}
L_{a}^{s} = \rho \, \ell_a \sim N^\frac12 
\hspace{8mm} , \hspace{8mm} 
L_{i}^{s} =  \rho \, \ell_i \sim N^\frac12
\hspace{8mm} \textrm{ and } \hspace{8mm}
L_{7}^{s} = \rho \, \ell_7 \sim N^\frac52  \ .
\end{equation}
so that $\,L_{7}^{s}\,$ becomes of the same order as $\,\tau_{0}^{-1} L\,$ in (\ref{scalings_one-forms_unmagnetised}). In addition, the internal geometry turns out to be anisotropic in agreement again with the general considerations put forward in \cite{Tringas:2025uyg}.

The flux choice in (\ref{example:flux_scalings_unmagnetised}) yields the scalings in (\ref{example:flux_scalings_unmagnetised_scalings}) for the four AdS$_{3}$ vacua in Table~\ref{Table:fields_vacua_unmagnetised}. However, since \textbf{vac 30} and \textbf{vac 31} feature tachyons at small values of $\,\alpha_0\,$ (see Figure~\ref{Fig:scalar_masses_plot}), it is important to determine the scaling of $\,\alpha_0\,$ with $\,N$. A direct evaluation gives $\,\alpha_0 \sim N^{-7}$, so tachyons will appear for \textbf{vac 30} and \textbf{vac 31} in the limit of large $\,N$, hence destabilising the vacua.\footnote{Alternatively, if one requires that $\,\alpha_0 \gg 1\,$ so that \textbf{vac 30} and \textbf{vac 31} are perturbatively stable, then it can be shown that $\,(2\pi\ell_{s})^{2} \, (2N_{\textrm{O}5} - N_{\textrm{D}5}) >0\,$ and scales with $\,N$. This clashes with achieving parametrically-controlled scale separation ($N\rightarrow \infty$) due to the upper bound for the number of O$5$-planes in the compactification.} On the contrary, \textbf{vac 32} and \textbf{vac 33} remain perturbatively stable in the scale-separated regime.

\subsection{Flux vacua with abelian magnetic fluxes on the D$5$-branes}

\begin{table}[]
\begin{center}
\scalebox{0.90}{
\renewcommand{\arraystretch}{1.7}
\begin{tabular}{!{\vrule width 1pt}c!{\vrule width 1pt}c!{\vrule width 1pt}c!{\vrule width 1pt}c!{\vrule width 1pt}c!{\vrule width 1pt}c!{\vrule width 1pt}c!{\vrule width 1pt}c!{\vrule width 1pt}c!{\vrule width 1pt}c!{\vrule width 1pt}c!{\vrule width 1pt}}
\noalign{\hrule height 1.5pt}
     ID & $V_{0} \equiv \left\langle V \right\rangle$ & $  \text{Im} R $ & $\text{Im} S$ & $\text{Im} T$ & $\text{Im} U$ & $\alpha$ & $\psi$ \\ 
\noalign{\hrule height 1pt}
       \cellcolor[HTML]{E25822}$ \textbf{vac~35} $ & \multirow{2}{*}{$-\frac{9g^2 \omega_2^8 f_{31}^6}{2^{19}f_{32}^8 \mathfrak{f}^4}$} & \multirow{2}{*}{$ \pm\frac{16 f_{32}^2 f_{33} \mathfrak{f}}{\left( \omega_2^5 f_{31}^3 \right)^{\frac12}} $} & \multirow{2}{*}{$ \pm \frac{3 f_{33} \left( \omega_2^3 f_{31} \right)^{\frac12}}{8 f_{32}^2 \mathfrak{f}}$} & \multirow{2}{*}{ $ \mp\frac{4 f_{32}^2 \mathfrak{f}}{\left( \omega_2^3 f_{31}^3 \right)^{\frac12}} $} & \multirow{2}{*}{$ \mp\frac{\left( \omega_2 f_{31} \right)^{\frac12}}{2 \mathfrak{f}} $} & \multirow{2}{*}{ $\frac{f_7}{3\mathfrak{f}}$}  & \multirow{2}{*}{$0$}
     \\
     \cline{1-1} 
       \cellcolor[HTML]{E25822}$\textbf{vac~36} $  &   &  &  & & &  & 
     \\
     \hline
           \cellcolor[HTML]{E25822}$ \textbf{vac~37} $ & \multirow{2}{*}{$-\frac{9g^2 \omega_2^8 f_{31}^6}{2^{19}f_{32}^8 \mathfrak{f}^4}$} & \multirow{2}{*}{$ \pm\frac{16 f_{32}^2 f_{33} \mathfrak{f}}{3\left( \omega_2^5 f_{31}^3 \right)^{\frac12}} $} & \multirow{2}{*}{$ \pm \frac{f_{33} \left( \omega_2^3 f_{31} \right)^{\frac12}}{8 f_{32}^2 \mathfrak{f}}$}  &\multirow{2}{*}{ $ \mp\frac{4 f_{32}^2 \mathfrak{f}}{\left( \omega_2^3 f_{31}^3 \right)^{\frac12}} $}  & \multirow{2}{*}{$ \mp\frac{\left( \omega_2 f_{31} \right)^{\frac12}}{2 \mathfrak{f}} $} & \multirow{2}{*}{ $\frac{f_7}{3\mathfrak{f}}$}  & \multirow{2}{*}{$0$}
     \\
     \cline{1-1} 
       \cellcolor[HTML]{E25822}$\textbf{vac~38} $  &   &  &  & & &  & 
     \\
 \noalign{\hrule height 1pt}
           \cellcolor[HTML]{E25822}$ \textbf{vac~39} $ & \multirow{2}{*}{$-\frac{g^2 \zeta(c_1,c_2) \omega_2^8 f_{31}^6}{2^{14} f_{32}^8 \mathfrak{f}^4}$}  & \multirow{2}{*}{$ \pm\frac{4d^2 f_{32}^2 f_{33} \mathfrak{f}}{c_2^3\left( \omega_2^5 f_{31}^3 \right)^{\frac12}} $} & \multirow{2}{*}{$ \pm \frac{c_2^3f_{33} \left( \omega_2^3 f_{31} \right)^{\frac12}}{2c_1^2 f_{32}^2 \mathfrak{f}}$}  &\multirow{2}{*}{ $ \mp\frac{4c_1^3 f_{32}^2 \mathfrak{f}}{c_2^3\left( \omega_2^3 f_{31}^3 \right)^{\frac12}} $}  & \multirow{2}{*}{$ \mp\frac{c_2\left( \omega_2 f_{31} \right)^{\frac12}}{2c_1 \mathfrak{f}} $} & \multirow{2}{*}{ $\frac{f_7}{3\mathfrak{f}}$}  & \multirow{2}{*}{$0$}
     \\
     \cline{1-1} 
       \cellcolor[HTML]{E25822}$ \textbf{vac~40} $  &   &  &  & & &  & 
     \\
\noalign{\hrule height 1.5pt}
\end{tabular}}
\caption{Moduli VEV's for the magnetised \textbf{vac~35} to \textbf{vac~40} of Table \ref{Table:flux_vacua_magnetised} after undoing the ``go to the origin" trick.}
\label{Table:fields_vacua_magnetised}
\end{center}
\end{table}

Let us finally consider \textbf{vac~35} to \textbf{vac~40} in Table \ref{Table:flux_vacua_magnetised} which are perturbatively stable, stabilise the closed-string moduli, and are compatible with $\,\mathfrak{f}\neq 0\,$ and $\mathfrak{g} = 0$. Moving again to the standard picture in which moduli VEV's are expressed in terms of free fluxes, we encounter the vacuum configurations presented in Table~\ref{Table:fields_vacua_magnetised}. In there, when presenting \textbf{vac~39,40}, we have introduced the quantity
\begin{equation}
\zeta(c_1,c_2) = \frac{c_2^{12}(3-3c_1-3c_1^2-8c_1^4)}{c_1^{12}} \ ,
\end{equation}
where the (exact) constants $\,c_{1}\,$ and $\,c_{2}\,$ were given in (\ref{const_vac_39,40_1})-(\ref{const_vac_39,40_2}).

Parametrically-controlled scale separation can be achieved for all the vacua in Table~\ref{Table:fields_vacua_magnetised}. A concrete example is provided by the flux choice
\begin{equation}
\label{example:flux_scalings_magnetised}
\mathfrak{f} =  a b N^4 
\hspace{2mm} , \hspace{2mm} 
\omega_2 =  b 
\hspace{2mm} , \hspace{2mm}
\omega_3 = b 
\hspace{2mm} , \hspace{2mm}
f_{31} = 2 b 
\hspace{2mm} , \hspace{2mm}
f_{32} = -  a b N^5 
\hspace{2mm} , \hspace{2mm}
f_{33} = -  b 
\hspace{2mm} , \hspace{2mm}
f_{7} = \textrm{free}  \ ,
\end{equation}
with $a$, $b$, $f_{7}$, $N \in \mathbb{N}$ so that the fluxes are quantised. Note that $\,f_{7}\,$ is a free parameter that sets the value of $\,\alpha\,$ and does not scale with $\,N$. It does not affect the stabilisation of the closed-string sector, nor the value of the scalar potential the vacuum. For any of the vacua in Table~\ref{Table:fields_vacua_magnetised}, the flux choice in (\ref{example:flux_scalings_magnetised}) gives rise to scalings of the form
\begin{equation}
\label{example:flux_scalings_magnetised_scalings}
g_s \sim N^{-1} 
\hspace{7mm} , \hspace{7mm} 
\text{vol}_7 \sim N^{19}
\hspace{7mm} , \hspace{7mm}
\tau_{0}^{-1} L \sim N^{7} \ ,
\end{equation}
while keeping $\,(2\pi \ell_{s})^{2} \, \big(2N_{\textrm{O}5} - N_{\textrm{D}5} \big) = 3 b^2\,$ independent of $\,N$. Then, taking the scaling parameter $\,N\,$ large produces scale separation and weak coupling, \textit{i.e.}, $(\tau_{0}^{-1} L)^{7} \gg \textrm{vol}_{7}\,$ and $\,g_{s} \ll 1$. The characteristic sizes of the would-be one-cycles in \eqref{L_string_def} read
\begin{equation}
\label{scalings_one-forms_magnetised}
L_{a}^{s} = \rho \, \ell_a \sim N^{2}
\hspace{8mm} , \hspace{8mm} 
L_{i}^{s} =  \rho \, \ell_i \sim N^{2}
\hspace{8mm} \textrm{ and } \hspace{8mm}
L_{7}^{s} = \rho \, \ell_7 \sim  N^{7}  \ ,
\end{equation}
so that $\,L_{7}^{s}\,$ becomes once again of the same order as $\,\tau_{0}^{-1} L\,$ in (\ref{example:flux_scalings_magnetised_scalings}). The internal geometry turns out to be anisotropic, again in agreement with \cite{Tringas:2025uyg}.

\section{Conclusions}
\label{sec:conclusions}

In this work we have constructed three-dimensional orientifold flux vacua of type IIB supergravity that incorporate dynamical open-string degrees of freedom associated with the presence of D$5$-branes in the compactification scheme. Focusing on the class of orientifold reductions put forward in \cite{Arboleya:2024vnp}, which include just a single type of D$5$-branes (and O$5$-planes), we enhanced the so-called RSTU-models described there to include also the dynamics of the open-string sector, along the lines of \cite{Danielsson:2017max,Balaguer:2024cyb}. We carried out the explicit dimensional reduction including the open-string fields, and reformulated the three-dimensional supergravity resulting from the compactification as a half-maximal gauged supergravity using the embedding tensor formalism of \cite{Nicolai:2001ac,deWit:2003ja}. When combined with algebraic geometry techniques \cite{DGPS}, this reformulation allowed us to completely chart the landscape of flux vacua, to assess the number of supersymmetries preserved at a given vacuum, and to obtain the complete spectrum of scalar and gravitino fluctuations about such a vacuum within half-maximal supergravity.

We found $\,42\,$ independent families of 3D flux vacua, which include not only the open-string generalisation of the closed-string vacua of \cite{Arboleya:2024vnp,Arboleya:2025ocb} (Table~\ref{Table:flux_vacua_paper}), but also new families of vacua that appear as a consequence of the open-string dynamics (Table \ref{Table:flux_vacua_unmagnetised} and Table \ref{Table:flux_vacua_magnetised}). Regarding the former, we showed that the non-supersymmetric vacua in \cite{Arboleya:2024vnp,Arboleya:2025ocb} maintain, contrary to expectations, their perturbative stability within half-maximal supergravity also when the open-string dynamics is taken into account. Amongst these vacua, \textbf{vac~10} and \textbf{vac~11} turned out to be particularly interesting, as they feature two remarkable properties: \textit{i}) they stabilise moduli and feature scale separation between the overall volume of the seven-dimensional internal space and the characteristic size of AdS$_{3}$, while remaining in the regime of validity of type IIB supergravity, and with quantised fluxes. \textit{ii}) all the would-be dual CFT$_2$ operators exhibit integer-valued conformal dimensions (as first noted in \cite{Conlon:2021cjk,Apers:2022tfm} for the scale-separated DGKT AdS$_{4}$ flux vacua \cite{DeWolfe:2005uu}) and are not relevant, thus describing (strongly-coupled) dead-end CFT's \cite{Nakayama:2015bwa}. Notably, these two properties persist upon the inclusion of open-string moduli. However, before speculating further, it is important to emphasise again that the existence of a holographic CFT$_{2}$ dual to the AdS$_{3}$ flux vacua presented in this work is not guaranteed at all. The situation is even worse for \textbf{vac~10} and \textbf{vac~11} which feature scale separation. To the best of our knowledge, no example of an AdS/CFT pair featuring scale separation has been constructed, and it may even be impossible \cite{Collins:2022nux}. A first step towards constructing such an AdS$_{3}$/CFT$_2$ pair (if it exists) would be to identify the brane configurations underlying our AdS$_3$ flux vacua (see \cite{Apers:2025pon} for some recent progress backtracking the scale-separated DGKT AdS$_{4}$ flux vacua \cite{DeWolfe:2005uu}). Given the non-zero $\,F_{(7)}\,$ and $\,F_{(3)}\,$ in our AdS$_{3}$ flux vacua, these would presumably correspond to some variant of D$1$/D$5$ system probing a singularity in 10D. While certainly interesting, these questions lie beyond the scope of the present work.

In this work we have used the overall volume of the internal space as a proxy for the masses of the KK modes in the compactification. The examples of scale-separated AdS$_{3}$ flux vacua we presented in this work have $\,\tau_{0}^{-1} L \gg (\textrm{vol}_{7})^{\frac{1}{7}}$, and we referred to them as scale-separated in this restricted sense. A potential issue is that, in all our examples, the size of a would-be one-cycle in the internal space -- we denote it $\,L^{s}_7\,$ in (\ref{L_string_def}) -- becomes comparable to the characteristic size of the external spacetime, as in the supersymmetric examples of \cite{VanHemelryck:2025qok}. However, our setup is compatible with the class of $\,\textrm{G}_{2}\,$ orientifold reductions discussed in \cite{Emelin:2021gzx}, where the internal space lacks both one-cycles and five-cycles. Only three-cycles and four-cycles exist -- together with the seven-dimensional volume form -- and, in our examples, all of them stay parametrically scale-separated with respect to the size of the external spacetime. It would then be very interesting to perform a detailed analysis of the KK spectrum about the scale-separated (in the restricted sense) AdS$_{3}$ vacua presented in this work (and in \cite{Arboleya:2024vnp,Arboleya:2025ocb} and \cite{VanHemelryck:2025qok}) in order to establish whether or not the KK modes are physically decoupled from the degrees of freedom of the effective three-dimensional gauged supergravity. Moreover, this would set the question about the higher-dimensional stability of our non-supersymmetric AdS$_{3}$ flux vacua. Such a detailed KK spectrometry could be carried out using exceptional field theory techniques \cite{Malek:2019eaz} (see \cite{Eloy:2020uix,Eloy:2023acy,Eloy:2024lwn} for some explicit examples), suitably adapted to the half-maximal context of type II orientifold reductions. Again, this goes beyond the scope of the present work.

Finally, throughout this paper, we have employed smeared sources in our analysis. Our approach assumes that smeared sources are a consistent ingredient, and that they provide a valid approximation to an underlying fully localised solution in string theory. We should then be able to see this full-fledged solution as a small perturbation around the smeared background. A natural next step would then be to investigate the backreaction of the O$5$-planes order by order in $\,g_{s}\,$ along the lines of \cite{Junghans:2020acz, Junghans:2023yue, Emelin:2022cac, Emelin:2024vug}, and assess the validity of the smeared approximation. We hope to come back to all these questions in future work.

\section*{Acknowledgements}

We would like to thank Valentina Bevilacqua, Giuseppe Dibitetto, Ant\'on Faedo, Ricardo Stuardo and Vincent Van Hemelryck for valuable discussions and explanations. The work of \'AA, AG and MM is supported by the Spanish national grant MCIU-22-PID2021-123021NB-I00. GS thanks the Physics Department of the University of Oviedo and the ICTEA for the kind hospitality during the completion of this work.

\appendix

\section{$\mathcal{N}=8$, $D=3$ supergravity with matter multiplets}
\label{Appendix:half-maximal_sugra}

Supergravity theories in three dimensions (3D) with half-maximal ($\mathcal{N}=8$) supersymmetry have been systematically studied in \cite{Nicolai:2001ac,deWit:2003ja} and, more recently, also in \cite{Deger:2019tem}. The most general 3D $\mathcal{N}=8$ supergravity theory can be constructed upon application of the so-called gauging procedure \cite{Nicolai:2001ac} to the ungauged theory of \cite{Marcus:1983hb}, which features an $\,\textrm{SO}(8,8+\mathfrak{N})\,$ global symmetry (also known as the duality group). The number $\,\mathfrak{N} \in \mathbb{N}\,$ denotes the arbitrary number of matter multiplets to which the supergravity multiplet can be coupled. Performing a gauging amounts to promote a subgroup $\,\textrm{G}_{g} \subset \textrm{SO}(8,8+\mathfrak{N})\,$ from global to local. The theory then becomes a gauged supergravity, with a scalar potential and scalar-dependent fermionic mass terms induced by the gauging.

The field content of the half-maximal supergravities consists of the $\mathcal{N}=8$ supergravity multiplet and the $\mathfrak{N}$ matter multiplets. The supergravity multiplet contains the dreibein $\,e_{\mu}{}^{\mathsf{a}}\,$ and eight gravitini $\psi_{\mu}{}^{\mathsf{A}}$ (with $\mathsf{A}=1,\ldots,8$). The index $\,\mathsf{A}\,$ transforms in the spinorial representation of the R-symmetry group $\textrm{SO}(8)_\textrm{R}$. Each matter multiplet contains eight real scalar fields $\phi^{\mathsf{I}}$ (spin-$0$) and eight two-component Majorana fermions $\chi^{\dot{\mathsf A}}$ (spin-$\frac12$), where $\,\mathsf{I}=1,\ldots,8\,$ and $\,\dot{\mathsf{A}}=1,\ldots,8\,$ transform in the vectorial and conjugate-spinorial representations of $\textrm{SO}(8)_\textrm{R}$, respectively. The scalar fields in the theory serve as coordinates on the coset space
\begin{equation}
\label{general_coset}
\mathcal{M}_{\textrm{scalar}} = \frac{\textrm{SO}(8,8+\mathfrak{N})}{\textrm{SO}(8)\times\textrm{SO}(8+\mathfrak{N})} \ ,
\end{equation}
thus comprising $64+8\mathfrak{N}$ scalars. The Lagrangian of the $\mathcal{N}=8$ gauged supergravities in 3D can be found, for example, in ref.~\cite{Deger:2019tem}. We will closely follow conventions therein. The bosonic part of the Lagrangian is of the form
\begin{equation}
\label{L_bosonic}
\mathcal{L}_{\textrm{bos.}} = - \frac{1}{4} \, e \,  R - \frac{1}{32} \, e \, D_\mu M_{\mathcal{M}\mathcal{N}} \, D^{\mu} M^{\mathcal{M}\mathcal{N}} - e \, V(M) + \mathcal{L}_{\textrm{CS}}  \ ,
\end{equation}
where the scalar-dependent matrix $\,M_{\mathcal{M}\mathcal{N}}\,$ can be constructed in terms of a coset representative
\begin{equation}
\label{general_coset_representative}
 \mathcal{V}_{\mathcal{M}}{}^{\underline{\mathcal{P}}}(\phi) \in \frac{\textrm{SO}(8,8+\mathfrak{N})}{\textrm{SO}(8) \times \textrm{SO}(8+\mathfrak{N})} \ ,
\end{equation}
as
\begin{equation}
M_{\mathcal{M} \mathcal{N}} = \mathcal{V}_{\mathcal{M}}{}^{\underline{\mathcal{P}}} \,\,  \mathcal{V}_{\mathcal{N}}{}^{\underline{\mathcal{Q}}} \,\, \delta_{\underline{\mathcal{P} \mathcal{Q}}} \ .   
\end{equation}
The indices $\,\mathcal{M}\,$ and $\,\underline{\mathcal{M}}\,$ denote the fundamental (or vectorial) indices of $\,\textrm{SO}(8,8+\mathfrak{N})\,$ and $\,\textrm{SO}(8) \times \textrm{SO}(8+\mathfrak{N})$, respectively.

As a consequence of the gauging procedure, the bosonic Lagrangian (\ref{L_bosonic}) contains a scalar potential $\,V(M)\,$ and covariant derivatives $\,D_{\mu}M\,$ for the scalar fields. The covariant derivatives involve the vector fields spanning $\,\textrm{G}_{g}\,$ which, in three dimensions, are (on-shell) dual to the scalars themselves. Namely, the vectors are nonlocal and nonlinear functions
of the scalar fields. At the level of the Lagrangian, the duality relations between vectors and scalars are implemented by the Chern--Simons (CS) term in (\ref{L_bosonic}) (we refer to \cite{Nicolai:2001ac,Deger:2019tem} for more details on this construction).

\subsection{Embedding tensor and covariant derivatives}

The gauging procedure is encoded in the so-called embedding tensor \cite{Nicolai:2001ac}. This tensor, $\,\Theta$, specifies the precise embedding of the subgroup $\,\textrm{G}_{g} \subset \textrm{SO}(8,8+\mathfrak{N})\,$ of the global symmetry group that is promoted to a local symmetry of the Lagrangian. The embedding tensor can be written as
\begin{equation}
\label{ET_N=8}
\Theta_{\mathcal{M}\mathcal{N}|\mathcal{P}\mathcal{Q}} = \theta_{\mathcal{MNPQ}} + 2 \left( \eta_{\mathcal{M}[\mathcal{P}} \theta_{\mathcal{Q}]\mathcal{N}} - \eta_{\mathcal{N}[\mathcal{P}} \theta_{\mathcal{Q}]\mathcal{M}} \right) + 2 \eta_{\mathcal{M}[\mathcal{P}} \eta_{\mathcal{Q}]\mathcal{N}} \theta \ ,
\end{equation}
in terms of three irreducible representations of $\textrm{SO}(8,8+\mathfrak{N})$. Setting $\,\mathfrak{N}=3$, in agreement with our choice of YM gauge group in (\ref{G_YM_SO(3)}), these three irrep's are are $\,\theta_{\mathcal{MNPQ}} = \theta_{[\mathcal{MNPQ}]} \in {\bf 3876}$, $\,\theta_{\mathcal{MN}} = \theta_{(\mathcal{MN})} \in {\bf 189}\,$ (with $\,\theta_{\mathcal{M}}{}^{\mathcal{M}} = 0$) and $\,\theta \in {\bf 1}$.
The $\textrm{SO}(8,8+\mathfrak{N})$ indices can be raised and lowered through a metric $\eta_{\mathcal{M}\mathcal{N}}$, which, while always adopting a Lorentzian basis along the $\textrm{SO}(\mathfrak{N})$ directions, takes the form
\begin{equation}
\label{eta_matrices}
\eta_{\mathcal{M N}}\Big|_{\textrm{Ltz}} = \left( \begin{matrix}
    -\mathbb{I}_8 & 0 & 0\\
    0 & \mathbb{I}_8 & 0\\
    0 & 0 &  \mathbb{I}_{\mathfrak{N}}
\end{matrix} \right)
\hspace{10mm} \textrm{ or } \hspace{10mm}
\eta_{\mathcal{M N}} \Big|_{\textrm{LC}} = \left( \begin{matrix}
    0 & \mathbb{I}_8 & 0\\
    \mathbb{I}_8 & 0 & 0\\
    0 & 0 & \mathbb{I}_{\mathfrak{N}}
\end{matrix} \right) \ ,
\end{equation}
depending on the choice of the Lorentzian (Ltz) or light-cone (LC) basis for the $\textrm{SO}(8,8)$ directions. Using the Lorentzian basis in (\ref{eta_matrices}), the $\,\textrm{SO}(8,8+\mathfrak{N})\,$ vectorial index $\,\mathcal{M}\,$ of the $\,\bf{16+\mathfrak{N}}\,$ splits in the vectorial indices of $\,\textrm{SO}(8)\,$ and $\,\textrm{SO}(8+\mathfrak{N})\,$ as $\,\mathcal{M} = (\mathfrak{I},\mathfrak{r})$. Using, instead, the light-cone basis along the $\,\textrm{SO}(8,8)\,$ directions (see (\ref{eta_matrices})), the vectorial index $\,\mathcal{M}\,$ decomposes as $\,\mathcal{M} = (A, \bar{A}, I)$, where $\,A, \bar{A} = 1, \ldots, 8\,$ and $\,I = 1, \ldots, \mathfrak{N}$. Note that the Ltz basis and the LC basis are related one to the other by an $\,\textrm{SO}(16+\mathfrak{N})\,$ rotation as
\begin{equation}
\label{LV_vs_Ltz}
\eta_{\mathcal{MN}}\Big|_{\textrm{Ltz (LC)}} = R_{\mathcal{M}}{}^{\mathcal{P}} \, R_{\mathcal{N}}{}^{\mathcal{Q}}  \,\,  \eta_{\mathcal{PQ}}\Big|_{\textrm{LC (Ltz)}}
\hspace{6mm} \textrm{with} \hspace{6mm}
R_{\mathcal{M}}{}^{\mathcal{N}} = \frac{1}{\sqrt{2}} 
\left( \begin{matrix}
    -\mathbb{I}_8 & \mathbb{I}_8 & 0\\
    \mathbb{I}_8 & \mathbb{I}_8 & 0\\
    0 & 0 & \mathbb{I}_{\mathfrak{N}}
\end{matrix} \right) \ .
\end{equation}

The set of generators $\,X_{\mathcal{MN}}\,$ that are promoted from global to local symmetries, and therefore enter the gauge connection, are selected by the embedding tensor $\,\Theta$. More concretely, the generators of $\,\textrm{G}_{g} \subset \textrm{SO}(8,8+\mathfrak{N})\,$ are given by
\begin{equation}
\label{XMN_tensor}
X_{\mathcal{MN}} = \Theta_{\mathcal{M}\mathcal{N}|\mathcal{P}\mathcal{Q}} \, L^{\mathcal{PQ}} \ ,
\end{equation}
with 
\begin{equation}
\label{Generators_SO(8,8+N)}
\left( L^{\mathcal{PQ}} \right)_{\mathcal{R}}{}^{\mathcal{T}} = 2 \, \delta_{\mathcal{R}}^{[\mathcal{P}} \, \eta^{\mathcal{Q}]\mathcal{T}}
\hspace{10mm} \textrm{and} \hspace{10mm}
\left( L^{\mathcal{PQ}} \right)_{\mathcal{RS}}{}^{\mathcal{TU}} = 8 \, \delta_{[\mathcal{R}}^{[\mathcal{P}} \,  \eta^{\mathcal{Q}][\mathcal{T}} \, \delta_{\mathcal{S}]}^{\mathcal{U}]}\ ,
\end{equation}
being the generators of $\,\textrm{SO}(8,8+\mathfrak{N})\,$ in the fundamental and adjoint representations, respectively. The generators (\ref{XMN_tensor}) enter the gauge connection and specify the covariant derivatives in (\ref{L_bosonic}). These are given by
\begin{equation}
\label{cov_der_3D}
D_{\mu}M_{\mathcal{MN}} = \partial_{\mu}M_{\mathcal{MN}} + 2 \, g \, A_{\mu}{}^{\mathcal{PQ}} \, X_{\mathcal{PQ}|(\mathcal{M}}{}^{\mathcal{R}} \, M_{\mathcal{N})\mathcal{R}} \ ,
\end{equation}
where $\,g\,$ is the gauge coupling and $\,A_{\mu}{}^{\mathcal{MN}}=A_{\mu}{}^{[\mathcal{MN}]}\,$ are the vectors  (on-shell dual to scalars) formally introduced to implement the gauging procedure.

Using the adjoint representation, \textit{i.e.} $X_{\mathcal{MNRS}}{}^{\mathcal{TU}}  \equiv \Theta_{\mathcal{M}\mathcal{N}|\mathcal{P}\mathcal{Q}} \, \left( L^{\mathcal{PQ}} \right)_{\mathcal{RS}}{}^{\mathcal{TU}}$, the gauge group generators in (\ref{XMN_tensor}) determine the gauge brackets
\begin{equation}
\label{brackets_N=8}
\left[ X_{\mathcal{MN}} , X_{\mathcal{PQ}}  \right] = -\frac12 X_{\mathcal{MNPQ}}{}^{\mathcal{RS}} \, X_{\mathcal{RS}} \ ,
\end{equation}
whose closure requires a set of quadratic constraints (QC's) of the form
\begin{equation}
\label{QC_N=8}
\Theta_{\mathcal{KL}|[\mathcal{M}}{}^{\mathcal{R}} \, \Theta_{\mathcal{N}]\mathcal{R}|\mathcal{P}\mathcal{Q}} + \Theta_{\mathcal{K}\mathcal{L}|[\mathcal{P}}{}^{\mathcal{R}} \, \Theta_{\mathcal{Q}]\mathcal{R}|\mathcal{M}\mathcal{N}} = 0 \ .
\end{equation}
Any embedding tensor of the form (\ref{ET_N=8}) that satisfies (\ref{QC_N=8}) generates a consistent gauged supergravity in three dimensions with half-maximal supersymmetry.

\subsection{Fermion masses and scalar potential}

As already mentioned, the gauging procedure gives rise to scalar-dependent masses for the fermionic fields in the theory as well as to a scalar potential \cite{Nicolai:2001ac}. Concretely,
\begin{equation}
\label{L_mass_fermi}
g^{-1} \, \mathcal{L}_{\textrm{fermi mass}} = \frac{1}{2} \, e \, A_{1}^{\mathsf{A}\mathsf{B}} \, \bar{\psi}_{\mu}{}^{\mathsf{A}} \gamma^{\mu\nu} \psi_{\nu}{}^{\mathsf{B}} + i \, e \, A_2^{\mathsf{A \, \dot{B}} {\mathfrak r}} \, \bar{\chi}^{\dot{\mathsf{B}}{\mathfrak r}} \, \gamma^{\mu} \, \psi_{\mu}^{\mathsf{A}} +  \frac{1}{2} \, e \, A_3^{\mathsf{\dot A} {\mathfrak r} \mathsf{\dot B} {\mathfrak s}} \,  \bar{\chi}^{\dot{\mathsf{A}} {\mathfrak r}}  \, \chi^{\dot{\mathsf{B}} {\mathfrak s}} \ ,
\end{equation}
and
\begin{equation}
\label{V_N=8}
g^{-2} \, V = 
-\frac{1}{4} A_1^{{\mathsf A \mathsf B}} A_1^{{\mathsf A \mathsf B}} + \frac{1}{8} A_2^{{\mathsf A \, \mathsf{\dot B}} {\mathfrak s}} A_2^{{\mathsf A \, \mathsf{\dot B}} {\mathfrak s}} \ ,
\end{equation}
where the scalar-dependent quantities $A_1^{{\mathsf A \mathsf B}}$ and $A_2^{{\mathsf A \, \mathsf{\dot B}} {\mathfrak r}}$ determine, respectively, the masses of the gravitini $\psi_\mu{}^{\mathsf A}$ and their quadratic interaction with the matter multiplets' fermions $\chi^{\dot{\mathsf A}{\mathfrak r}}$. The masses of the fermions $\chi^{\dot{\mathsf A}{\mathfrak r}}$ are, instead, described by a third associated quantity $A_3^{\mathsf{\dot A} {\mathfrak r} \mathsf{\dot B} {\mathfrak s}}$.

The explicit expressions of the three fermionic mass terms $\,A_1^{{\mathsf A \mathsf B}}$, $\,A_2^{{\mathsf  A \, \mathsf{\dot B}} {\mathfrak s}}\,$ and $\,A_3^{\mathsf{\dot A} {\mathfrak r} \mathsf{\dot B} {\mathfrak s}}\,\,$ are obtained as follows. Using the Lorentzian basis in (\ref{eta_matrices}), in which the index $\,\mathcal{M}\,$ of $\,\textrm{SO}(8,8+\mathfrak{N})\,$ splits as $\,\mathcal{M} = (\mathfrak{I},\mathfrak{r})$, one first constructs the coset representative
\begin{equation}
\label{Ltz_coset}
\mathcal{V}_{\mathcal{M}}{}^{\underline{\mathcal{P}}}(\phi) = e^{\phi_{\mathfrak{I r}} \left(L^{\mathfrak{I r}}\right)_{\mathcal{M}}{}^{\underline{\mathcal{P}}}} \ ,
\end{equation}
where $\,\phi_{\mathfrak{I r}}\,$ are the $\,64+8\mathfrak{N}\,$ physical scalars associated with the non-compact generators $\,L^{\mathfrak{I r}}\,$ of the duality group $\,\textrm{SO}(8,8+\mathfrak{N})$.\footnote{The generators $\,L^{\mathfrak{I J}}\,$ and $\,L^{\mathfrak{r s}}\,$ span the compact space $\textrm{SO}(8)\times\textrm{SO}(8+\mathfrak{N})$ in the denominator of (\ref{general_coset}), and have no physical scalars associated.} The coset representative (\ref{Ltz_coset}) is then used to construct the so-called $T$-tensor, which is nothing but the embedding tensor \eqref{ET_N=8} dressed up with the scalars in the theory. It has components
\begin{equation}
\label{T-tensor_components}
\begin{array}{rcl}
T_{\underline{\mathcal{MNPQ}}} &=&  (\mathcal{V}^{-1})_{\underline{\mathcal M}}{}^{\mathcal R} \, (\mathcal{V}^{-1})_{\underline{\mathcal N}}{}^{\mathcal S} \, (\mathcal{V}^{-1})_{\underline{\mathcal P}}{}^{\mathcal T} \, (\mathcal{V}^{-1})_{\underline{\mathcal Q}}{}^{\mathcal U} \, \theta_{\mathcal{RSTU}} \ , \\[2mm]
T_{\underline{\mathcal{MN}}} &=&  (\mathcal{V}^{-1})_{\underline{\mathcal M}}{}^{\mathcal R} \, (\mathcal{V}^{-1})_{\underline{\mathcal N}}{}^{\mathcal S} \,  \theta_{\mathcal RS} \ , \\[2mm]
T &=& \theta \ .
\end{array}
\end{equation}
Finally, using the $\textrm{SO}(8)_{\textrm{R}}$-invariant $\gamma^{(p)}$-forms, the fermionic mass terms are obtained upon suitable projection of the $T$-tensor into irreducible representations of the R-symmetry group. In particular,
\begin{equation}
\label{Fermion-Shifts}
\begin{array}{rcl}
A_1^{\mathsf{AB}} &=& -\frac{1}{48} \, [\gamma^{\mathfrak{IJKL}}]_{\mathsf{A B}} \,\, T_{\underline{\mathfrak{IJKL}}} - \frac14 \delta^{\mathsf{A B}} \left( T_{\underline{\mathfrak{II}}} - 4T \right) \ , \\[3mm]
A_2^{\mathsf{A \, \dot{B}} {\mathfrak r}}  &=& -\frac{1}{12} \,  [\gamma^{\mathfrak{IJK}}]_\mathsf{A\dot{B}} \,\, T_{\underline{\mathfrak{IJK}{\mathfrak r}}} - \frac{1}{2} [\gamma^{\mathfrak{I}}]_{\mathsf{A\dot{B}}} \,\, T_{\underline{\mathfrak{I} {\mathfrak r}}} \ , \\[3mm]
A_3^{\mathsf{ \dot A} {\mathfrak r} \mathsf{ \dot B} {\mathfrak s}} &=& \frac{1}{48} \, \delta^{\mathfrak{rs}} [\gamma^{\mathfrak{IJKL}}]_{ \mathsf{\dot A} \mathsf{ \dot B}}\,\, T_{\underline{\mathfrak{IJKL}}} + \frac{1}{2} \,[\gamma^{\mathfrak{IJ}}]_{\mathsf{\dot A}\mathsf{\dot B}} \,\, T_{\underline{\mathfrak{IJ}\mathfrak{rs}}} -2 \, \delta^{\mathsf{ \dot A}\mathsf{ \dot B}} \left[ \delta^{\mathfrak{rs}} \Big( T  - \frac{1}{8} T_{\underline{\mathfrak{qq}}} \Big) +  T_{\underline{\mathfrak{rs}}} \right] \ .
\end{array}
\end{equation}
Equipped with the fermionic mass terms in (\ref{Fermion-Shifts}), the scalar potential in (\ref{V_N=8}) is obtained straightforwardly.

\newpage

\section{Gravitino masses}
\label{app_gravitini_masses}

Here we report the masses of the gravitini for the vacuum configurations of Table \ref{Table:flux_vacua_paper}, Table \ref{Table:flux_vacua_unmagnetised} and Table \ref{Table:flux_vacua_magnetised}. They are presented in Table \ref{Table:flux_vacua_paper_gravitini}, Table \ref{Table:flux_vacua_unmagnetised_gravitini} and Table \ref{Table:flux_vacua_magnetised_gravitini}, respectively.

\vspace{25mm}

\begin{table}[h]
\begin{center}
\scalebox{1}{
\renewcommand{\arraystretch}{1.5}
\begin{tabular}{!{\vrule width 1.5pt}c!{\vrule width 1pt}c!{\vrule width 1pt}c!{\vrule width 1pt}cccccc!{\vrule width 1pt}cc!{\vrule width 1pt}ccc!{\vrule width 1pt}c!{\vrule width 1pt}c!{\vrule width 1.5pt}}
\noalign{\hrule height 1.5pt}
     ID & Gravitini spectrum  \\
\noalign{\hrule height 1pt}
     $ \textbf{vac~1} $ & $g^{-2} m^2_{3/2} = \left[\frac{(\kappa \pm \xi)^{2}}{16}\right]_{(3)} , \left[\frac{9(\kappa \pm \xi)^{2}}{16}\right]_{(1)}$  \\[2mm] 
     \hline 
     $ \textbf{vac~2} $ & \multirow{2}{*}{$g^{-2} m^2_{3/2} = \left(\frac{9 \kappa^{2}}{16}\right)_{(2)} , \left(\frac{\kappa^{2}}{16}\right)_{(6)}$}  \\ 
     \cline{1-1} 
     $ \textbf{vac~3} $ &  \\ 
\noalign{\hrule height 1pt}
     $ \textbf{vac~4} $ & $m^2_{3/2} L^{2} = 1_{(4)} , 9_{(4)}$
     \\[1mm] 
    \hline
     $ \textbf{vac~5} $ &  $m^2_{3/2} L^{2} = 9_{(4)} , 25_{(4)}$ 
     \\[1mm]
\noalign{\hrule height 1pt}
     $ \textbf{vac~6} $ & $m^2_{3/2} L^{2} = 1_{(3)} , 9_{(4)} , 25_{(1)}$  \\[1mm] 
     \hline 
     $ \textbf{vac~7} $ & $m^2_{3/2} L^{2} = 1_{(1)} , 9_{(4)} , 25_{(3)}$  \\[1mm]
\noalign{\hrule height 1pt}
     $ \textbf{vac~8} $ & $m^2_{3/2} L^{2} = 1_{(1)} , 9_{(1)} , 25_{(3)}  , 49_{(3)}$ \\[1mm] 
     \hline 
     $ \textbf{vac~9} $ &  $m^2_{3/2} L^{2} = 9_{(4)} , 25_{(4)}$ \\[1mm]
\noalign{\hrule height 1pt}
     \cellcolor[HTML]{A3C1AD}$ \textbf{vac~10} $ & $m^2_{3/2} L^{2} = 9_{(3)} , 49_{(3)} , 81_{(1)}  , 169_{(1)}$   \\[1mm]
     \hline 
     \cellcolor[HTML]{A3C1AD}$ \textbf{vac~11} $ & $m^2_{3/2} L^{2} = 25_{(6)} , 49_{(1)} , 225_{(1)}$   \\[1mm]
\noalign{\hrule height 1pt} 
     $ \textbf{vac~12} $ &  $m^2_{3/2} L^{2} = 1_{(4)} , 9_{(1)} , 25_{(3)}$  \\[1mm]
     \hline 
     $ \textbf{vac~13} $ & $m^2_{3/2} L^{2} = 1_{(1)} , 9_{(4)} , 49_{(3)}$   \\[1mm]
     \hline
     $ \textbf{vac~14} $ &  $m^2_{3/2} L^{2} = 9_{(7)} , 25_{(1)}$  \\[1mm]
     \hline
     $ \textbf{vac~15} $ &  $m^2_{3/2} L^{2} = 9_{(1)} , 25_{(7)}$ \\[1mm]
\noalign{\hrule height 1.5pt}
\end{tabular}}
\caption{Gravitini masses at the flux vacua of Table~\ref{Table:flux_vacua_paper}. The subscript in $n_{(s)}$ denotes the multiplicity of the mass $n$ in the spectrum. For the AdS$_{3}$ vacua we have normalised the spectrum using the AdS$_{3}$ radius $L^2=-2/V_{0}$.}
\label{Table:flux_vacua_paper_gravitini}
\end{center}
\end{table}

\begin{table}[]
\begin{center}
\scalebox{0.9}{
\renewcommand{\arraystretch}{1.8}
\begin{tabular}{!{\vrule width 1.5pt}c!{\vrule width 1pt}c!{\vrule width 1pt}c!{\vrule width 1pt}cccccc!{\vrule width 1pt}cc!{\vrule width 1pt}ccc!{\vrule width 1pt}c!{\vrule width 1pt}c!{\vrule width 1.5pt}}
\noalign{\hrule height 1.5pt}
     ID & Gravitini spectrum  \\
\noalign{\hrule height 1pt}
     $ \textbf{vac~16} $ & $m^2_{3/2} L^{2} = 1_{(3)} , 9_{(4)} , 25_{(1)}$
     \\[1mm] 
     \hline
     $ \textbf{vac~17} $ &  $m^2_{3/2} L^{2} = 1_{(1)} , 9_{(4)} , 25_{(3)} $ 
     \\
\noalign{\hrule height 1pt}
      $ \textbf{vac~18} $ & $m^2_{3/2} L^{2} = 1_{(1)} , 9_{(1)} , 25_{(3)} , 49_{(3)}$
     \\[1mm] 
     \hline
     $ \textbf{vac~19} $ &  $m^2_{3/2} L^{2} = 9_{(4)} , 25_{(4)}$ 
     \\
\noalign{\hrule height 1pt}
     $ \textbf{vac~20} $ & $m^2_{3/2} L^{2} = 1_{(4)} , 9_{(1)} , 25_{(3)}$
     \\[1mm] 
    \hline
     $ \textbf{vac~21} $ &  $m^2_{3/2} L^{2} = 1_{(1)} , 9_{(4)} , 49_{(3)} $ 
     \\[1mm]
     \hline
     $ \textbf{vac~22} $ & $m^2_{3/2} L^{2} = 9_{(7)} , \ 25_{(1)}$  \\[1mm] 
     \hline 
     $ \textbf{vac~23} $ & $m^2_{3/2} L^{2} = 9_{(1)}, \ 25_{(7)}$  \\
      \noalign{\hrule height 1pt}
    $ \textbf{vac~24} $ & $ m^2_{3/2} L^{2} = 1_{(4)}, \ 9_{(4)}  $
     \\[1mm] 
    \hline
     $ \textbf{vac~25} $ &  $ m^2_{3/2} L^{2} = 9_{(4)}, \ 25_{(4)} $ \\
\noalign{\hrule height 1pt}
    $ \textbf{vac~26} $ & $  m^2_{3/2} L^{2} = \left[ \frac{\left( 4+4\alpha^2 \pm \sqrt{4+2\alpha^2 + \alpha^4}\right)^2}{4+2\alpha^2 + \alpha^4} \right]_{(4)} $
     \\[1mm] 
    \hline
     $ \textbf{vac~27} $ &  $ \begin{array}{ccl} m^2_{3/2} L^{2} &=& \left[ \frac{\left( 8+6\alpha^2\pm\sqrt{4+2\alpha^2 + \alpha^4}\right)^2}{4+2\alpha^2 + \alpha^4} \right]_{(2)} ,
    \\ & &  \left[ \frac{\left( 8+4\alpha^2 \pm \sqrt{4 + 5\alpha^4 + 2\alpha^2 \left( 1-2\sqrt{4+2\alpha^2+\alpha^4} \right)} \right)^2}{4+2\alpha^2 + \alpha^4} \right]_{(1)} , \\ & & \left[ \frac{\left( 8+4\alpha^2 \pm \sqrt{4 + 5\alpha^4 + 2\alpha^2 \left( 1+2\sqrt{4+2\alpha^2+\alpha^4} \right)} \right)^2}{4+2\alpha^2 + \alpha^4} \right]_{(1)} 
    \end{array} $ \\[5mm]
\noalign{\hrule height 1pt}
    $ \textbf{vac~28} $ & $\begin{array}{ccl} m^2_{3/2} L^{2} &=& \left[ \frac{\left( 4+4\alpha^2-\sqrt{4+2\alpha^2 + \alpha^4}\right)^2}{4+2\alpha^2 + \alpha^4} \right]_{(4)} , \left[ \frac{\left( 8+6\alpha^2+\sqrt{4+2\alpha^2 + \alpha^4}\right)^2}{4+2\alpha^2 + \alpha^4} \right]_{(2)} ,
    \\ & &  \left[ \frac{\left( 8+4\alpha^2 \pm \sqrt{4 + 5\alpha^4 + 2\alpha^2 \left( 1+2\sqrt{4+2\alpha^2+\alpha^4} \right)} \right)^2}{4+2\alpha^2 + \alpha^4} \right]_{(1)}
    \end{array}$
     \\[1mm] 
    \hline
     $ \textbf{vac~29} $ &  $\begin{array}{ccl} m^2_{3/2} L^{2} &=& \left[ \frac{\left( 4+4\alpha^2+\sqrt{4+2\alpha^2 + \alpha^4}\right)^2}{4+2\alpha^2 + \alpha^4} \right]_{(4)} , \left[ \frac{\left( 8+6\alpha^2-\sqrt{4+2\alpha^2 + \alpha^4}\right)^2}{4+2\alpha^2 + \alpha^4} \right]_{(2)} ,
    \\ & &  \left[ \frac{\left( 8+4\alpha^2 \pm \sqrt{4 + 5\alpha^4 + 2\alpha^2 \left( 1-2\sqrt{4+2\alpha^2+\alpha^4} \right)} \right)^2}{4+2\alpha^2 + \alpha^4} \right]_{(1)}
    \end{array}$
     \\
\noalign{\hrule height 1pt}
 \cellcolor[HTML]{89CFF0}$ \textbf{vac~30} $ &  \multirow{2}{*}{$\begin{array}{ccl}
        m_{3/2}^2 L^2 &=&  \left[237+\frac{256}{x_2^4}\pm\frac{64\sqrt{2x1}-96}{\sqrt{2x_1}x_2^3}-\frac{488\sqrt{2x_1}+12}{x_2^2\sqrt{2x_1}}\mp\frac{60\sqrt{2x_1}-90}{x_2\sqrt{2x_1}}\right]_{(1)}\, ,  \\
        & & \left[ \frac{1}{9}\left(373+\frac{256}{x_2^4}\pm\frac{192\sqrt{2x_1}+96}{\sqrt{2x_1}x_2^3}-\frac{584\sqrt{2x_1}-36}{\sqrt{2x_1}x_2^2}\mp\frac{228\sqrt{2x_1}+114}{\sqrt{2x_1}x_2}\right)\right]_{(3)}
        \end{array}$} \\
        
     \cline{1-1} 
      \cellcolor[HTML]{89CFF0}$ \textbf{vac~31} $ & \\[2mm]
     \noalign{\hrule height 1pt}
 \cellcolor[HTML]{89CFF0}$ \textbf{vac~32} $ &  \multirow{2}{*}{$\begin{array}{ccl}
        m_{3/2}^2 L^2 &=&  \left[237+\frac{256}{x_2^4}\pm\frac{64\sqrt{2x1}+96}{\sqrt{2x_1}x_2^3}-\frac{488\sqrt{2x_1}-12}{x_2^2\sqrt{2x_1}}\mp\frac{60\sqrt{2x_1}+90}{x_2\sqrt{2x_1}}\right]_{(1)}\, ,  \\
        & & \left[ \frac{1}{9}\left(373+\frac{256}{x_2^4}\pm\frac{192\sqrt{2x_1}-96}{\sqrt{2x_1}x_2^3}-\frac{584\sqrt{2x_1}+36}{\sqrt{2x_1}x_2^2}\mp\frac{228\sqrt{2x_1}-114}{\sqrt{2x_1}x_2}\right)\right]_{(3)}
        \end{array}$} \\[2mm]
        
     \cline{1-1} 
     \cellcolor[HTML]{89CFF0}$\textbf{vac~33} $ & \\
\noalign{\hrule height 1pt}
$ \textbf{vac~34} $ &  $\begin{array}{ccl}
        m_{3/2}^2 L^2 &=&  \left[\frac{2 \left(y_1^2-\left(4 y_2\pm y_3-2\right) y_1\pm y_2 y_3\right){}^2}{y_1 \left(3 y_1^3+\left(6 y_2^2+4 y_2+4\right) y_1+4 y_2^3\right)}\right]_{(3)} \, , \left[\frac{2 \left(y_1 \left(2-3 y_1+4 y_2\right)\pm\left(3 y_1+y_2\right) y_3\right){}^2}{y_1 \left(3 y_1^3+\left(6 y_2^2+4 y_2+4\right) y_1+4 y_2^3\right)}\right]_{(1)}   \end{array}$\\[2mm]
\noalign{\hrule height 1.5pt}
\end{tabular}}
\caption{Gravitini normalised masses at the AdS$_{3}$ flux vacua of Table~\ref{Table:flux_vacua_unmagnetised}. The subscript in $n_{(s)}$ denotes the multiplicity of the mass $n$ in the spectrum.}
\label{Table:flux_vacua_unmagnetised_gravitini}
\end{center}
\end{table}

\begin{table}[]
\begin{center}
\scalebox{0.9}{
\renewcommand{\arraystretch}{1.9}
\begin{tabular}{!{\vrule width 1.5pt}c!{\vrule width 1pt}c!{\vrule width 1pt}c!{\vrule width 1pt}cccccc!{\vrule width 1pt}cc!{\vrule width 1pt}ccc!{\vrule width 1pt}c!{\vrule width 1pt}c!{\vrule width 1.5pt}}
\noalign{\hrule height 1.5pt}
     ID & Gravitini spectrum  \\
\noalign{\hrule height 1pt}
     \cellcolor[HTML]{E25822}$ \textbf{vac~35} $ & \multirow{4}{*}{$ m^2_{3/2} L^{2} = \left( \frac{373 \,\pm\, 76\sqrt{3}}{9}\right)_{(3)} \,\,\, , \,\,\,  3\left(79\pm 20 \sqrt{3} \right)_{(1)}  $} \\
     \cline{1-1}
     \cellcolor[HTML]{E25822}$ \textbf{vac~36} $ & \\
     \cline{1-1}
     \cellcolor[HTML]{E25822}$ \textbf{vac~37} $ & \\
     \cline{1-1}
     \cellcolor[HTML]{E25822}$ \textbf{vac~38} $ &  \\[1mm]
     \noalign{\hrule height 1.5pt}
 \cellcolor[HTML]{E25822}$ \textbf{vac~39} $ & \multirow{2}{*}{$\begin{array}{ccl}
     m^2_{3/2} L^{2} &=& \left[ \frac{2\left( 1+2c_1+2c_1^2\pm c_2\right)^2}{ 3-3c_1-3c_1^2-8c_1^4 } \right]_{(3)}, \left[ \frac{2\left( 3+6c_1-2c_1^2\pm 3c_2\right)^2}{3-3c_1-3c_1^2-8c_1^4} \right]_{(1)}
     \end{array}$} \\
     \cline{1-1} 
     \cellcolor[HTML]{E25822}$ \textbf{vac~40} $ &  \\
\noalign{\hrule height 1pt}
     $ \textbf{vac~41} $ & \multirow{2}{*}{$\begin{array}{ccl}
     m^2_{3/2} L^{2} &=& \left[ \frac{\left( 21 + 31 c_3 + 10 c_3^2\right)^2}{ 108 + 252 c_3 + 189 c_3^2 + 60 c_3^3 + 4 c_3^4 } \right]_{(3)}, \left[ \frac{9\left( 9 + 11 c_3 + 2 c_3^2\right)^2}{ 108 + 252 c_3 + 189 c_3^2 + 60 c_3^3 + 4 c_3^4 } \right]_{(1)} , \\[2mm]
      & &
      \left[ \frac{9\left( 7 + 7 c_3 + 2 c_3^2\right)^2}{ 108 + 252 c_3 + 189 c_3^2 + 60 c_3^3 + 4 c_3^4 }  \right]_{(3)}, \left[ \frac{\left( 27 + 27 c_3 + 10 c_3^2\right)^2}{ 108 + 252 c_3 + 189 c_3^2 + 60 c_3^3 + 4 c_3^4 } \right]_{(1)} 
     \end{array}$} \\[4mm]
     \cline{1-1} 
     $ \textbf{vac~42} $ &  \\
\noalign{\hrule height 1.5pt}
\end{tabular}}
\caption{Gravitini normalised masses at the AdS$_{3}$ flux vacua of Table~\ref{Table:flux_vacua_magnetised}. The subscript in $n_{(s)}$ denotes the multiplicity of the mass $n$ in the spectrum.}
\label{Table:flux_vacua_magnetised_gravitini}
\end{center}
\end{table}

\newpage

\bibliographystyle{JHEP}
\bibliography{references}

\end{document}